\pdfoutput=1
\documentclass[aps,pre,twocolumn,showpacs,amsmath,amssymb]{revtex4-1}
\usepackage{graphicx,dcolumn,bm,verbatim,hyperref,subfigure,amsthm,hyperref,color,cleveref}
\usepackage{longtable}

\theoremstyle{plain}  

\makeatletter
\newcommand{\printfnsymbol}[1]{%
  \textsuperscript{\@fnsymbol{#1}}%
}
\makeatother

\begin{document}


\title{Stochastic Replica Voting Machine Prediction of Stable Cubic and Double Perovskite Materials and Binary Alloys} 
\author{T. Mazaheri} \email{Equal Contribution}
\affiliation{Department of Physics, Washington University in St.
Louis, MO 63160, USA}
\author{A. S. Thind}
\email{Equal Contribution}
\affiliation{Institute of Materials Science and Engineering, Washington University in St. Louis, St. Louis, MO 63130, USA}
\author{Bo Sun} \email{Equal Contribution}
\affiliation{Department of Physics, Washington University in St.
Louis, MO 63160, USA}
\author{J. Scher-Zagier}
\affiliation{Department of Physics, Washington University in St.
Louis, MO 63160, USA}
\author{D. Magee}
\affiliation{Department of Physics, Washington University in St.
Louis, MO 63160, USA}
\affiliation{MIT Lincoln Laboratory, Lexington, Massachusetts 02421, USA}
\author{P. Ronhovde}
\affiliation{Department of Physics, Washington University in St.
Louis, MO 63160, USA}
\affiliation{Department of Physical Sciences, The University of Findlay, 1000 N. Main St., Findlay, Ohio 45840, USA}
\author{T. Lookman}
\affiliation{Theoretical Division, Los Alamos National Laboratory, MS-B262, Los Alamos, New Mexico 87545, USA}
\author{R. Mishra}
\email{rmishra@wustl.edu}
\affiliation{Department of Mechanical Engineering and Materials Science, and Institute of Materials Science and Engineering, Washington University in St. Louis, St. Louis, MO 63130, USA}
\author{Z. Nussinov}
\email{zohar@wuphys.wustl.edu}

\affiliation{Department of Physics, Washington University in St. Louis,
Campus Box 1105, 1 Brookings Drive, St. Louis, Missouri 63130, USA}%

\date{\today}

\begin{abstract}

A  machine learning approach that we term  the ``Stochastic Replica Voting Machine'' (SRVM) algorithm is presented and applied to a binary and a 3-class classification problems in materials science. Here, we employ SRVM to predict candidate compounds capable of forming stable perovskites and double perovskites and further classify binary ($AB$) solids. The results of our binary and ternary classifications compared well to those obtained by SVM and neural network algorithms. 
\end{abstract}


\maketitle{}

\section{Introduction}

Recently, there has been a flurry of activity involving the use of machine learning, an important subfield of artificial intelligence, in the study of materials 
and complex physical systems, e.g.,  \cite{Muller,2011,2012a,statistical,2012b,2016,eun-ah,2017a,2017b,maciek}.
Data mining techniques enable a rapid search through millions of candidate compounds in order to identify promising technological materials and to potentially predict their detailed properties. Such a task may require far more significant efforts when performed experimentally \cite{Uconn,double perovskites} via the traditional trial and error approach. Machine learning can make such searches far more efficient by systematically pointing to promising materials that may then be fabricated and tested experimentally. In this publication, we will focus on two material types: perovskites and binary alloys. 

\begin{figure}
   \includegraphics[width=\linewidth]{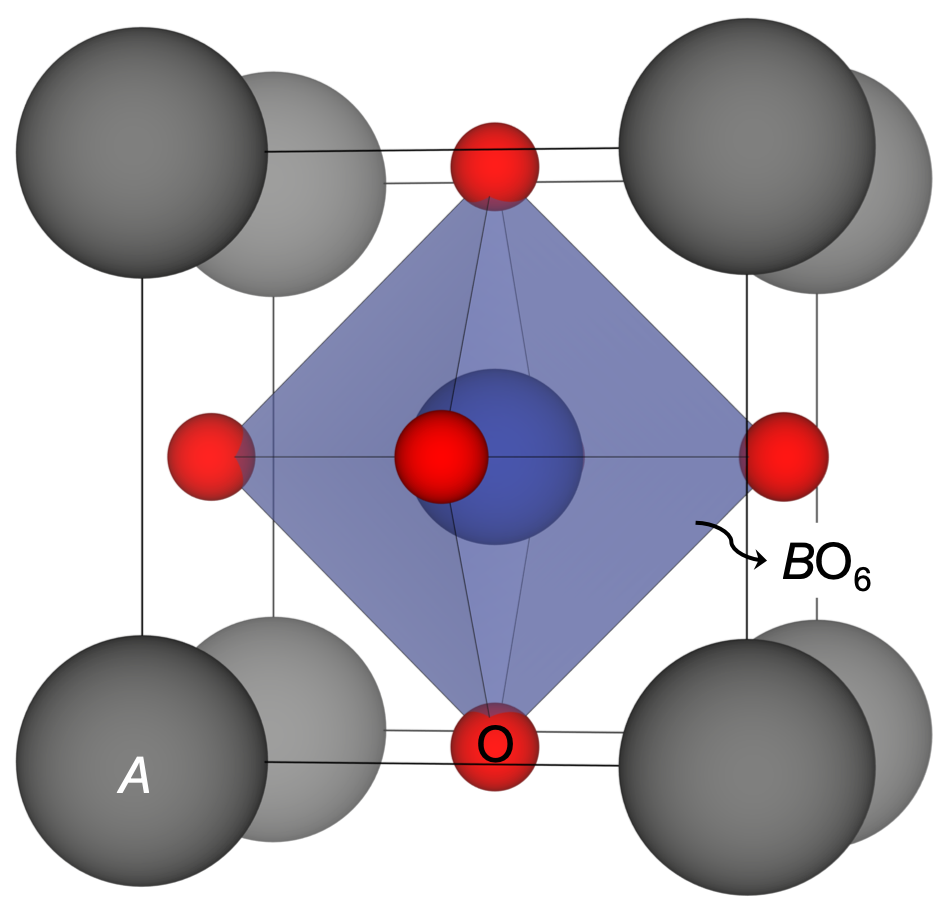}
  \caption{ The structure of an ABO$_3$ Perovskite. 
  A quintessential material having this structure is CaTiO$_3$.}
  \label{fig:Perovskite}
\end{figure}

Perovskites (��named after Russian nobleman and mineralogist Lev Perovski) ��are a large 
class of compounds having an ABX$_{3}$ stoichiometry, where A and B are cations and X is an anion \cite{BO-1,Perovskite Structure}. Numerous technologically important materials display the perovskite structure shown in Fig. \ref{fig:Perovskite}  Some examples include certain high-temperature superconductors, semiconductors for high-efficiency photovoltaic cells  \cite{Perovskite solar}, light-emitting diodes 
,lasers, and solid-oxide fuel cells (see, e.g., Refs. \cite{BO-2,BO-3}).

\begin{figure}[ht!]
\centering
\includegraphics[width=90mm]{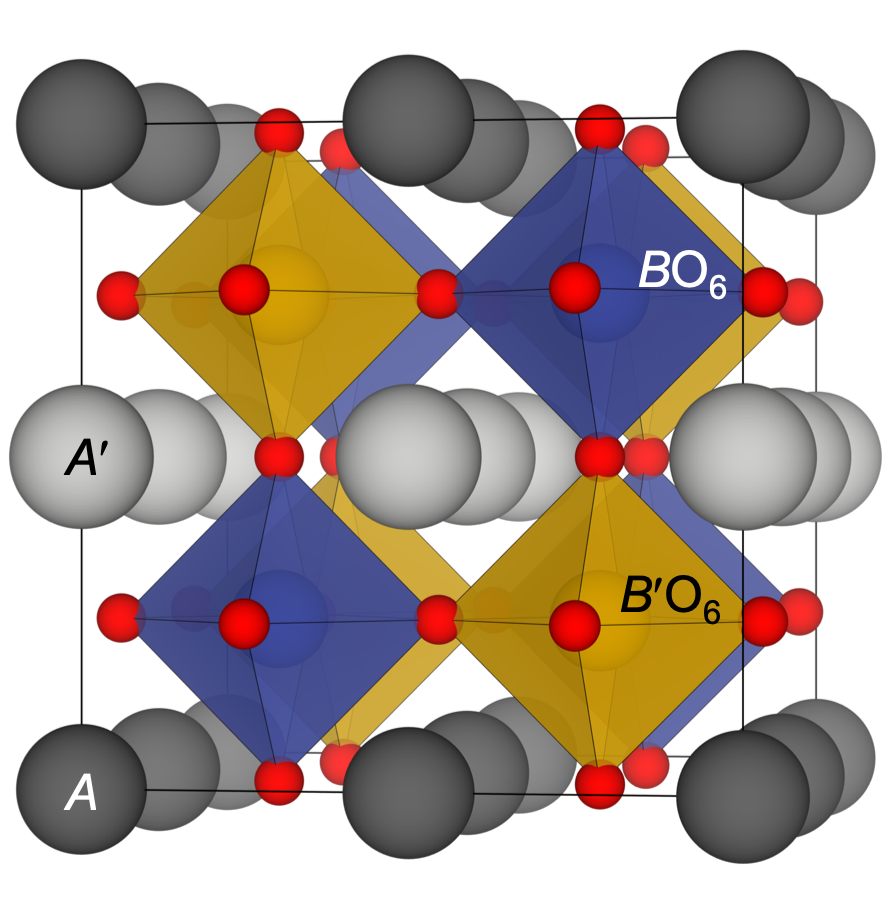}
\caption{A double perovskite Structure with two different cations at both the A-site and the B-site.  
 \label{double-perovskite-figure}}
\end{figure}

In the examples that we will study here, X will be an oxygen anion. Following a standard convention, the A atoms are defined to be the larger of the two cations. An ideal perovskite has a cubic crystal structure that is formed by corner-sharing BO$_{6}$ octahedra as seen in Fig. \ref{fig:Perovskite}. As is seen in this figure, A ions lie at the corners of the cube while the B and O ions are, respectively, located at the body-center and face-centers of the cube. The BO$_{6}$ octahedra can distort and neighboring octahedra can tilt and rotate. This lowers the cubic crystal symmetry but accommodates a large combination of cations from the Periodic Table.
To ensure stability, the relative size of the A and B cations must, typically, satisfy certain criteria \cite{formability}. (Additional illuminating relations between the atomic radii and structure are found in \cite{GdFeO3}.) More complex double perovskites \cite{double perovskites} (see Fig. \ref{double-perovskite-figure}) exhibit the same architecture yet with a larger unit cell; these compounds are of the generic chemical composition A$'$A$''$B$'$B$''$O$_{6}$ (or, more generally,
A$'_{y'}$A$''_{2-y'}$B$'_{z'}$B$''_{2-z'}$O$_{6}$ with $0 <y',z'<2$). Fig. \ref{double-perovskite-figure} provides an illustration of a double perovskite with two different A cations and B cations. There are over 80 elements having, at least, one stable nuclide. Thus, a priori, numerous combinations of these elements may potentially realize stable perovskite or double perovskite structures. To experimentally determine the stability of the vast number of these candidate perovskites would be an arduous if not impossible task. In recent years, materials scientists have turned to machine-learning models, often combined with first-principles total energy calculations based on density-functional theory (DFT), to predict the stability of new theoretical compounds. Along similar lines, in the current work, we will introduce an algorithm that takes in different elements as inputs and predicts whether or not their combination will result in a stable perovskite or double perovskite.  

\begin{figure}
 \includegraphics[width=\linewidth]{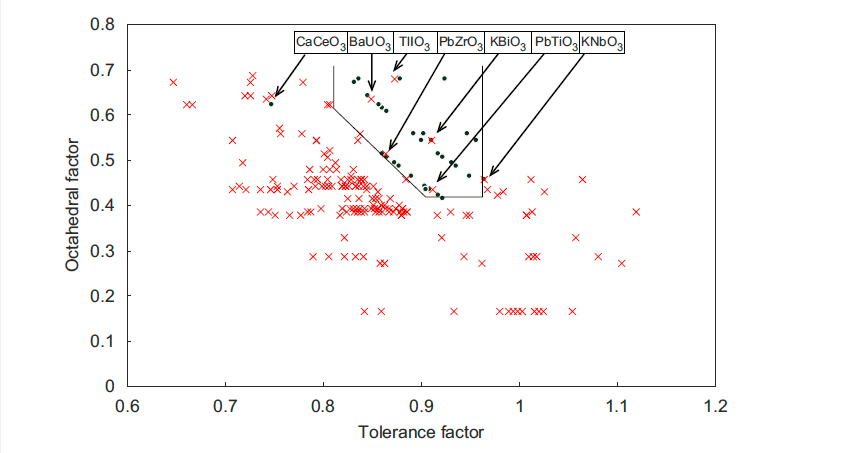}
 \caption{Color online. Reproduced from \cite{formability}. Classification of cubic perovskite oxides. Candidate perovskite compounds are displayed according to their tolerance and octahedral factors. (Black) dots indicate stable perovskite compounds; compositions marked by (red) crosses indicates unstable compounds. 
 A goal of our work is to predict which materials might be stable.} 
  \label{222}
 \end{figure}

Our algorithm takes, as an input, the data from known combinations of A and B cations that are capable of forming a perovskite structure, see, e.g., \cite{formability} and Fig. \ref{222} for perovskite formability and Appendix \ref{Sec:D} and Figure \ref{double-perovskite-figure} for double perovskite. From these data, the algorithm learns which conditions should be met for the different elements in order to allow them to form the perovskite structure. This is the so-called ``training process''. Following the training phase, the algorithm may test other combinations of the ions; the algorithm will then yield a ``Yes" answer for a predicted stable perovskite structure and yield ``No" for compositions that are predicted to form an unstable perovskite structure. In the parlance of machine learning, we are training a new binary classifier over a set of known data. Once this training is complete, we then apply the trained classifier to investigate hitherto unknown chemical compositions in order to assess their formability as stable perovskites. We will follow the prevalent practice of classifying the stability of candidate perovskite materials by two well studied ratios: (i) the ``tolerance factor" and (ii) the ``octahedral factor".  We will further study other features including electronegativity. 

In the current work, we will introduce and summarize our new algorithm (general details are further discussed in Ref. \cite{patrick}), and demonstrate its utility for the classification (viable formability) of (1) perovskite-type compounds and (2) binary octet alloys. In both cases, we achieve high accuracy. Our method enables the prediction of new
stable perovskites and the properties of binary compounds. Other works, e.g., \cite{machine1+,machine2+,Turab} study various aspects of perovskites with existing machine learning algorithms. In the current work, we employed a new and very general machine learning algorithm (whose details will be reported on in \cite{patrick}) and delineated new phase boundaries in the two classification problems that we investigated. 

Our bare binary classifier can be trivially extended to non binary (multi-class) problems via, e.g., the ``One-Versus-Rest'' approach \cite{Ryan Rifkin}. We will detail a 3-class problem when investigating binary (``AB'') alloys. 

The remainder of this work is organized as follows: In Section \ref{rvm1}, we provide a description of our algorithm. In subsections \ref{Gaussian} and \ref{binary} therein, we will, respectively, outline the Gaussian and multinomial variants of our algorithm. We will next explain (Section \ref{Perovskite}), how we train the algorithm to ascertain perovskite formability. In Section \ref{sec:NND}, we apply a neural network analysis based version of our approach to the study of double perovskites  when only the tolerance and octahedral factors are provided.  In Section \ref{Section:5feature}, we will apply both neural network, and Gaussian kernel based version of our algorithm to the study of double perovskite formability when five features (that further incorporate ionic electronegatives) are included. With these analysis we may make predictions as to which candidate systems might be stable perovskites. In Section \ref{sec:DFT}, we contrast the SRVM predictions for stable double perovskites with detailed DFT calculations for the enthalpy of formation. Finally, in Section \ref{CAB}, we will invoke the ``One-Versus-All'' approach to a ternary classification problem involving $AB$ solids \cite{AB}.

\section{The Stochastic Replica Voting Machine algorithm}
\label{rvm1}

As befits its name, our ``Stochastic Replica Voting Machine'' (SRVM) algorithm relies on a voting procedure among stochastically generated classifiers. As we will explain, these individual classifiers are defined by a kernel that may be of any type: e.g., a sum of Gaussians or a multinomial. Initially, we ``train'' the system to predict the correct answer. The trained system may then subsequently predict
the outcome given initial inputs. Training is performed by adjusting the kernel of each individual classifier such that it reproduces known results. The voting of classifiers is then given new data and
a vote is taken amongst the predictions of the individual classifiers. 

The input (``training set'') data for $N$ items that need to be classified is given in terms of a set of a vectors $\{\vec{v}_{i}\}_{i=1}^{N}$ defining the features of the items and their corresponding classification $\rho_{i}$. If the classification is amongst $q$ different groups, then classification function is a Potts spin variable whose value $\rho_{i} =1, 2, \cdots, q$ denotes the group that item $i$ correctly belongs to. Potts variables may be generally used as a classification index in numerous arenas, e.g., \cite{Shai,Ronhovde1,Ronhovde2,Ronhovde3}.  The features of each item are combined into a vector $ \vec{v} = (v_{1}, v_{2}, \cdots, v_{d})$. 
Thus, the Cartesian components of each vector $\vec{v}_{i}$ are equal to the values of all parameters of the input data associated with item $i$ (e.g., the values of the individual atomic radii of the ions forming in a candidate perovskite material). If numerous features
are given for each data point $i$, then the dimensionality ($d$) of the vectors $\vec{v}_i$ will be high. 
The goal of machine learning is to make an educated guess (a ``prediction'') as to what
the corresponding classification outcome will be for a new vector $\vec{v}$ for which there is no a priori correct classification known outcome. Since no additional information is available, the predicted outcome $\rho$ can only be some function $F$ of all supplied input: the features defining $\vec{v}$ and all known training set data. That is, the underlying assumption of any 
machine learning approach is that
\begin{eqnarray}
\rho(\vec{v}) = F(\vec{v};\{\vec{v}_{i}\}_{i=1}^{N}, \{\rho_{i}\}_{i=1}^{N}).  
\end{eqnarray}
The natural question is ``how may we determine the correct or `optimal' function $F$''? Numerous machine learning approaches exist. We briefly comment on two of these. In one important subclass of these, known as ``Support Vector Machines'' (SVM), e.g., \cite{SVM,SVM1}, $F$ is implicitly ascertained by inequalities applied to assumed specific function types. In neural network based machine learning \cite{neuron1,neuron2}, in particular in ``deep learning'' \cite{DL}, the function $F$ is formed by a particular hierarchal recursive structure. Our approach (SRVM) is, in some regards, more rudimentary yet, as we will explain, may extend these and other prevalent models. To illustrate its basic premise, we will consider the binary (i.e., $q=2$) classification problem. Here, $\rho_{i} = 1,2$ and thus $\tilde{\rho}_{i}  \equiv (2\rho_{i} -3) = \pm 1$ naturally classifies any data point $\vec{v}$ into one of two groups 
(labelled by $\tilde{\rho}_{i} = 1$ and $\tilde{\rho}_{i} =-1$). We define $\tilde{F} \equiv (2F-3)$ and initially
consider $\tilde{F}$ to be an outcome of a vote amongst the predictions of a large {\it voting} of general continuous stochastic functions $\{G_{a}\}_{a=1}^{r}$ (that we need not be of different types) where $r$ is the number
of ``replicas'' in this voting. We will, principally, focus on two broad types of stochastic functions. We will examine what occurs (a) if  $\{G_{a}\}_{a=1}^{r}$ are expressible as sums of random basis functions. We then turn to 
(b) functions $\{G_{a}\}_{a=1}^{r}$ generated by a weighted averages of ``neural network type function''; by the latter, we allude to functional composition of linear and Fermi function (also known as a ``sigmoid'' function in the machine learning community). Apart from SVM, related machine learning approaches include``voting'' methods \cite{voting} which employ randomly generated data sets, ``boosting'' \cite{boost} which aims to combine different weaker algorithms into a stronger learners, and ``decision tree learners''  \cite{tree} and ``random forest'' methods \cite{forest} which employ decision trees to combine the results of various classifiers and often do so while choosing different subspaces of the features or given data. While our algorithm enjoys many commonalities with these and other approaches, we underscore that its essential character is that of a stochastic average over different randomly generated functions. Basically, we perform (a discrete sum version of) ``functional integration'' over different randomly generated fits to the given data set in order to suggest the most likely outcomes for given vectors $\vec{v}$. In performing these ``functional integration'' averages, the SRVM algorithm that we briefly introduce below does not discard any data points nor features to generate lower dimensional subspaces on which an voting of classifiers is trained on. 

We start with stochastic functions of type (a) and expand in terms of simple (equivalent type) non-orthogonal randomly chosen basis functions. A simple choice for the functions $G_a$ (that will be investigated in the current work) is one in which they are a sum of $R$ random Gaussian functions \cite{explain-Gauss}.
Thus, we set
\begin{eqnarray}
\label{GaussG}
G_{a} = \sum_{j=1}^{R} c_{j a} e^{-(\vec{v}-\vec{v}_{ja})^{2}/(2 \sigma_{ja}^2)},
\end{eqnarray}
where $\{c_{ja}\}_{j=1}^{R}$ are coefficients that we will discuss momentarily.
In the most minimal form of $G_a$, all standard deviations $\sigma_{ja}$ are set to a uniform fixed
value, $\sigma_{ja} = \sigma$. The centers  $\{\vec{v}_{ja}\}$ of the Gaussians are randomly
chosen in the volume spanned by the feature space. Thus, for each of the $r$ functions $\{G_a\}_{a=1}^{r}$, we randomly choose $R$ ``anchor points'' in the feature space volume to be $\{\vec{v}_{ja}\}_{j=1}^{R}$.
The location of these anchor points differs from replica to replica. That is, we define each replica ``A'' by a different stochastic set of vectors $\{\vec{v}_{ja}\}_{j=1}^{R}$. 
More comprehensive than the specific choice of random Gaussians in Eq. (\ref{GaussG}), 
the functions $G_a$ may be generally chosen to be 
of the form 
\begin{eqnarray}
G_a = \sum_{j=1}^{R} c_{ja} K_{a}^{j}(\vec{v}).
 \label{kernelG}
 \end{eqnarray}
 Here, the kernel (or basis) functions $K_{a}^{j}$ could be any arbitrary stochastic functions.  For the Gaussian form of Eq. (\ref{GaussG}),
 $K_{a}^{j}(\vec{v})= e^{-(\vec{v}-\vec{v}_{ja})^{2}/(2 \sigma_{ja}^2)}$.
Other general kernels $K$, different from a Gaussian function, may, of course, be considered. For instance, another natural (yet typically computationally expensive) choice for the kernel $K_{a}$ that we will return to in the current work  
 (reasonable when the outcome likelihood is expected to be analytic 
 as a function of the features) is that of multinomials in the Cartesian components of $\vec{v}$.  

 During the training phase, we optimize the values of the coefficients $\{c_{ja}\}_{j=1}^{R}$ given the known outcome for the training points $i=1,2, \cdots, N$ such that $G_{a}(\vec{v}_{i})$ matches the correct classification $\tilde{\rho}_{i}$. The number $R$ of the coefficients required in order to achieve high prediction accuracy, is typically smaller than the number of training data points, $R <N$ (in most instances, in fact, $R \ll N$). The optimal value of $R$ depends on the nature of data as well as the size of data and should be chosen carefully to avoid over-fitting. For each replica, $a=1, 2, \cdots, r$, the given training data set translates into linear equations for $\{c_{ja}\}_{j=1}^{R}$. Thus, Eq. (\ref{kernelG}) explicitly reads
$ G_a(\vec{v}= \vec{v}_{i}) = \sum_{j=1}^{R} K_{a}^{ij} c_{ja}$, 
 where $K_{a}^{ij} \equiv K_a^j(\vec{v}=\vec{v}_{i})$. This embodies a set of overdetermined (since $N > R$) linear equations for the coefficients $\{c_{ja}\}$. 
 For each of the replicas $a=1, 2, \cdots, r$, the above relation can be trivially cast as an explicit matrix equation,
$ \hat{G}_a=  \hat{K}_{a} \hat{c}_a$.
 Here, $\hat{G}_a$ and $\hat{c}_a$ are two column vectors of, respectively, lengths $N$ and $R$ whose entries are, respectively, $\{ G_{a}(\vec{v} = \vec{v}_{i})\}_{i=1}^{N}$ and $\{c_{ja}\}_{j=1}^{R}$. 
The elements of the rectangular $N \times R$ dimensional matrix $K_a$ are, as defined above, given by $(\hat{K}_{a})^{ij} \equiv K_{a}^{j}(\vec{v}=\vec{v}_i)$. The coefficients $\hat{c}_{a}$ minimizing the cost function or ``energy'' defined by the square sum
$|| \hat{G}_{a} - \hat{K}_{a} \hat{c}_{a} ||^2$ are given by 
\begin{eqnarray}
\label{Kinverse}
c_{ja} = \sum_{i}(\hat{K}_{a}^{-1})_{ji} G_{ia}.
\end{eqnarray}

\begin{figure}
 \includegraphics[width=\linewidth]{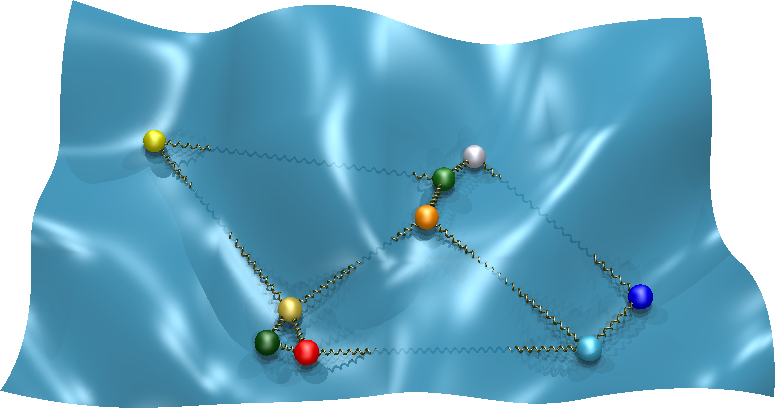}
 \caption{A schematic representation of replica ``interactions''. The spheres depict individual replicas that navigate an ``energy landscape'' looking for stable minima while simultaneously interacting with one another. As compared to a single solver. In the simplest setting, these ``interactions'' may correspond to a vote amongst their predictions and notably when seeing at which parameter values, the predictions of the replicas are most uniform and thus robust.
 These ``interacting'' replicas may more readily avoid false minima and converge on the stable low energy solutions leading more stable and accurate predictions. In the algorithm that we outline in the current work, the simplest multi-replica vote of Eq. (\ref{sgnsgn}) is employed. We will further optimize the function types to fit the by seeing when the replica predictions are most uniform.}
  \label{peter}
 \end{figure}

Here, the rectangular matrix $\hat{K}_a^{-1}$ (with elements $(\hat{K}_{a}^{-1})_{ji}$)
is the pseudoinverse of $\hat{K}_a$. Thus, in the training phase, the goal is to find the coefficient vectors $\hat{c}_{a}$ for each of the replicas $a=1,2, \cdots, r$. 
With the above values of $c_{ja}$ in tow, we may now predict
the classification of a new ``test'' item $\vec{v}$ different from all prior training data points (i.e., $\vec{v} \neq \vec{v}_{i}$ for $1 \le i \le N$). 
That is, we may compute the classification of $\vec{v}$ as predicted by the $r$ independent replicated stochastic functions,
$\{sgn(G_{a}(\vec{v}))\}_{a=1}^{r}$ (where $sgn$ denotes the sign function) and then perform a vote amongst all of these classifiers.  
The vote then yields the final prediction of the SVRM,
\begin{eqnarray}
\label{sgnsgn}
\tilde{\rho}(\vec{v})= sgn (\sum_{a=1}^{r} sgn(G_{a}(\vec{v}))).
\end{eqnarray}
For the $q=2$ classification problem that we have considered thus far, the inner $sgn$ function in Eq. (\ref{sgnsgn}) may be replaced by other appropriately chosen symmetric functions $\tilde{\rho}(\vec{v})=W(\{G_{a}(\vec{v})\})$ with the condition that may only assume the two values $\pm 1$ (corresponding to the two possible classes to which an item $\vec{v}$ may belong to). The voting in Eq. (\ref{sgnsgn}) emulates a more general multi-replica ``interaction'' sketched in Fig. (\ref{peter}). In this schematic, each sphere denotes an individual replica. Together, the voting of replicas may better hone in on optimal predictions for the classification of $\vec{v}$. 

Putting all of the pieces together, Eqs. (\ref{kernelG},\ref{Kinverse},\ref{sgnsgn}) nearly completely define the SRVM program. The kernels $K_{a}^{j}$ may, a priori, stochastically be chosen to be of any particular functional form. 
Of course, if a theory exists then the functional form of $G_a$ may be more efficiently designed. In the absence of any such information,
one may simply examine the predictions for random kernels $K_a^j$. There are three remaining inter-related natural questions: \newline

{\bf{(1)}} Is there a particular metric to determine the confidence with which the results are predicted? \newline

{\bf{(2)}} How do we determine the `optimal' number $r$ of the replicas to be used? \newline

{\bf{(3)}} Similarly, what sets the number $R$ of kernel functions in Eq. (\ref{kernelG})? \newline

As we will describe, the answer to all questions may be determined by examining the overlap of the predictions of the different stochastic replica functions $\{G_{a}\}_{a=1}^{r}$.
Throughout the current work, we will employ a simple variant of the overlap ${\cal{O}}(\vec{v})$ associated with any point $\vec{v}$ whose classification is predicted by the $r$ replicas $\{G_{1}, G_{2}, \cdots, G_{r}\}$, namely 
\begin{eqnarray}
\label{overlap}
{\cal{O}}(\vec{v}) \equiv \frac{1}{r} | \sum_{a=1}^{r}   sgn(G_{a}(\vec{v}))|.
\end{eqnarray}
With this definition, we first explicitly turn to question {\bf{(1)}}. If all replicas yield identical predictions (and thus ${\cal{O}}$ is close to unity), 
then (as is intuitively natural and we verified by numerical experiments), this common predicted answer is likely correct. 
Analogously, if the replicas are far from a unanimous agreement about the correct classification (and, consequently, ${\cal{O}}$ is much smaller than one) then the predicted answer cannot be trusted
with high confidence. The above rule of thumb enables us to scan the parameters $r$ and $R$ to find values that are likely to yield optimal accuracy (questions {\bf{(2)}} and {\bf{(3)}} above).
Typically as the number of replicas $r$ increases so does the accuracy. However, larger values of $r$ entail increasing computational costs with no real benefit.  We thus seek sufficiently large $r$ 
that enable high accuracy. By contrast, when the number of anchor points (or more general basis functions) $R$ becomes too large, overfitting leads to increasing errors. 
There are optimal values of $R$ that are sufficiently large to capture the characteristics of the data yet not so big that overfitting occurs. 
In reality, we may fix $r$ and $R$ to specific values and examine the replica overlap to ascertain whether the predicted values may be trusted \cite{patrick}.
That is, when the overlap ${\cal{O}}$ is averaged over all new data points $\vec{v}$ (whose correct classification is not a priori known and that need to be classified by the algorithm)
is high, then the consensus reflected by the average ${\cal{O}}$ will suggest that the current parameters $r$ and $R$ defining Eqs. (\ref{kernelG}, \ref{overlap}) enable a correct prediction of the classification problem. 

A variant that we will touch on later is that of ``an expansion in a box''. For typical basis functions $K_{a}^{j}$, the functional form 
of Eq. (\ref{kernelG}) assumes that the outcome is a generally smooth function of $\vec{v}$. 
 If the system exhibits ``phase transitions'' as a functions of the features $(v_{1}, v_{2}, \cdots, v_{d})$
 then such an assumption is void. Instead, one may fit the training data with a particular function
 of the form of Eq. (\ref{kernelG}) with specific coefficients $\{c_{ja}\}$ only when $\vec{v}$ lies in
 a particular volume, $\vec{v} \in \Omega$; different regions will correspond to different functional forms
 (i.e., the coefficients $\{c_{ja}\}$ may change from one region of $\vec{v}$-space to
 another). Here, the expansion will be valid only in a particular ``box''. The function $G_a$
 will be allowed to change as $\vec{v}$ goes from being in one domain $\Omega$ to another. 
Thus, in each of the domains $\{\Omega_{b}\}$ comprising the system (in which the 
system is assumed to be ``analytic'') there will be a different function $G_{ab}$
(specified by coefficients $c_{jab}$). 
In these cases, a natural question is how to ascertain phase transitions and effectively employ the existence of these volumes. 
Our approach here is once again that of noting when the overlap between different replicas is highest.
That is, given a particular test point $\vec{v}$, we may train the system with all 
data that lies in a volume $\Omega$ (a ``box'')  that encloses $\vec{v}$. We then
see when, as a function of the size $||\Omega||$, the overlap ${\cal{O}}(\vec{v})$
between the replicas for the predicted outcome at point $\vec{v}$
will be the highest. We employed this approach when the overlap between 
the various replicas was small and the our original classification outcome was less certain. 

The accuracy of machine learning classification algorithms is typically tested by randomly fitting a fraction $z$ of the known data (i.e., using these data for ``{\it training}'') and then
seeing how well the algorithm correctly predicts the classification of the remaining data that are not used as training but rather
supplied to the algorithm only as new vectors $\vec{v}$ whose correct classification is known yet not given to the user but is to be predicted by the algorithm.
This process (or training with a fraction $z$
of the data and {\it testing} the predictions on the remaining fraction of $(1-z)$) is repeated over and over again with different ways of splitting the known data 
into two subgroups of relative numbers set by a parameter $z$,
\begin{eqnarray}
\label{test:}
\mbox{training data points}&:& \mbox{test data points} \nonumber
\\ =
 z&:&(1-z).
\end{eqnarray}
The accuracy of the predicted classification is then averaged over the many ways of splitting the data with this ratio between the size of the number of training data points and the tested points kept fixed. 
In the accuracy tests that we will report on, we will follow the prevalent practice of choosing $z=0.8$.

\subsection{Gaussian kernels}
\label{Gaussian}

In what follows, we provide an explicit example in which the value of $R$ (the number of basis functions) is determined.
In the current context, we seek to find the optimal number $R$ of anchor points for the Gaussian of Eq. (\ref{GaussG}). Towards this end, we may plot the average overlap ${\cal{O}}$ between different replicas as a function of the number of replicas $r$ and the number of anchor points $R$.
This overlap enables us to determine the optimal values of $r$ and $R$ for which ${\cal{O}}$ obtains its maximum (or, more generally, its maxima).
In general, optimal parameter values (such as $R$) used in the SRVM model (also for function types are than Gaussian kernels) may be found by examining when the predictions of the different replicas are most robust.
SRVM does not merely average over the predictions of different replicas. Rather, as sketched in Figure \ref{peter}, the replicas ``interact'' with one another so as to make their collective predictions uniform. In the current setting, 
this corresponds to a choice of $R$ that maximizes the inter-replica overlap ${\cal{O}}$.  

\begin{figure}[h]
 \includegraphics[width=\linewidth]{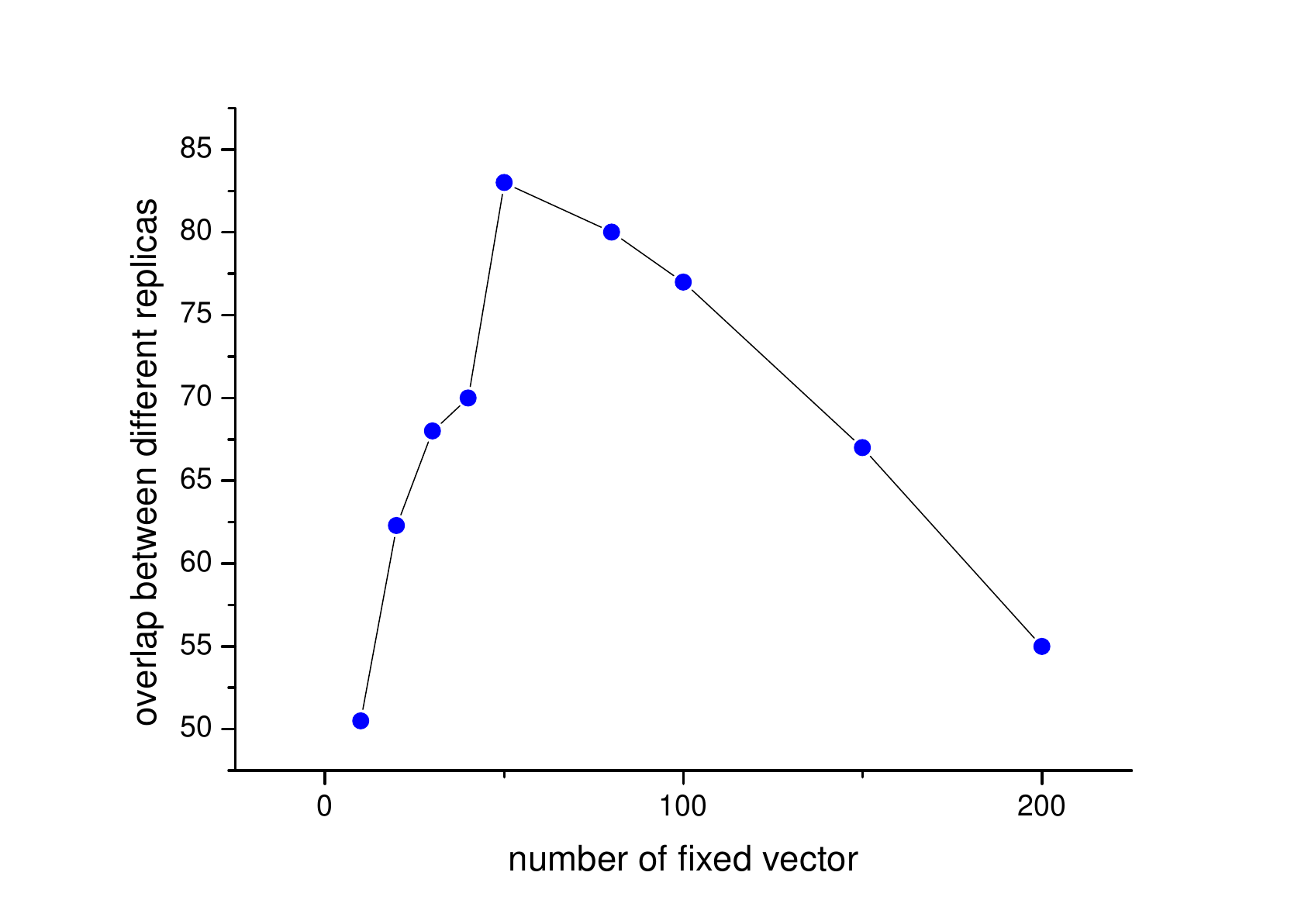}
 \caption{The overlap of Eq. (\ref{overlap}) averaged over $r=11$ replicas and different test points $\vec{v}'$ that appear in different cross-validation partitions for the stability of candidate perovskites.  Here, this average overlap (given in percentage) is computed for the predictions of different Gaussian type (Eq. (\ref{GaussG})) replicas
having $R$ fixed anchor points. The average inter-replica overlap is largest when $R \approx 60$ (suggesting that the most accurate  predictions are obtained for such values of $R$).}
  \label{overlap2}
 \end{figure}

To explicitly illustrate the basic premise, we examine the data of the perovskite classification problem that we will turn to in greater detail later on. 
For the time being, we probe how the average of the overlap ${\cal{O}}$ varies as a function of the number of basis functions used (or anchor points in the case of
the Gaussian kernel of Eq. (\ref{GaussG})). (As remarked earlier (see discussion after Eq. (\ref{GaussG})), the anchor points are randomly placed in the feature space.) 
As Fig. (\ref{overlap2}) illustrates, the overlap between different replicas is maximal for $R \approx 60$ anchor points. Since the inter-replica overlap
is maximal for this value of $R$, we suspect using this number of anchor points would result in the optimal accuracy. The average accuracy that we reached with the Gaussian kernel for determining stable Perovskite oxides was 94.19\%. This accuracy
may be contrasted with the performance of a current state of the art SVM package \cite{SVM package} employing radial basis (i.e., Gaussian) functions; the SVM method yielded a mean accuracy of 92.53\%. 

\subsection{Multinomial kernels}
\label{binary}

As we alluded to earlier, another set of kernels in Eq. (\ref{kernelG}) is afforded by a $d-$component vector $\vec{j}$ defining monomials,
\begin{equation}
\label{monomial}
K^{\vec{j}}(\vec{v}) =  
v_{1}^{j_{1}} v_{2}^{j_{2}}...v_{d}^{j_{d}}.
\end{equation}
Here, $v_{k}$ are the Cartesian components ($1 \le k \le d$). There are a variety of ways to produce multinomial based replica. For instance, different rotations in parameter space may lead to independent multinomials. A general rotation $v_{k} \to U^{h}_{kk'} v_{k'} \equiv v_{kh}$ with $U^{a}$ a random rotation matrix, will transform the monomial of Eq. (\ref{monomial}) into multinomial in which the sum of all powers 
in each of the individual monomials
\begin{eqnarray}
J \equiv \sum_{k=1}^{d} j_{k} 
\end{eqnarray}
is unchanged relative to its value in Eq. (\ref{monomial}). Thus, if we choose a basis of monomials $\{ K_{a}^{\vec{j}}(\vec{v}) \}$ with $0 \le j_{k} \le p$ 
(with a general natural number $p$)
for all $1 \le k \le d$ in one coordinate system $v_{k}$ then an independent basis of monomials is afforded by
\begin{eqnarray}
\label{monomial}
K_{a}^{\vec{j}} =  
v_{1a}^{j_{1}} v_{2a}^{j_{2}}...v_{da}^{j_{d}},
\end{eqnarray}
with $j_{k} \le p$. This is so as the highest power of each of the Cartesian coordinates is $p < J$.  
In Eq. (\ref{monomial}), $\{v_{hk}\}_{k=1}^{d}$ are the coordinates in the rotated basis generated by $U^{a}$. 
Eq. (\ref{kernelG}) may be used to concoct several replica functions $G_{a}=  \sum_{j=1}^{R} c_{ja} K_{a}^{\vec{j}}(\vec{v})$. 


\subsection{Neural network type functions}
\label{NN:sec}

We next turn to functions of the sort associated with neural networks \cite{neuron1,neuron2}. These functions are, typically, nested composites of linear transformations followed by Fermi type functions. The input to the node $k$ in layer $\alpha = 0,1,2, \cdots , N_{f}$ is given by a linear combination
\begin{eqnarray}
\label{ek}
\epsilon^{(\alpha)}_k &&=  L_{\alpha}(\{\zeta^{(\alpha -1)}_{k'}\}, \{c_{k}^{(\alpha-1)}\}) \nonumber
\\ &&\equiv - (\sum_{k'} w^{(\alpha-1))}_{kk'} \zeta^{(\alpha -1)}_{k'})- c^{(\alpha-1)}_{k},
\end{eqnarray}
with $w^{(\alpha-1)}_{kk'}$ and $c^{(\alpha -1)}_{k}$ being constants. For $N_{f} > \alpha' >0$, the function $\zeta^{(\alpha')}_{k'}$ is of the Fermi (or sigmoid) type, 
\begin{equation}
\label{zk}
\zeta^{(\alpha')}_{k'}=  f(\epsilon^{(\alpha')}_{k'}) =  \frac{1}{1+ e^{ \epsilon^{(\alpha')}_{k'}}}.
\end{equation}
The variables $ \zeta^{(0)}_{k'} = v_{k'}$ are input features. All  $(N_{f}-1)$ ``layers'' $\alpha \neq 0,1$ are often termed ``hidden layers''. 
In the last, output. layer there is only one value of $\epsilon^{(N_f})$ (i.e., $k=1$ only); here, $\zeta^{(N_{f})} = \theta(-\epsilon^{(N_f}))$ with $\theta(z) = (1+ sgn(z))/2$ being the Heaviside function. 
That is, a value of $\zeta^{(N_{f})} = 1$ corresponds to a positive classification in the scheme of the earlier subsection ($\tilde{\rho} (\vec{v}) = 1$) while $\zeta^{(N_{f})} = 0$ corresponds of a prediction of an assignment to the ``no'' class
$(\tilde{\rho}(\vec{v}) = -1$). The coefficients $w^{(\alpha)}_{kk'}$ form rectangular matrices with a number of rows equal to the number of ``nodes'' or ``neurons'' $n_{\alpha}$ (i.e., $k=1,2, \cdots, n_{\alpha}$) in layer $\alpha$ and a number of columns set by the number of nodes $n_{\alpha -1}$ in the preceding layer. The initial layer is that of the input values (i.e., $v_{k} = \zeta^{(0)}_{k}$ are the features). In typical neural nets, the single $\zeta^{(N_{f})}$ appearing in the final layer provides the prediction sought after in a classification problem. Specifically, in binary classification problems, if $\zeta_f >0.5$ then the predicted classification of $\vec{v}$ is of one type (e.g., ``yes'') and if $\zeta_f <0.5$ the prediction is that $\vec{v}$ belongs to the other class. Eqs. (\ref{ek}, \ref{zk}) schematically correspond to the composite of $2 N_{f}$ individual $f$ and $L$ functions 
\begin{eqnarray}
\theta (L_{N_{f}}(f(L_{N_{f}-1}( f(\cdots L_{1}(\vec{v}) \cdots ) ))))
\end{eqnarray}
All  $(N_{f}-1)$ ``layers'' apart from the first are transformation on the initial input vector $\vec{v}$ are often termed ``hidden layers''. Starting from random initial parameters $\{w^{(\alpha)}_{kk'}\}$ and $\{c^{(\alpha)}_{k}\}$, we will optimize the parameters of such neural net functions to fit the data. Different initializations do not necessarily lead to a unique set of these parameters when iteratively optimized (via gradient ``back propagation'' steps)  to fit the training data. Different neural architectures having a different number of layers $N_{f}$ and the set $(n_{1}, n_{2}, \cdots, n_{N_{f}})$ specifying the number of nodes in each iterative layer also constitute different functions of the same variety. As in Section \ref{Gaussian}, we may find the ``optimal'' neural net function types (including contending architectures) by seeing which types of architecture lend themselves to uniform consistent predictions amongst different replicas.

\subsection{Ternary and multi-class SRVM}
\label{ternary}

Thus far, we focused on binary classification (wherein the sign (Eq. (\ref{sgnsgn})) decided to which of two categories a particular point $\vec{v}$ should belong to).
There is, of course, more to life than only binary classification. In order to classify $\vec{v}$ into one of $p>2$ groups, various constructs are possible. 
One, very rudimentary, design is to iteratively classify as a point $\vec{v}$ as belonging (or not) to any one of the classes $q=1,2, \cdots, p$. 
Such a rudimentary approach emulates the well known ``One-Versus-all''  (OVR)  \cite{Ryan Rifkin} technique; this is the what we will adopt in the current
work when we will classify AB solids into one of $p=3$ groups (Section \ref{CAB}). Specifically, we will start by predicting the results of an input vector $\vec{v}$ for each of the $q$ possible output values with the SRVM algorithm that we introduced in
 the earlier subsections. Similar to the binary classification, in order construct the pseudo-inverse for the i-th output value bivariate algorithm, we will set the result of a data point as +1 if it output the i-th output value, and to be -1 otherwise. Instead of just taking the sign of the outputted results, we compared the raw values from results.
 That is, if the output associated with the vector $\vec{v}_{i}$ as tested against candidate classes $q=1,2, \cdots, p$ had the highest incidence of positive values for a particular class $q=q'$ then the vector $\vec{v}_{i}$ was classified as belonging to group $q'$.

\section{Cubic Perovskite Formability }
\label{Perovskite}

In this section we employ SRVM to predict whether candidate $ABO_{3}$ compounds form stable perovskite structures. The training data that we used \cite{formability} has $d=2$ features briefly noted in the Introduction: \newline
 (i) The ``tolerance factor", 
\begin{eqnarray}
\label{x1eq}
v_{1}  \equiv \frac{r_{A}+ r_{X}}{ \sqrt{2}(r_{B}+r_{X})}
\end{eqnarray}
where $r_{i=A,B,X}$ denote, respectively, the radii of the A, B, and X ions, and \newline
(ii) The "octahedral factor" defined as the ratio 
\begin{eqnarray}
\label{x2eq}
v_2 \equiv \frac{r_{B}}{r_{X}}.
\end{eqnarray}

\begin{figure}[h]
 \includegraphics[width=\linewidth]{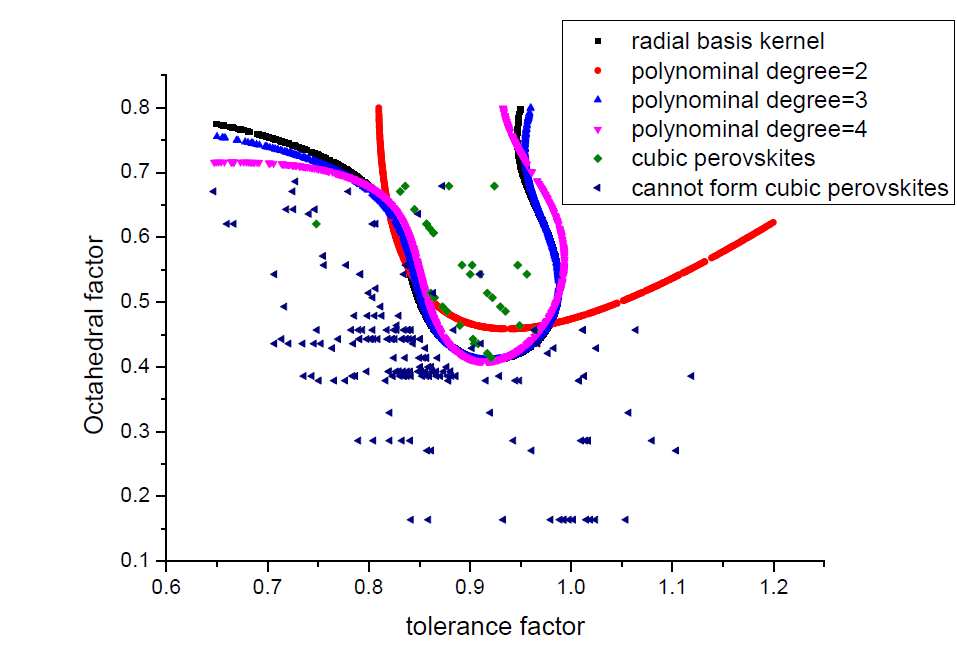}
 \caption{Classification results using different
SVM kernels employing the Libsvm-3.0 package. \cite{SVM package}}
  \label{333}
 \end{figure}

 \begin{figure}[h]
 \includegraphics[width=\linewidth]{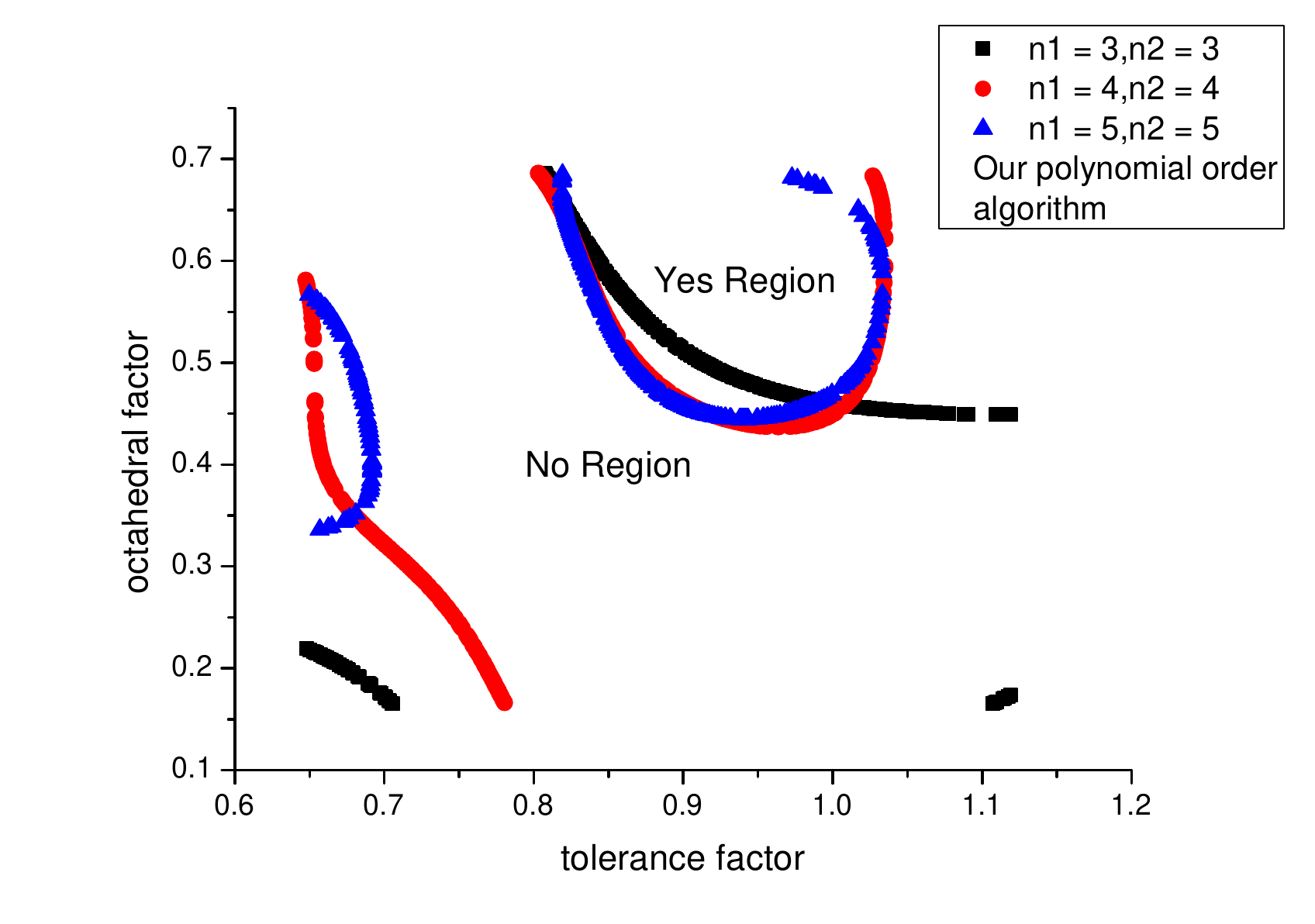}
 \caption{The viable region in the tolerance factor ($v_1$)- octahedral factor ($v_2$) plane for materials that may form cubic perovskite structure as ascertained by a multinomial order kernel in the SRVM method . Here we employed multinomials of three different orders (3, 4, and 5). The common region in which all multinomials predict formability of a cubic perovskite structure is marked as the ``Yes Region''. Similarly, the region where all three replica predict the lack of stable perovskites is denoted
 as the ``No Region".}
  \label{Polynomial order Formability}
 \end{figure}

 \begin{figure}
 \includegraphics[width=\linewidth]{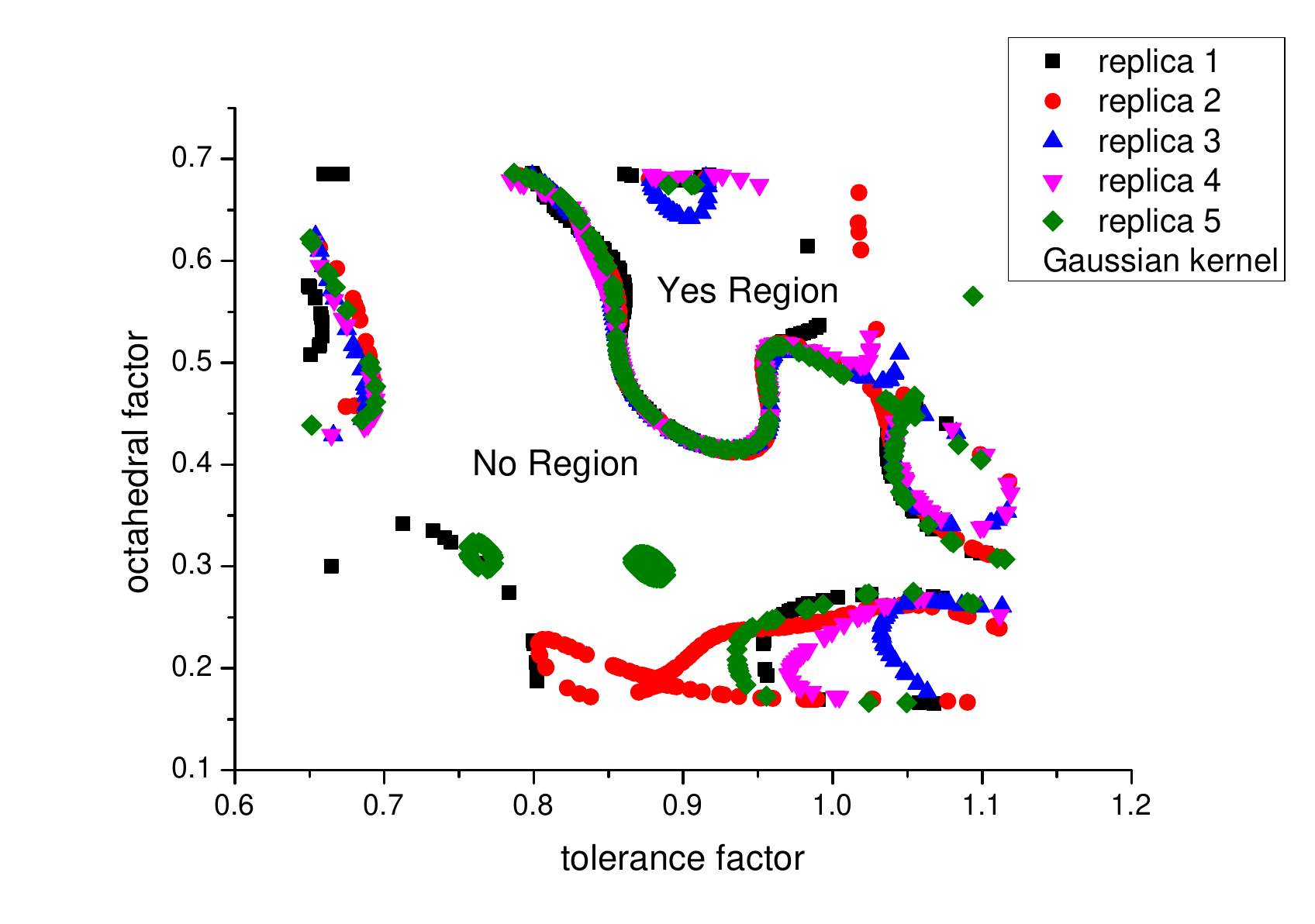}
 \caption{The predicted formability of the cubic perovskite structure as provided by the Gaussian kernels of Eq. (\ref{GaussG}) for five different replicas. These different replicas are produced by randomly choosing $R=50$ fixed vectors in Eq. (\ref{GaussG}) (see text).}
  \label{Gaussian Formability}
 \end{figure}

 \begin{figure}[h]
 \includegraphics[width=\linewidth]{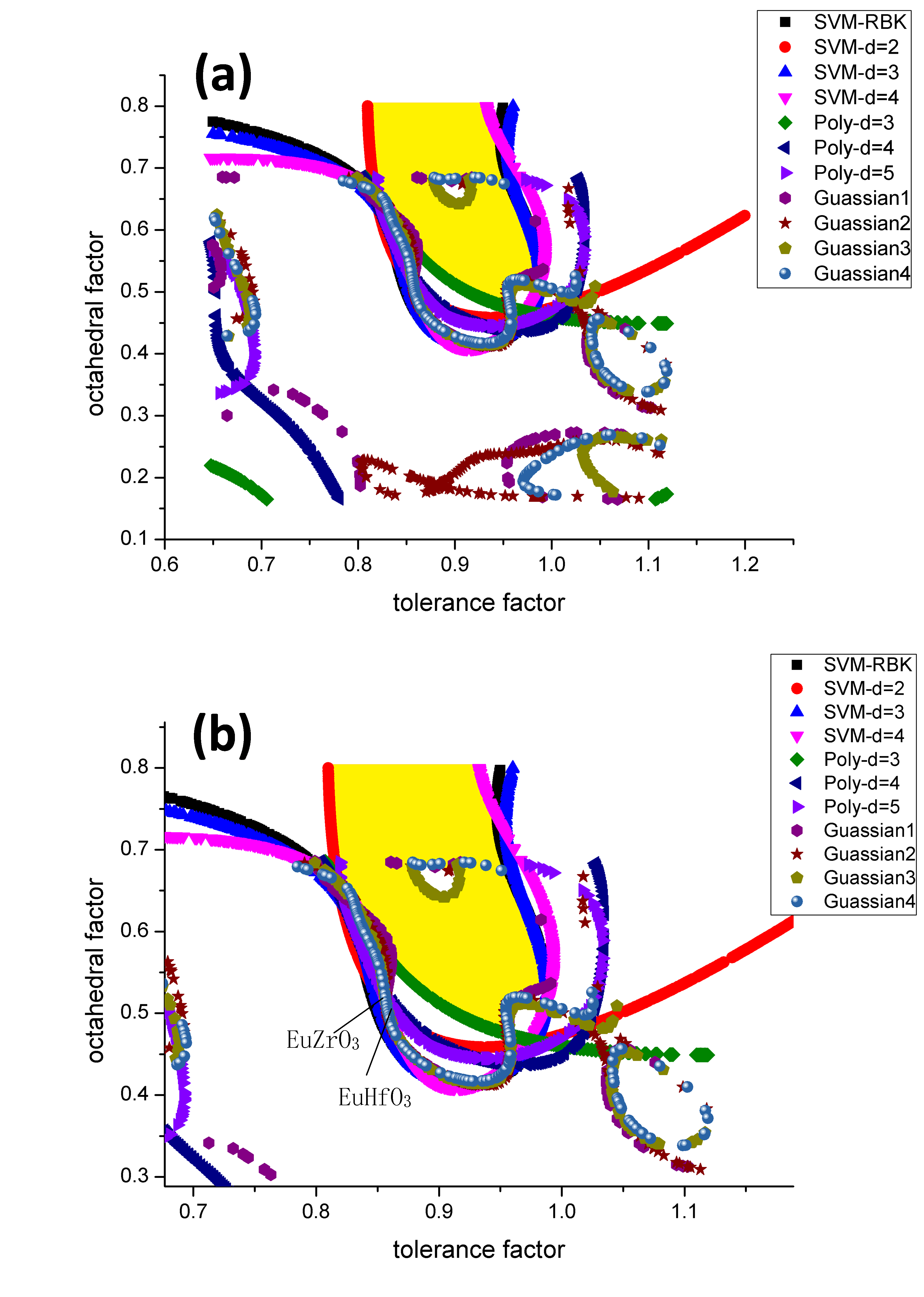}
 \caption{The formability of cubic perovskite structure at two different resolutions. The yellow region is that in which all methods
 (SVM, SRVM with both multinomial and Gaussian kernels) predict that cubic perovskite structures will form. In panel (a), we show the entire region of 
 measured tolerance and octahedral ratios. Panel (b) provides a zoomed viewed. Two possible candidate compounds that can from cubic perovskite structure are  highlighted: EuHfO$_3$ and EuZrO$_3$.}
  \label{777}
 \end{figure}

The data in \cite{formability} consist of 223 candidate compounds of the $ABO_{3}$ type. Of these compositions, 34 correspond to stable perovskite structures and the rest are unstable structures. (After removing duplicate compounds that share the same tolerance factor and octahedral factor, 188 data points remain, 29 of which form stable cubic perovskite structure.) Once the training is performed with input data, we use it to make the binary prediction regarding the stability of the contending perovskite compounds. 
Following Eq. (\ref{test:}), we repeatedly partitioned the data into two random subgroups with $z=0.8$. Several partitions with this ratio were generated by the standard 
cross-validation method in which the data are divided into nearly five equal parts. Subsequently, four of these five sets are then used together to train the algorithm and the remaining one fifth of the data is used as a resource of test data to see how accurate the predictions of the algorithm are. The set that is used
as the test data is cycled through (being chosen to be all of the five nearly equal parts of the data). The accuracy is then averaged over the predictions made over the five groups when these are used as test data. The accuracy is further averaged over different random partitions into five groups. Both for comparison as well as in order to obtain a more comprehensive picture, aside from our own SRVM algorithm, we also employed both the standard Gaussian and polynomial kernels in the well known SVM 
method \cite{SVM,SVM1}. In Figure \ref{333}, we provide the results that we obtained by applying the SVM algorithm for different kernels. In this figure, the region above the drawn curves (associated with individual SVM kernels) is predicted to correspond to stable perovskite structures; in the parameter region below these curves,
no stable perovskite materials are anticipated. In line with our main thesis (that of inferring a likely outcome from multiple independent kernels), the region that is above all drawn curves corresponds to a domain in the $v_{1}v_{2}$ plane in which we may expect (with high confidence) stable perovskite structures. Similarly,
in Figures \ref{Polynomial order Formability}, \ref{Gaussian Formability}, we display the results obtained by our SRVM algorithm
for, respectively, the multinomial and Gaussian kernels respectively (see Section \ref{rvm1} and the discussion following
Eq. \ref{GaussG} for a description of replicas in the Gaussian case). The designations of ``Yes'' and ``No'' reflect the predictions of the algorithm 
regarding the viability of putative compounds of an $ABO_{3}$-type composition to form stable perovskite structures. In Figure \ref{777},
we overlay (with different levels of resolution in the two panels) the predictions of the SVM method and our SRVM algorithm with multiple kernels/replicas. 
The shaded region in Figure \ref{777} is the one in which all methods/replicas/functions predict that stable perovskite structure should form. 
With this region in hand, all candidate ABO$_{3}$ materials (of the correct chemistry to allow such a composition) with tolerance and octahedral factors that lie in the shaded area are predicted to be stable perovskites. Some compositions lie near the boundary and do not enable (insofar as our approach is concerned)
a definite prediction regarding perovskite structures that do not appear in the data set that we used for training and validation \cite{formability}. Two such candidates are EuZrO$_3$ ($v_1 =0.857,~v_2=0.514$)
and EuHfO$_3$ ($v_1=0.861,~v_2= 0.507$). The location of these new potential stable perovskite structures is highlighted in panel (b) 
of Figure \ref{777}.  With $z=0.8$, the SVM algorithm achieved an accuracy of 92.52\%. By contrast, the SRVM algorithm obtained an accuracy of 94.14 \% with a  multinomial kernel (here two different multinomials (where different order multinomials were used as replicas) and we further employed the ``expansion in the box'' construction); SRVM achieved an accuracy of 94.19\% with a Gaussian kernel (here we employed 11 replicas each having randomly chosen anchor points). These two candidates are indeed known to constitute stable perovskites \cite{viallet,prasan}.

\section{Formability of bismuth-based double perovskites using tolerance and octahedral factors}
\label{sec:NND}
We next turn to predictions regarding the formability of more complex double perovskites (see, e.g., Fig. \ref{double-perovskite-figure}). Specifically, we are interested in finding bismuth-based oxide double perovskites that can described with a general formula of $A'A"B'$BiO$_{6}$ with Bi occupying half of the B-site cations. Our interest in Bi-based oxide double perovskites is motivated by the goal to achieve stable and non-toxic alternatives for lead-halide perovskites that have recently emerged as high-performance semiconductors with applications in solar cells and light-emitting diodes, but are plagued with instability and toxicity issues. \cite{Rohan-1,Rohan-2,Rohan-3}

 
   For the double perovskites, we define the octahedral and tolerance factors by Eqs. (\ref{x1eq}, \ref{x2eq}):  
\begin{eqnarray}
\label{rabab}
r_{A} \equiv \frac{r_{A'} + r_{A''}}{2} , \nonumber
\\ r_{B} \equiv \frac{r_{B'}+r_{B''}}{2}.
\end{eqnarray}
Towards this end, we will examine what occurs if the functions $\{G_{a}\}_{a=1}^{r}$ are either a sum of randomly generated Gaussians or neural-network-type functions. We will examine both  (i) a data set involving only double perovskites as well as (ii) combined data for both cubic and double perovskites. The double perovskite data set that we first study (see Appendix \ref{Sec:D}) consists of 72 candidate compositions. 
Of these, 57 are stable and 15 are unstable double perovskites. Similar to Section \ref{Perovskite}, these data have only two features: the tolerance and octahedral factors $v_{1,2}$. These two factors are calculated from the Shannon radii $r$ of the ions forming the double perovskite. The octahedral and tolerance factors of double perovskites are defined just as they were for the perovskites (Eqs. (\ref{x1eq},\ref{x2eq})). Now, however, we replace the radius of the A-site (B-site) cation with the average radii of the A-site and A'-site  (B-site and B'-site) cations involved. 

\subsection{Neural net analysis of combined double perovskite and cubic perovskite data}
\label{NND2}

We examine the 295 instances that include data on both (i) double perovskite data of Appendix \ref{Sec:D} and (ii) the ternary perovskite formability data \cite{formability} (that we employed in Section \ref{Perovskite}). Additional details regarding these data sets have been relegated to the Appendix. We study these materials using the ``neuralnet'' library \cite{neuralnet,neuralnet1,neuralnet2,neuralnet3,neuralnet4}. We randomized the input data and performed a ten-cross validation analysis  for a neural net consisting of one layer and three nodes. We run an SRVM version of this neural net: the initial input weights (i.e., those set before back propagation is applied) are chosen randomly each time that we run the neural net model. Details concerning this analysis are provided in Appendix \ref{Sec:B}.

\subsection{SRVM-Neural net predictions for new stable double perovskites}
\label{srvm-predict-2}

Armed with a proof of principle of the reliability of the ensemble of neural network functions for the combined formability data \cite{formability} and the double perovskite data of Appendix \ref{Sec:D}, we now apply it to suggest hitherto unexplored candidate perovskite compounds. Predictions for the stability of the candidate double perovskite compounds are reported, rather expansively, in Table  \ref{Stable double perovskite } of Appendix \ref{Sec:C}. Herein, the stability probabilities for the various compounds are computed as equal to the ratio of (the number of replicas that yield a positive outcome)/(the total number of replicas employed). 

\section{Double perovskite analysis with electronegativities and ionic Shannon radii}
\label{Section:5feature}

We now turn to the full double perovskites in the data set of Appendix \ref{Sec:D} and analyze these (sans the ternary perovskite data). In addition to the tolerance and octahedral factors of Eqs. (\ref{x1eq}, \ref{x2eq}), we will now add basic electronegativity features that directly touch on the interactions underlying the stability of these systems. Two of these new features are the electronegativity of the X (in the materials that we consider, oxygen anion) sites relative to average electronegativity $n_{A}$ and $n_{B}$ of, respectively, the ions at the A and B sites,
\begin{eqnarray}
\label{v34}
v_{3} = n_{X}- \frac{n_{A'} + n_{A''}}{2} \equiv n_{X} - n_{A} \nonumber
\\ v_{4} = n_{X} - \frac{n_{B'} + n_{B''}}{2} \equiv n_{X} - n_{B}.
\end{eqnarray}
In practice, in order to compute $n_{A,B}$ we need to know the four electronegativities $n_{A'}, n_{A''}, n_{B'}$ and $n_{B''}$. We will employ the two features of Eq. (\ref{v34}) since these may be extended also to the ternary perovskites ABO$_{3}$. Since the B-O bond is covalent while the A-O interactions are more ionic, we anticipate $v_{3}$ to be more important that $v_{4}$.  A new feature suggested by \cite{Scheffler_new} (dubbed therein as 
$\tau$) which we included was a nontrivial function of the average electronegativity of the A ions and the average Shannon radii $r_{A,B}$ (or, equivalently, the tolerance and octahedral factors of Eqs. (\ref{x1eq}, \ref{x2eq})) with the substitution of Eq. (\ref{rabab}),
\begin{eqnarray}
\label{eqtau}
v_{5} = \tau \equiv \frac{r_{X}}{r_{B}} - n_{A} ( n_{A} - \frac{r_{A}/r_{B}}{\ln (r_{A}/r_{B})}).
\end{eqnarray}
The input features of known stable and unstable double perovskites and our results are, respectively, provided in Tables \ref{Stable double perovskite} and \ref{non stable double perovskite}
of Appendix \ref{Sec:D}. Our corresponding predictions are given in Table \ref{Screened Predictions} of Appendix \ref{sec:e}.  In Appendix \ref{Sec:E}, we provide specifics concerning neural net functions used to fit the data.  We return to a simple Gaussian SRVM of Sections (\ref{Gaussian}, \ref{Perovskite}) and apply it to an analysis of the double perovskite data with the above mentioned five features $v_{1 \le i \le 5}$. 
A vote within sets of $r=31$ replicas was performed 10 times to yield final predictions. In \cref{SRVM1,SRVM2,SRVM3,SRVM4,SRVM5,SRVM6,SRVM7,SRVM8,SRVM9}, 
we only display $r=9$ such constructed replica functions. Each of the replica functions is of the form of Eq. (\ref{GaussG}) with $R=17$ anchor points.
The centers$\{\vec{v}_{ja}\}$  of the Gaussians and the corresponding coefficients $\{c_{ja}\}_{j=1}^{R}$ are listed in Tables \cref{SRVM1,SRVM2,SRVM3,SRVM4,SRVM5,SRVM6,SRVM7,SRVM8,SRVM9} of Appendix \ref{Sec:F}. 
All standard deviations $\sigma_{ja}$ are set to unity, $\sigma_{ja} = \sigma=1$. Unlike Section \ref{Perovskite}, we did not employ the overlap to optimize the number of anchor points. The resulting 5-fold cross validated accuracy for the double perovskite data of Appendix \ref{Sec:D} with the $d=5$ features using the SRVM-Gaussian algorithm is 92\%. In Table \ref{Stable double perovskite}, we provide predictions for the stability of the screened candidate compounds. In Table \ref{Screened Predictions}, we also present the results of SVM with radial basis kernels are applied and neural networks of different architecture. The five-fold cross validated accuracy for this model is, similarly, also 92\%. 

\section{Theoretical Formation Enthalpy of Selected Double Perovskites}
\label{sec:DFT}

As described in Sections \ref{sec:NND} and \ref{Section:5feature}, we employed SRVM to predict the formability of hypothetical Bi-based double perovskite oxides with a general formula of A'A"B'BiO$_{6}$, where Bi occupies half of the B-site cations. From a simple charge balance, 30,357 total double perovskite oxides of the form $A'A"B'$BiO$_{6}$ are possible. We considered all the cations up to Bi in the periodic table for A', $A"$ and B'-sites, while excluding lanthanides and radioactive Tc as possible candidates. Moreover, all possible oxidation states for each cation were included during charge balance while keeping the oxidation state of Bi fixed at $+3$.
 
Due to the limited number of experimentally synthesized Bi-based double perovskite oxides, we use a two-fold approach for training our SRVM/SVM models. First, we devised a quick screening criterion to screen some hypothetical compounds based on the atomic features of the 28 unique A'A"B'BiO$_{6}$ double perovskites (with Bi at the B-site) that have been reported in inorganic crystal structure database (ICSD) \cite{ICSD}. The atomic features that we used for initial screening are the tolerance and octahedral factors of Eqs. (\ref{x1eq}, \ref{x2eq}). We used Slater's empirical atomic radii \cite{Slater}  for calculating these factors. This was done since the tolerance and  octahedral facors are independent of oxidation state and coordination number, which are unknown for a hypothetical perovskite. We found that these factors follow a linear regression model with an R-squared value of 0.84, for the experimental A'A"B'BiO$_{6}$ compounds. Moreover, we find that 144 hypothetical A'A"B'BiO$_{6}$ compounds adhere to this linear regression model. 

Next, using density-functional theory (DFT), we optimized the crystal structure of each screened compound, to calculate $\Delta H_f$ and evaluate its potential formability. The perovskite framework undergoes cooperative tilting of the BX$_6$ octahedra to optimize the coordination environment of A and B-site cations \cite{cations}. We consider all viable octahedral tilt patterns for investigating the ground state of a given stoichiometry. These possible tilt patterns and their corresponding space-group symmetries are summarized in Tables \ref{table:s1} and \ref{table:s2} \cite{Knapp,Howard1,Howard2}. For a stoichiometry to be considered stable, its calculated $\Delta H_f$  should be negative. However, compounds with positive $\Delta H_f$ are quite common and can be experimentally synthesized by optimizing experimental conditions. For example, a recent survey of the DFT-calculated $\Delta H_f$ of all existing binary oxides reported $\sim$ 90th percentile of the compounds lie within 94 meV/atom above the ground state polymorph \cite{WSun}. Therefore, we expect compounds with $\Delta H_f < 100$ meV/atom to be formable under suitable experimental conditions. We label each of the 144 compounds as stable ($\Delta H_f < 100$ meV/atom) and unstable ($\Delta H_f > 100$ meV/atom).
	
We used experimentally synthesized Bi-based double perovskite oxides (with Bi at the  B-site), non-perovskite oxides, and the compounds labeled as stable and unstable from initial screening to train our SVM models. Here non-double perovskite oxide corresponds to a compound following the stoichiometry of a double perovskite but adopts a non-perovskite crystal structure, listed in the ICSD \cite{ICSD}. As described in Section \ref{Section:5feature},
We used five different SVM models followed by a voting procedure (SRVM) to predict the formability of a hypothetical Bi-double perovskite oxide. The models employed in this classification process use the following atomic features- the tolerance factor, the octahedral factor, the average A--site electronegativity, and the and average B-site electronegativity. We calculate the octahedral and tolerance factors using Slater's empirical atomic radii \cite{Slater} as well as Shannon's ionic radii \cite{rtshannon}. We consider a hypothetical Bi-double perovskite oxide to be formable if at least one of the five SVMs predicts it to be stable. 
	
	Out of a total 30,357 hypothetical $A'A"B'$BiO$_6$ compounds, our SRVM algorithm voted 9,795 compounds to be stable, where 8,115 of the 9795 compounds are predicted to be stable by all the five SVM models. We found 1,680 out of 9,795 compounds are voted as stable by at least one SVM model. To further assess the formability of compounds that were voted by our SRVM algorithm to be stable, we selected a few compounds to perform structural optimization and subsequent formation enthalpy analysis using DFT. At this point, it is not possible to optimize all the 9,795 compounds using DFT due to their computational expense. We optimized the crystal structure of the selected compounds using DFT and calculate their formation enthalpy ($\Delta H_f$). As mentioned earlier, the perovskite framework can undergo cooperative tilting of the BO$_6$ octahedra to optimize the coordination environment of A and B-site cations \cite{cations}. We consider all possible octahedral tilt patterns for investigating the ground state of a given stoichiometry. The DFT calculations were performed using the projector-augmented-wave (PAW) method \cite{PAW} as implemented in the Vienna Ab-initio Simulation Package (VASP) \cite{VASP}. We employed generalized gradient approximation (GGA) as implemented in the Perdew-Burke-Ernzerhof (PBE) functional \cite{PBE} structure optimization. Layered and rock-salt ordering were imposed at the A and B-site, respectively. A plane-wave basis set with a cutoff of 400 eV was used for structure relaxation, and 520 eV for the final static total energy calculation step. Gamma-centered Monkhorst-Pack \cite{Monkhorst-Pack} k-points mesh was used for sampling the Brillouin zone, where k-points per reciprocal atom (KPPRA) was set to be $\sim$ 8000 for relaxation and the single-step static calculation. A Hubbard U parameter of 4 eV was used to account for localized d-electron interactions in Fe \cite{Fe-ref}. G-type anti-ferromagnetism was imposed during relaxation. The DFT calculations were carried out using the pseudopotentials and other DFT settings employed by OQMD \cite{OQMD}.

	The calculated formation enthalpy for each compound is the difference between the total energy of a given structure with the minimized free energy of chemical reactants, which lie at the convex hull, for the given stoichiometry as implemented within OQMD \cite{OQMD,OQMD'}. For example, Eq. (\ref{eq1}) shows the chemical reaction path for which the free energy is minimized for the reactants for KSrFeBiO$_6$. The formation enthalpy for KSrFeBiO$_6$ ($\Delta H_f$ (KBaTeBiO$_6$)) is then calculated using Eq. (\ref{eq2}), where E(KSrFeBiO$_6$) is the DFT total energy/formula unit of KSrFeBiO$_6$, while E(SrFeO$_3$) and E(KBiO$_3$) are DFT total energies/formula unit of SrFeO$_3$ and KBiO$_3$ respectively.
\begin{eqnarray}
\mbox{SrFeO}_{3}+ \mbox{KBiO}_{3} \to \mbox{KSrFeBiO}_{6},
\label{eq1}
\end{eqnarray}
\begin{eqnarray}
 \Delta H_{f}(\mbox{KSrFeSiO}_{6}) = && E(\mbox{KSrFeBiO}_{6})
\nonumber
 \\ && - [E(\mbox{SrFeO}_3) - E(\mbox{KBiO}_3)].
\label{eq2}
\end{eqnarray}



 \begin{table}[ht]
\caption{Space group symmetries and octahedral tilt patterns for A'A''B'B''O$_{6}$ compounds} 
\centering 
\begin{tabular}{c c c} 
\hline\hline 
Index  &space group &tilt system \\ [0.5ex] 
\hline 
2 & P${\overline{1}}$ & a$^-$b$^-$c$^-$  \\ 
4 & P2$_{1}$ & a$^-$a$^-$c$^+$ \\
5 & C2 & a$^-$b$^0$c$^+$  \\
11 & P2$_1$/m & a$^-$a$^-$c$^0$ \\
13 & P2/c & a$^+$b$^-$c$^0$ \\
16 &  P222 &  a$^{-}$b$^{-}$c$^{-}$ \\
49 & Pccm & a$^{+}$b$^{'0}$c$^{0}$ \\ 
81 & P4 & a$^{+}$a$^{+}$c$^{-}$ \\
85  & P4/n & a$^{0}$ a$^{0}$ c$^{-}$ \\ 
90 & P42$_{1}$2 & a$^{0}$ a$^{0}$  c$^{+}$\\
111 &  P42m & a$^{+}$  a$^{+}$ c$^{0}$ \\ 
129 & P4/nmm & a$^{0}$ a$^{0}$ c$^{0}$\\[1ex]
\hline 
\end{tabular}
\label{table:s1} 
\end{table}

\begin{table}[ht]
\caption{Space group symmetries and octahedral tilt patterns forA$_{2}$B'B''O$_{6}$ compounds} 
\centering 
\begin{tabular}{c c c} 
\hline\hline 
Index  &space group &tilt system \\ [0.5ex] 
\hline 
2 & P${\overline{1}}$ & a$^-$b$^-$c$^-$  \\ 
12 & C2/m & a$^0$b$^-$b$^-$ \\
14 & P2$_1$/c & a$^+$b$^-$b$^-$  \\
15 & C2/2 & a$^0$b$^-$c$^-$ \\
48 & Pnnn & a$^+$b$^+$c$^+$ \\
86 &  P4$_{2}$/n  &  a$^{+}$a$^{+}$c$^{-}$ \\
87 & I4/m & a$^{0}$a$^{'0}$c$^{-}$ \\ 
128 & P4/mnc & a$^{+}$a$^{+}$c$^{-}$ \\
134  &  P4$_{2}$/nnm  & a$^{0}$ b$^{-}$ b$^{+}$ \\ 
148 & R$\overline{3}$ & a$^{-}$ a$^{-}$  a$^{-}$\\
201 &  Pn$\overline{3}$ & a$^{-}$  a$^{+}$ a$^{+}$ \\ 
225 & Fm$\overline{3}$m & a$^{0}$ a$^{0}$ a$^{0}$\\[1ex]
\hline 
\end{tabular}
\label{table:s2} 
\end{table}

\begin{figure}[h]
 \includegraphics[width=.9\linewidth]{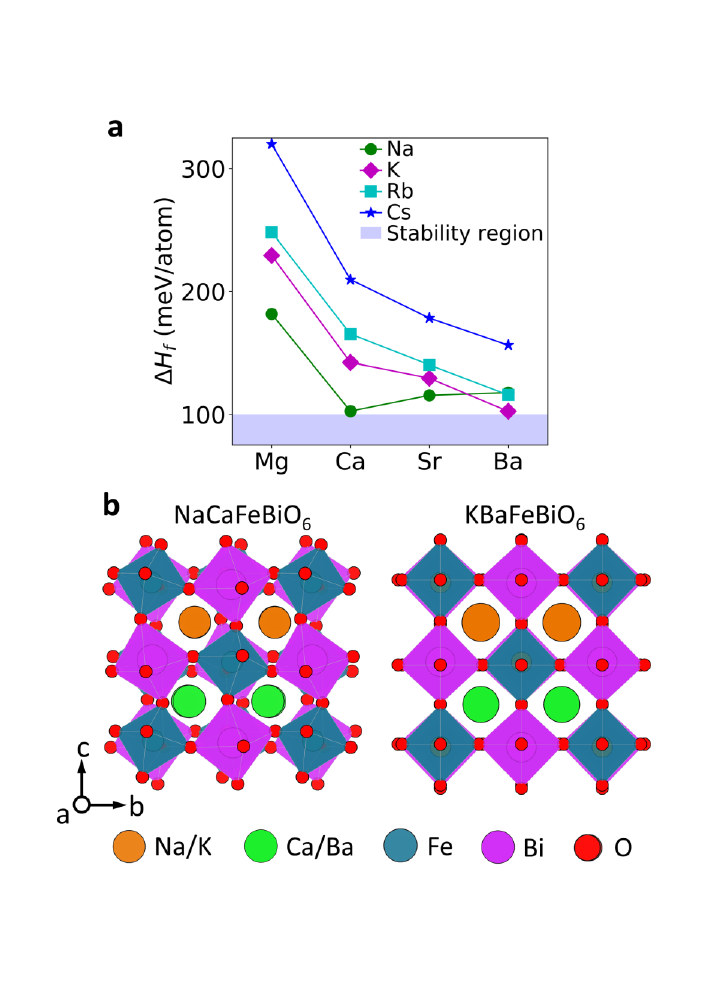}
 \caption{  a) Formation enthalpy for the ground state structure of the $A'A"$FeBiO$_6$ family of compounds. The X-axis represents the $A"$-site cation whereas A'-site cations are defined in the legend. b) Schematics showing the [100] projections of ground state structures of NaCaFeBiO$_6$ and KBaFeBiO$_6$.}
  \label{fig:dft}
 \end{figure}

We calculated the formation enthalpy $\Delta H_f$ of compounds with a general formula of A$'$A$"$FeBiO$_6$ using DFT, where A$'$ = Na, K, Rb and Cs and A$"$ = Mg, Ca, Sr and Ba. This family of compounds was voted to be stable by the SRVM algorithm. We exclude the Li+ and Be2+ cations as possible A-site candidates due to their extremely small size for the A-site cubo-octahedral cavities. Of the 16 possible compounds in the A$'$A$"$FeBiO$_6$ family, 13 compounds are predicted to be stable double-perovskites by all the five SVM models, while for the other three compounds (NaMgFeBiO$_6$, KMgFeBiO$_6$ and RbMgFeBiO$_6$) SVM-2 and SVM-5 predict them to be unstable with an overall vote predicting them to be stable. 
From an analysis of the formation enthalpy, we find that the A$'$A$"$FeBiO$_6$ family of double perovskite oxides is reasonably close to our formability criterion ($\Delta H_f <$ 100 meV/atom). As shown in Figure \ref{fig:dft}, we find that the ground state of all the 16 compounds has $\Delta H_f >$ 100 meV/atom. However, 8 of the 16 compounds have $\Delta H_f$ within (100 to 150) meV/atom. Figure \ref{fig:dft}b shows the [100] projections of ground state crystal structures of NaCaFeBiO$_6$ and KBaFeBiO$_6$. NaCaFeBiO$_6$ is predicted to have ground state crystal structure belonging to $P2_{1}$ space group  with $a^{-}a^{-}c^{+}$  tilt pattern, whereas the ground state of KBaFeBiO$_6$ is predicted to be the ideal double perovskite structure ($P4/nmm$) without any octahedral tilts.
The compounds having Mg at the A$"$-site exhibit the highest $\Delta H_f$. The average $\Delta H_f$ for compounds having Mg at the A$"$-site is 245 meV/atom. The small size of Mg makes it unsuitable for the large A-site cavities, even after octahedral tilts, resulting in a high $\Delta H_f$. Also, smaller size at the A-site results in higher degree of octahedral tilting. For instance, the ground state of RbMgFeBiO$_6$ has a space-group symmetry of P$\overline{1}$, which corresponds to a tilting pattern $a^{-}b^{-}c^{-}$ , resulting in octahedral tilting along all three crystallographic directions. Despite having $100 < \Delta H_f
< 150$ meV/atom, $A'A"$FeBiO$_6$ shows a good potential towards thermodynamic stability and we expect some of the compounds within this family can be synthesized by optimizing experimental conditions.
We remark that while we achieve very high cross-validation accuracies (the 92 \% cross-validation accuracy of Section \ref{Section:5feature}), the success of our method in predicting new stable double perovskite compounds (inasmuch as these can be ascertained via DFT calculations) is far lower. 

\section{Ternary classification of AB solids} 
\label{CAB}

 \begin{figure}[h]
 \includegraphics[width=\linewidth]{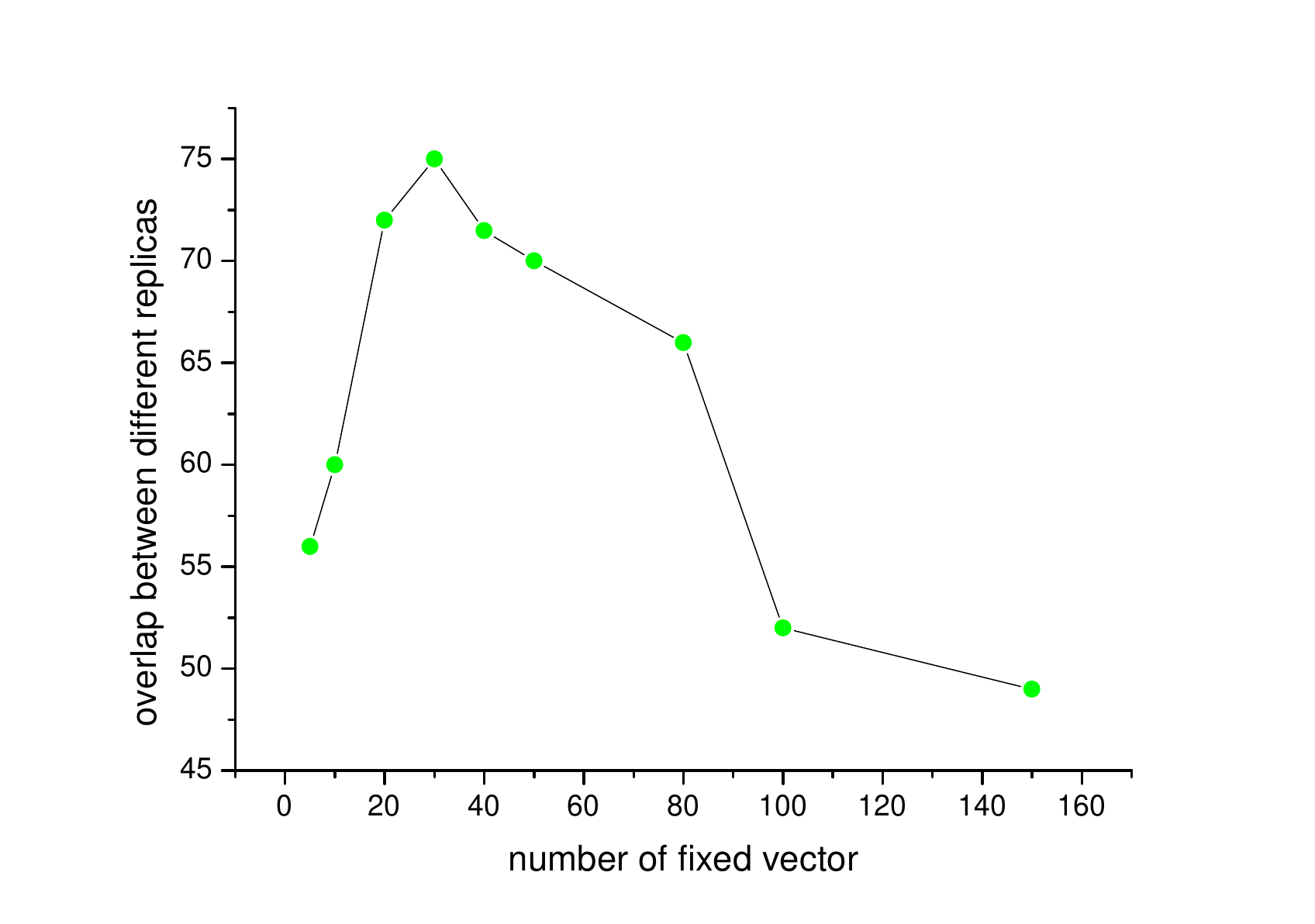}
 \caption{The average overlap (given as a percentage) as computed by Eq. (\ref{oaverage}) averaged over the different cross-validation between the $r=7$ different replicas. Here, replicas of the Gaussian replica model (Eq. (\ref{GaussG})) were employed for the ternary classification of the binary ($AB$) solids as a function of fixed anchor points $R$. 
The maximum overlap (suggesting an accurate prediction) appears circa $R \approx 40$.}
  \label{overlap4}
 \end{figure}

We next turn our attention, using the data of \cite{saad}, to the classification of binary octet solids \cite{AB} (these solids have the chemical composition $A^{n}B^{8-n}$ where $n$ denotes
the number of valence electrons) into one of $q=3$ groups
(denoted W, Z or R \cite{AB,saad}). Octet binary alloy crystals include technologically important semiconductors such as GaAs, GaN, and ZnO. The three classes that we analyze correspond to dominant zincblende (Z), wurtzite (W), or rocksalt (R) crystal structures that these alloys typically form. Similar to Section \ref{Perovskite}, we applied both the standard SVM technique with the multi-class variant our SRVM approach  
(see Section \ref{ternary}) with multinomial and Gaussian kernels. We employed two commonly used figures of merit \cite{CP} as features,
\begin{eqnarray}
 r_{\sigma} \equiv r_{s}^{A} + r_{p}^{A} - r_{s}^{B} - r_{p}^{B}, \nonumber
\\  r_{\pi} \equiv r_{p}^{A} - r_{s}^{A} + r_{p}^{B} - r_{s}^{B}.
\end{eqnarray}
Here, $r_{s}^{A}$, $r_{p}^{A}, r_{s}^{B}$ and $r_{p}^{B}$ denote the pertinent radii for an electron bound to the A or B ion that is in an $s$ or $p$ orbital.

\begin{figure}[h]
 \includegraphics[width=\linewidth]{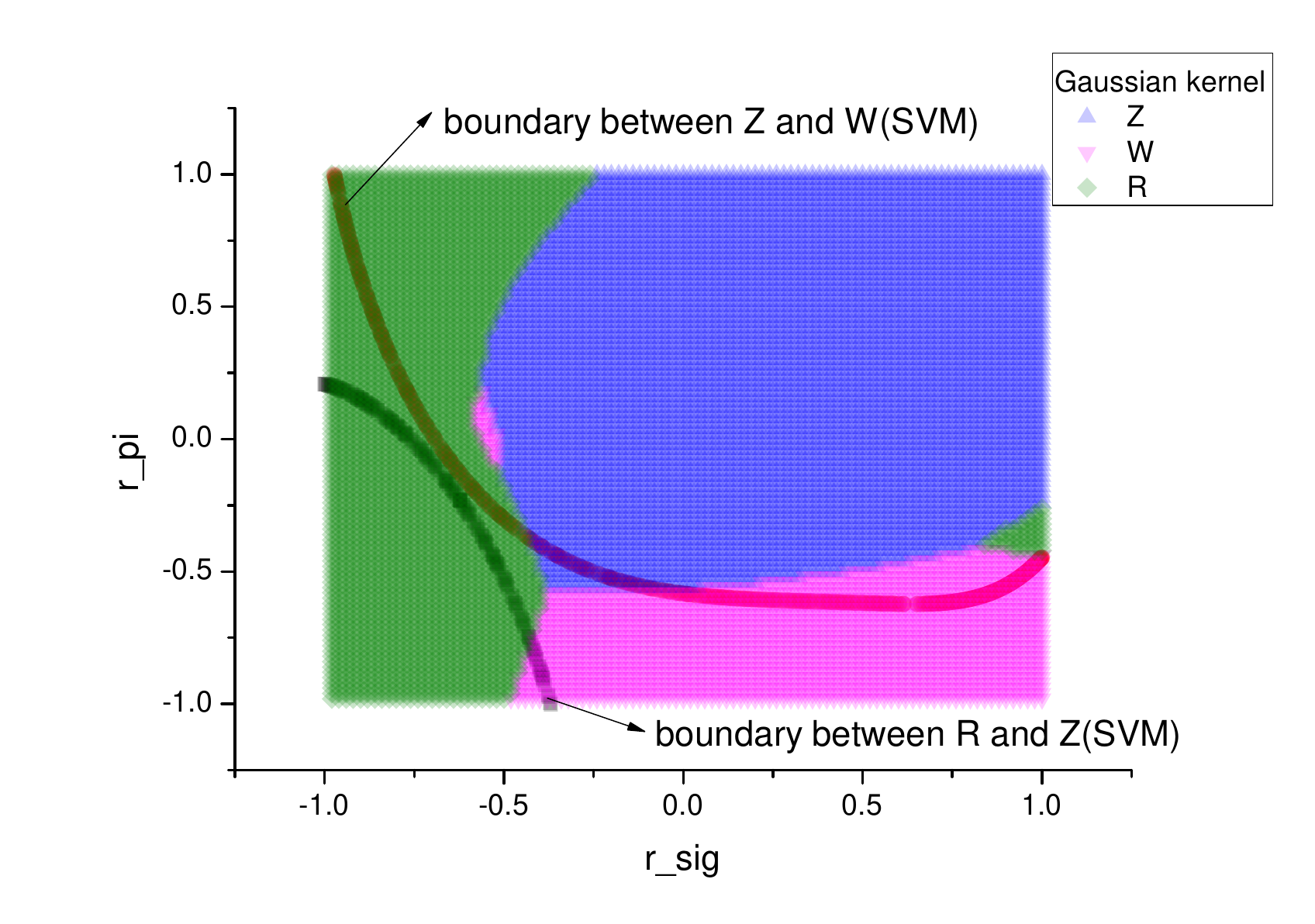}
 \caption{The phase diagram predicted by both the SVM algorithm (denoted by thick solid lines) and our method when using Gaussian kernels (different color solid regions).}
  \label{three_class_gaussian_fit}
 \end{figure}

In the Gaussian approach, we employed the sum of $R=30$ individual Gaussians (associated with different anchor points). The general behavior of the inter-replica
overlap is displayed in Figure \ref{overlap4} in which it is seen that the overlap becomes maximal at $R \sim 30$.  The final classification for each data point was determined by the group for which a given data point appeared most frequently out of the five replicas employed. Let $n^{\vec{v}'}_{w}$ be the number of replicas that
classify point $\vec{v}'$ into one of the three classes $w=0,1,2$. For a given cross-validation run, 
the average replica overlap for this three-classification problem is defined as
\begin{eqnarray}
\label{oaverage}
{\overline{{\cal{O}}}} \equiv \frac{1}{rN_{\vec{v}'}} \sum_{\vec{v}'} \max_{w} \{n_{w}^{\vec{v}'}\}. 
\end{eqnarray}
Here, $N_{\vec{v}'}$ is the number of test points $\vec{v}'$. The average of ${\overline{{\cal{O}}}}$ over the different cross-validation partition is plotted in Fig. (\ref{overlap4}) as a percentage. The cross-validation accuracies (as ascertained by Eq. (\ref{test:}) for $z=0.8$) that our SRVM algorithm obtained for the Gaussian and multinomial kernels were, respectively, $92.72\%$ and $90.90\%$. These values were lower than the accuracy achieved by an SVM algorithm with a radial kernel (that we found to be $94.54\%$). In Figure \ref{three_class_gaussian_fit}, we provide 
the phase boundaries (between the W, Z, and R phases) as ascertained by SVM (see the solid curves therein) alongside the boundaries determined by our SRVM method (the domains of the different phases as predicted by SRVM are marked by different colors).

\section{Conclusion}
\label{Conclusion}

In summary, we introduced and implemented a new classification algorithm to classify various materials and identify new promising compounds. 
In particular, we investigated (1) the formability of cubic and double perovskite type compounds (a binary classification problem) and (2) binary octet alloys (via ternary classification). A more detailed description of our new algorithm appears in a companion paper \cite{patrick}. Using this algorithm, we achieved a high accuracy in both problems. Combining our approach with other machine techniques, we suggest new candidate stable perovskites and properties of binary compounds.

{\bf{Note added in proof.}}
As our work was being finalized for publication, we back aware of a recent work by Park {\it et al.} \cite{ParkJCP} that explores, along very different lines, the applicability of machine learning for double perovskites.

\section{Acknowledgements}
This research was partially supported by the NSF CMMT under grant number 1411229. RM and AST acknowledge support from NSF Ceramics program through grant number 1806147. Computations in this work benefited from the use of the Extreme Science and Engineering Discovery Environment (XSEDE), which is supported by NSF grants ACI-1053575 and ACI-1548562.

\newpage

\appendix

\section{Screened Double Perovskite data}
\label{Sec:D}

Here, we provide the double perovskite data employed in Sections \ref{sec:NND} and \ref{Section:5feature}.  In Table \ref{Stable double perovskite}, we list values for stable double perovskites while Table \ref{non stable double perovskite} provides values for unstable double perovskite compositions. 
In both tables, the first two columns list the tolerance and octahedral factors of Eq. (\ref{x1eq}, \ref{x2eq}) (with the substitution of Eq. (\ref{rabab})) when the Shannon radii are employed. The third and fourth columns provide the difference in average electronegativity on the A sand B sites as compared to that of the Oxygen (Eq. (\ref{v34})). The entries on the last column is the function $\tau$ of Eq. (\ref{eqtau}).

\begin{longtable*}[t]{| c | c | c | c | c | c |}
 \caption{Experimental data for stable double perovskites} .\label{Stable double perovskite}\\
 
 \hline

 \hline
Formula & Shannon tolerance factor & Shannon octahedral factor & en-O - Ave-A & en-O - Ave-B & $\tau$  \\
 \hline
 \endfirsthead
 
 \hline

 \hline
Formula & Shannon tolerance factor & Shannon octahedral factor & en-O - AVe-A & en-O - Ave-B & $\tau$ \\
 \hline
 \endhead
 
 \hline
 \endfoot
 
 \hline

 \hline\hline
 \endlastfoot

    Ba2 Bi Y O6 & 0.960108 & 0.614815 & 2.55  & 1.82  & 3.481819 \\
    Ba2 Bi Dy O6 & 0.957473 & 0.619259 & 2.55  & 1.82  & 3.492014 \\
    Ba Bi0.75 Na0.25 O3 & 0.962315 & 0.611111 & 2.55  & 1.6925 & 3.473927 \\
    Ba2 Ru0.67 Bi1.33 O6 & 0.983529 & 0.576361 & 2.55  & 1.3597 & 3.426214 \\
    Ba2 Ce Bi O6 & 0.936481 & 0.655556 & 2.55  & 1.87  & 3.605649 \\
    Ba2 Pr Bi O6 & 0.94069 & 0.648148 & 2.55  & 1.865 & 3.577963 \\
    Ba2 Nd Bi O6 & 0.942172 & 0.645556 & 2.55  & 1.86  & 3.568829 \\
    Ba2 Tb Bi O6 & 0.95507 & 0.623333 & 2.55  & 1.88  & 3.502059 \\
    Sr2 (Bi1.4 Ca0.6) O6 & 0.89858 & 0.626293 & 2.49  & 1.726 & 3.993775 \\
    Ba2 Bi0.667 Te O6 & 1.060568 & 0.461856 & 2.55  & 1.371637 & 3.420366 \\
    Sr2 Sc Bi O6 & 0.881711 & 0.657407 & 2.49  & 1.75  & 4.225958 \\
    Ba2 Bi Ta O6 & 0.957911 & 0.618519 & 2.55  & 1.68  & 3.49026 \\
    Ba2 Bi Ir O6 & 0.973505 & 0.592593 & 2.55  & 1.33  & 3.442616 \\
    Ba2 La Bi O6 & 0.931895 & 0.663704 & 2.55  & 1.88  & 3.638862 \\
    Ba2 Sb Bi O6 & 0.96676 & 0.603704 & 2.55  & 1.405 & 3.45978 \\
    Ba ((Ce0.5 Bi0.5) O3) & 0.936481 & 0.655556 & 2.55  & 1.87  & 3.605649 \\
    Ba (Zr0.9 Bi0.1) O3 & 1.002652 & 0.546296 & 2.55  & 1.753 & 3.422888 \\
    Ba (Bi0.5 Pb0.5) O3 & 0.957911 & 0.618519 & 2.55  & 1.495 & 3.49026 \\
    Ba (Pr0.75 Bi0.25) O3 & 0.94654 & 0.637963 & 2.55  & 2.0875 & 3.543711 \\
    Ba2 Bi Sb O6 & 0.96676 & 0.603704 & 2.55  & 1.405 & 3.45978 \\
    Sr2 Bi Nd O6 & 0.888061 & 0.645556 & 2.49  & 1.86  & 4.129623 \\
    Ba Bi0.67 Ca0.33 O3 & 0.956073 & 0.62163 & 2.55  & 1.7566 & 3.497777 \\
    Sr2 Bi Lu O6 & 0.913135 & 0.60037 & 2.49  & 1.795 & 3.848118 \\
    Ba (Bi0.5 In0.5) O3 & 0.982646 & 0.577778 & 2.55  & 1.54  & 3.427237 \\
    (Sr0.44 K0.56) Bi O3 & 0.945895 & 0.606963 & 2.5628 & 1.42  & 3.844093 \\
    (Ba0.4 K0.6) (Bi O3) & 0.973089 & 0.602963 & 2.592 & 1.42  & 3.738023 \\
    (Sr0.4 K0.6) Bi O3 & 0.950869 & 0.602963 & 2.568 & 1.42  & 3.82319 \\
    (Ba0.58 K0.42) (Bi O3) & 0.960538 & 0.620963 & 2.5794 & 1.42  & 3.744803 \\
    (Bi0.8 Mg0.2) (Mg0.3 Ti0.65) O3 & 0.889275 & 0.451296 & 1.562 & 2.046 & 3.073151 \\
    Bi2 Ni Mn O6 & 0.903376 & 0.461111 & 1.42  & 1.71  & 2.104359 \\
    Bi (Mg0.5 Ti0.5) O3 & 0.885421 & 0.490741 & 1.42  & 2.015 & 2.353289 \\
    Bi2 Zn Ti O6 & 0.881043 & 0.498148 & 1.42  & 1.845 & 2.432745 \\
    Bi ((Fe0.85 Mn0.15) O3) & 0.893188 & 0.477778 & 1.42  & 1.652 & 2.231194 \\
    Bi (Fe0.5 Co0.5) O3 & 0.901092 & 0.464815 & 1.42  & 1.585 & 2.129705 \\
    Bi2 (Mn1.334 Ni0.666) O6 & 0.899947 & 0.466678 & 1.42  & 0.10024 & 2.14306 \\
    Bi (Zn0.5 V0.5) O3 & 0.886522 & 0.488889 & 1.42  & 1.8   & 2.334544 \\
    Bi (Fe0.7 Mn0.3) O3 & 0.893188 & 0.477778 & 1.42  & 1.694 & 2.231194 \\
    (Na0.5 Bi0.5) Ti O3 & 0.951249 & 0.448148 & 1.965 & 1.9   & 3.877894 \\
    (K0.5 Bi0.5) Ti O3 & 0.99646 & 0.448148 & 2.02  & 1.9   & 3.743897 \\
    (Sr0.6 Bi0.4) (Fe O3) & 0.968076 & 0.451111 & 2.062 & 1.61  & 3.164053 \\
    (Bi0.5 Ag0.5) (Ti O3) & 0.931355 & 0.448148 & 1.465 & 1.9   & 3.971692 \\
    (Bi0.7 Sr0.3) (Mn O3) & 0.93812 & 0.452222 & 1.741 & 1.89  & 2.633209 \\
    (La0.4 Bi0.6) (Mn O3) & 0.920125 & 0.477778 & 1.788 & 1.89  & 1.894613 \\
    (Pb0.9 Bi0.1) (Ti O3) & 1.012264 & 0.452963 & 1.555 & 1.9   & 3.560141 \\
    (Bi0.5 Sr0.5) (Cr O3) & 0.971471 & 0.431481 & 1.955 & 1.78  & 3.011132 \\
    (Bi0.5 Gd0.5) (Fe O3) & 0.882023 & 0.477778 & 1.83  & 1.61  & 2.412267 \\
    (Bi0.75 La0.25) (Fe O3) & 0.910023 & 0.477778 & 1.65  & 1.61  & 2.006529 \\
    (Bi0.85 Sm0.15) Fe O3 & 0.896909 & 0.477778 & 1.5475 & 1.61  & 2.176919 \\
    (Bi0.5 Sr0.5) (Mn O3) & 0.968964 & 0.435185 & 1.83  & 1.89  & 3.006043 \\
    (Bi0.6 Nd0.4) (Fe O3) & 0.907365 & 0.477778 & 1.772 & 1.61  & 2.038629 \\
    (Bi0.75 Pr0.25) Fe O3 & 0.893985 & 0.477778 & 1.6425 & 1.61  & 2.219324 \\
    (Bi0.5 Ca0.5) (Fe O3) & 0.937411 & 0.455556 & 1.93  & 1.61  & 3.097583 \\
    (Bi0.7 Dy0.3) (Fe O3) & 0.883937 & 0.477778 & 1.66  & 1.61  & 2.379138 \\
    (Bi0.8 La0.2) Ni O3 & 0.927579 & 0.444444 & 1.7025 & 1.53  & 1.88114 \\
    (Bi0.90 Y0.10) (Fe O3) & 0.88982 & 0.477778 & 1.5   & 1.61  & 2.282808 \\
    (Bi0.75 Ba0.25) (Fe O3) & 0.939238 & 0.466667 & 1.7025 & 1.61  & 2.462057 \\
    Ca0.75 Bi0.25 Mn O3 & 0.980781 & 0.413889 & 2.185 & 1.89  & 3.555244 \\
    Bi0.5 Pb0.5 Cr O3 & 0.980619 & 0.431481 & 1.495 & 1.78  & 2.981501 \\
    Tb0.5 Bi0.5 (Mn O3) & 0.879896 & 0.477778 & 1.88  & 1.89  & 2.450169 \\
   
 \end{longtable*}%

\begin{table*}[t]
  \centering
  \caption{Experimental data for unstable double perovskites}
    \begin{tabular}{|lllll | r | r | r| r | r |}
\hline
    Formula &       &       &       &       & Shannon tolerance factor & Shannon octahedral factor &  en-O - Ave-A & en-O - Ave-B & $\tau$ \\
\hline

 \hline

    Ca    & Ca    & Bi    & V     & O     & 0.890921 & 0.581481 & 2.44  & 1.615 & 4.104166 \\
    Ca    & Ca    & Bi    & As    & O     & 0.907932 & 0.551852 & 2.44  & 1.34  & 3.939962 \\
    Cd    & Cd    & Bi    & As    & O     & 0.897806 & 0.551852 & 1.75  & 1.34  & 4.043104 \\
    Mn    & Mn    & Bi    & P     & O     & 0.79485 & 0.522222 & 1.89  & 1.335 & 6.736051 \\
    Mg    & Mg    & Bi    & V     & O     & 0.741883 & 0.581481 & 2.13  & 1.615 & 15.78217 \\
    Ba    & Ba    & Bi    & Ru    & O     & 0.974638 & 0.590741 & 2.55  & 1.33  & 3.440226 \\
    Cu    & Cu    & Bi    & As    & O     & 0.702044 & 0.551852 & 1.54  & 1.34  & -98.5381 \\
    Bi    & Mn    & Fe    & Fe    & O     & 0.86904 & 0.455556 & 1.655 & 1.61  & 3.829308 \\
    Cu    & Cu    & Bi    & V     & O     & 0.688891 & 0.581481 & 1.54  & 1.615 & -27.8845 \\
    Cu    & Cu    & Bi    & P     & O     & 0.715709 & 0.522222 & 1.54  & 1.335 & 57.34434 \\
    Bi    & Bi    & Cd    & Ge    & O     & 0.852588 & 0.548148 & 1.42  & 1.59  & 3.17829 \\
    Pb    & Pb    & Bi    & P     & O     & 0.977218 & 0.522222 & 1.11  & 1.335 & 3.563378 \\
    Zn    & Zn    & Bi    & P     & O     & 0.774205 & 0.522222 & 1.79  & 1.335 & 8.370353 \\
    Mg    & Mg    & Bi    & P     & O     & 0.770764 & 0.522222 & 2.13  & 1.335 & 8.749942 \\
    Pb    & Pb    & Bi    & V     & O     & 0.940601 & 0.581481 & 1.11  & 1.615 & 3.643428 \\
\hline
\hline
    \end{tabular}%
  \label{non stable double perovskite}%
\end{table*}%

\section{Two feature neural network analysis of combined single and double perovskite data}
\label{Sec:B}

\begin{figure}[htb!]
\centering
\includegraphics[width=90mm]{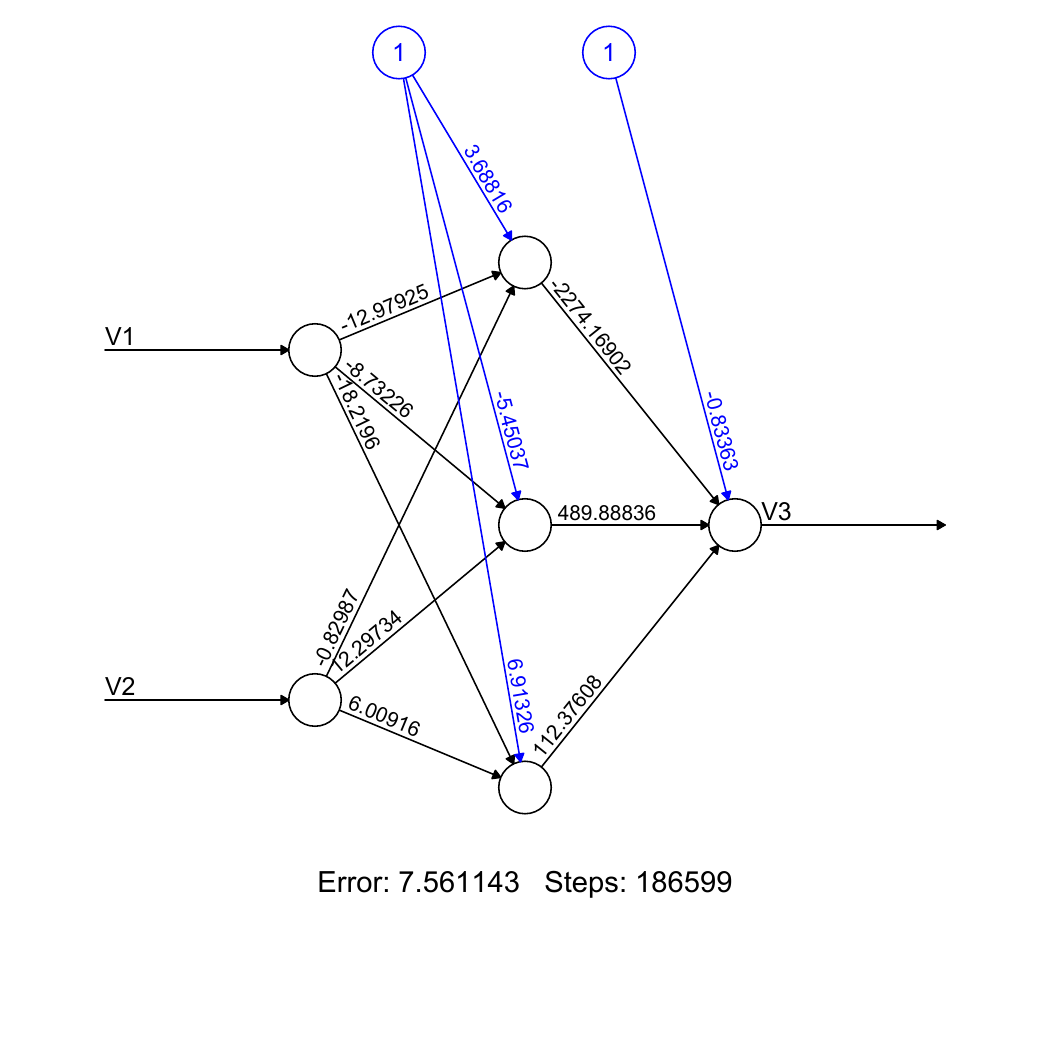}
\caption{A neural network model with three neurons in the hidden ($\alpha =2$) layer constructed for examining the combined double perovskite (Appendix \ref{Sec:D}) and cubic perovskite formability data \cite{formability} (Section \ref{NND2}). The neural network takes as input data the tolerance (V1) and octahedral factors (V2). The weights provided in the figure correspond to typical optimized values once the training is complete.  The weights on the dark links denote the constants $\{w^{(\alpha')}_{kk'}\}$ in Eq. (\ref{ek}); the numerical constants on the blue (color online) links from the external ``1" ovals denote the values of the constants $\{c^{(\alpha')}_{k}\}$. The output ``V3" is the class of the instances: 1 if the compound is stable and 0 if the compound is unstable. The quoted ``error''  in this figure is that of the root mean square error of the  ``neuralnet'' library package \cite{neuralnet,neuralnet1,neuralnet2,neuralnet3,neuralnet4}.
   \label{Table-1}}
\end{figure}

\begin{figure}[ht!]
\centering
\includegraphics[width=90mm]{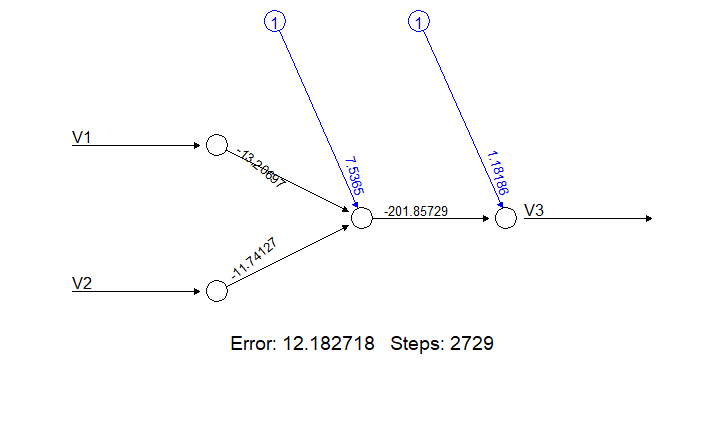}
\caption{A minimal neural network model constructed for examining the combined double perovskite (Appendix \ref{Sec:D}) and cubic perovskite formability data \cite{formability} (Section \ref{NND2}). The legend is the same as in Figure \ref{Table-1}. 
\label{Rplot1}}
\end{figure}

\begin{figure}[ht!]
\centering
\includegraphics[width=90mm]{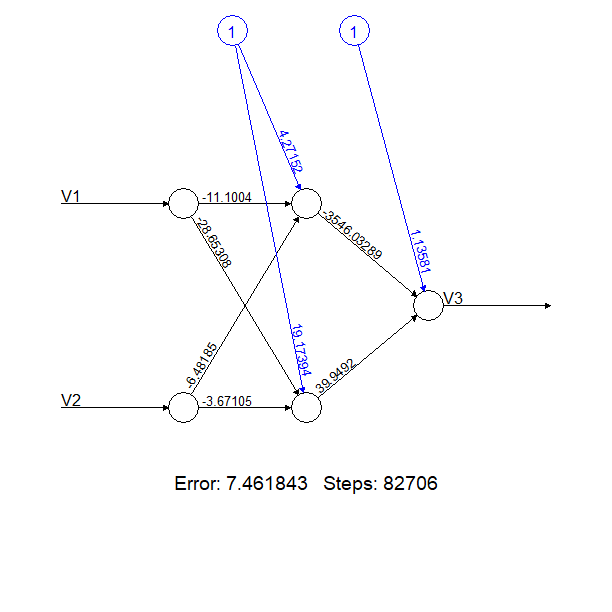}
\caption{A single hidden layer neural network model constructed for examining the combined double perovskite (Appendix \ref{Sec:D}) and cubic perovskite formability data \cite{formability} (Section \ref{NND2}). The legend is the same as in Figure \ref{Table-1}. 
 \label{rplot2}}
\end{figure}

\begin{figure}[ht!]
\centering
\includegraphics[width=90mm]{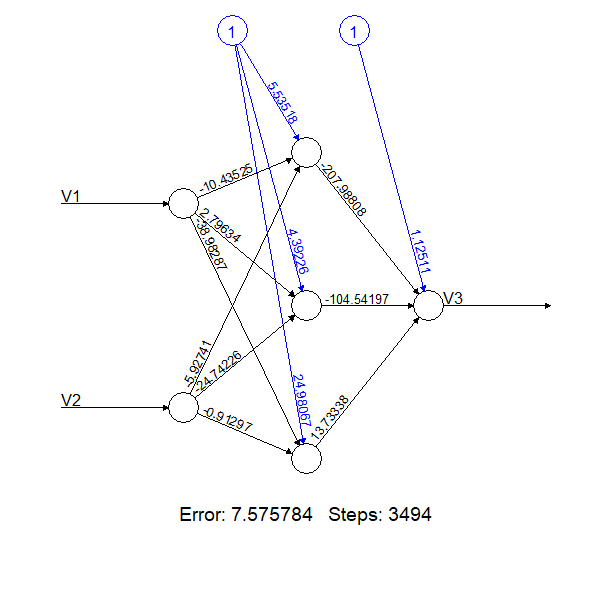}
\caption{A single hidden layer neural network constructed for examining the combined double perovskite (Appendix \ref{Sec:D}) and cubic perovskite formability data \cite{formability} (Section \ref{NND2}). The legend is the same as in Figure \ref{Table-1}.
\label{rplot3}}
\end{figure}

\begin{figure}[ht!]
\centering
\includegraphics[width=90mm]{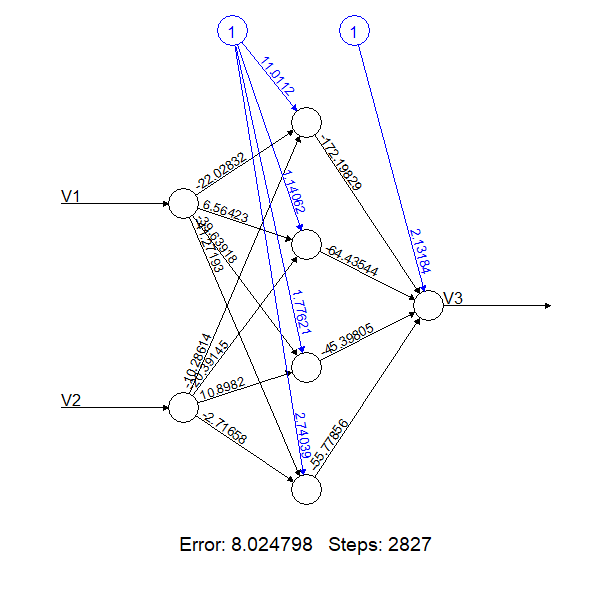}
\caption{A single hidden layer neural network for examining the combined double perovskite (Appendix \ref{Sec:D}) and cubic perovskite formability data \cite{formability} (Section \ref{NND2}). The legend is the same as in Figure \ref{Table-1}.
 \label{rplot4}}
\end{figure}

\begin{figure}[ht!]
\centering
\includegraphics[width=90mm]{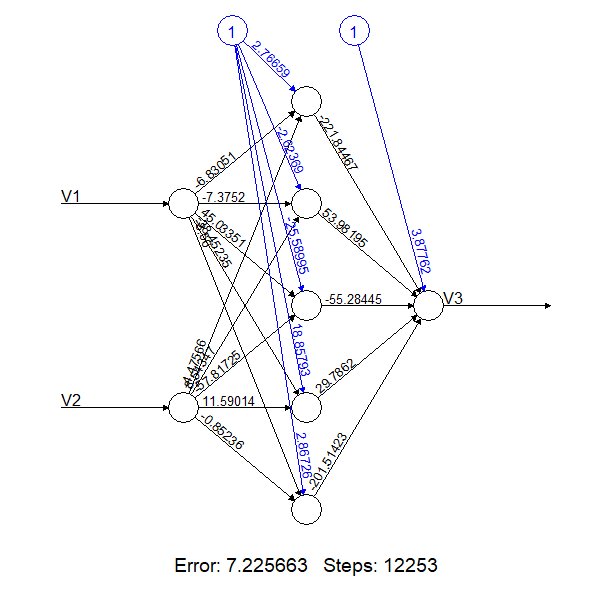}
\caption{A single hidden layer neural network model for examining the combined double perovskite (Appendix \ref{Sec:D}) and cubic perovskite formability data \cite{formability} (Section \ref{NND2}). The legend is the same as in Figure \ref{Table-1}.  \label{rplot5}}
\end{figure}

\begin{figure}[ht!]
\centering
\includegraphics[width=90mm]{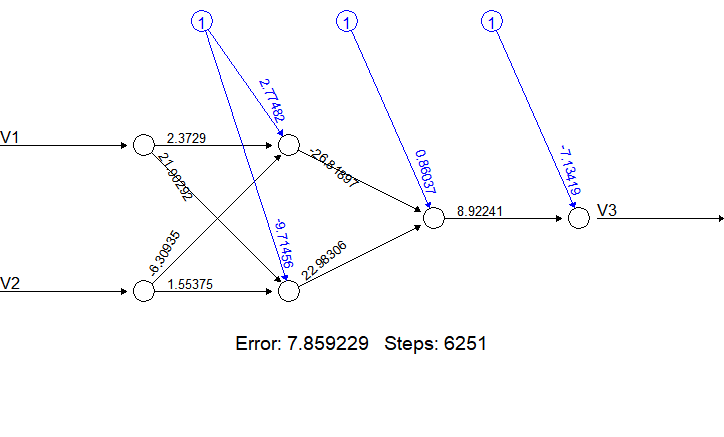}
\caption{A two hidden layer neural network designed for examining the combined double perovskite (Appendix \ref{Sec:D}) and cubic perovskite formability data \cite{formability} (Section \ref{NND2}). The legend is the same as in Figure \ref{Table-1}. \label{rplot2-1}}
\end{figure}

\begin{figure}[ht!]
\centering
\includegraphics[width=90mm]{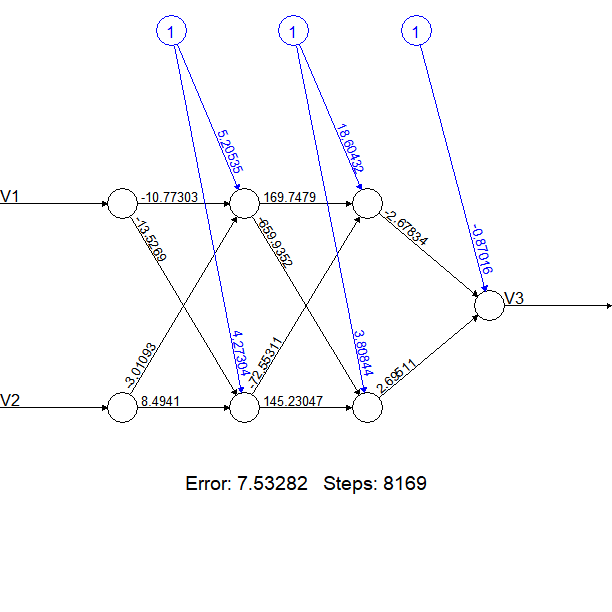}
\caption{A two hidden layer for examining the combined double perovskite (Appendix \ref{Sec:D}) and cubic perovskite formability data \cite{formability} (Section \ref{NND2}). The legend is the same as in Figure \ref{Table-1}. \label{rplot2-2}}
\end{figure}

\begin{figure}[ht!]
\centering
\includegraphics[width=90mm]{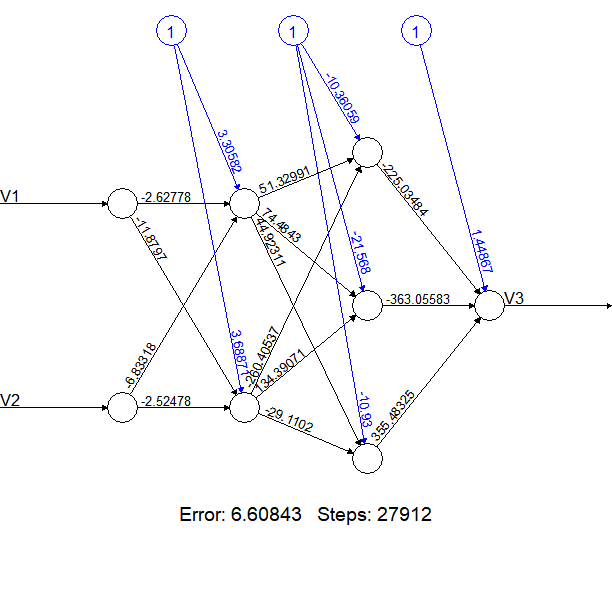}
\caption{A two hidden layer constructed for examining the combined double perovskite (Appendix \ref{Sec:D}) and cubic perovskite formability data \cite{formability} (Section \ref{NND2}). The legend is the same as in Figure \ref{Table-1}. \label{rplot2-3}}
\end{figure}

\begin{figure}[ht!]
\centering
\includegraphics[width=90mm]{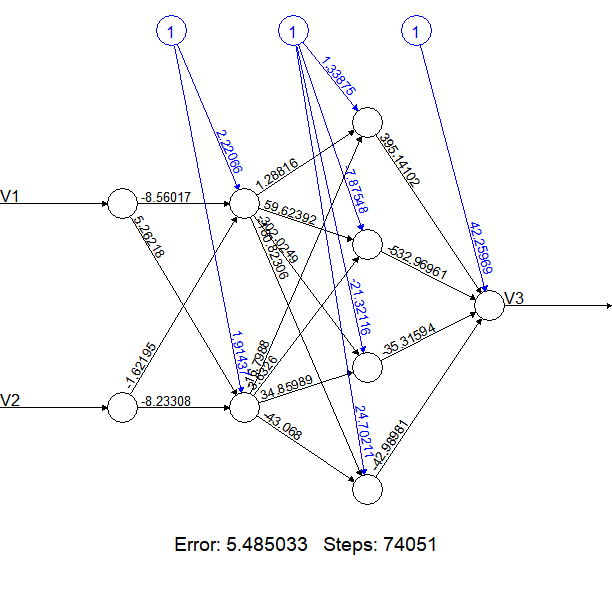}
\caption{A two hidden layer neural network for examining the combined double perovskite (Appendix \ref{Sec:D}) and cubic perovskite formability data \cite{formability} (Section \ref{NND2}). The legend is the same as in Figure \ref{Table-1}.
\label{rplot2-4}}
\end{figure}

\begin{figure}
\centering
\includegraphics[width=90mm]{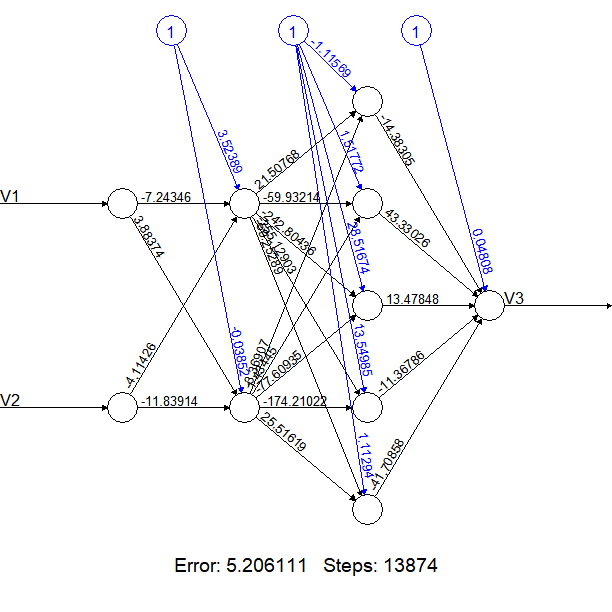}
\caption{A two hidden layer network constructed for examining the combined double perovskite (Appendix \ref{Sec:D}) and cubic perovskite formability data \cite{formability} (Section \ref{NND2}). The legend is the same as in Figure \ref{Table-1}. \label{rplot2-5}}
\end{figure}

In this Appendix, we sketch the functions that we obtained when employing the ``neuralnet'' library \cite{neuralnet,neuralnet1,neuralnet2,neuralnet3,neuralnet4} to build optimal neural net type functions (Section \ref{NN:sec}) when using only two features (octahedral and tolerance factors) and provide specific information underlying Section \ref{NND2}. A synopsis of our analysis using the neural net work of figure \ref{Table-1} appears in Table \ref{tab:3-nod-algorithm-result}. We peruse the results obtained when the architecture is altered. In all of the figures, we sketch the neural networks. ``V1" and ``V2" are the tolerance and octahedral factors respectively. The final output ``V3" provides the predicted classification of the individual input vectors $(v_1, v_2)$; if the compound is stable, $V3$ is set equal to one while if the compound is unstable, $V3 =0$. A synopsis of our results with different neural network architectures is provided in Table \ref{tab:3-nod-algorithm-result} (Figure \ref{Table-1}) and in Tables \ref{tab:neural net voting} and \ref{tab:neural net voting examination}  (for 
all \cref{Table-1,Rplot1,rplot2,rplot3,rplot4,rplot5,rplot2-1,rplot2-2,rplot2-3,rplot2-4,rplot2-5} and other additional networks not displayed). Not all neural networks that we trained were able to converge on general cross-validation tests (hence the sets of examined networks in Table \ref{tab:neural net voting} and Table  \ref{tab:neural net voting examination} for training and cross-validation tests are not identical yet both Tables contain results for the architectures of \cref{Table-1,Rplot1,rplot2,rplot3,rplot4,rplot5,rplot2-1,rplot2-2,rplot2-3,rplot2-4,rplot2-5}). As is seen in Table \ref{tab:neural net voting examination}, the cross-validation accuracy is not monotonic in the number of neurons and seems to be optimized when the number of neurons in the second layer is larger than five; this is an analog of the dependence of the overlap on the number anchor points for the Gaussian SRVM  that we analyzed in Sections \ref{Perovskite} and \ref{CAB}. The neural network parameters of Figure \ref{Table-1} correspond to a typical ``snapshot'' in which we optimized the parameters in a cross-validation test in which a fraction of the data were removed/
The parameters displayed in all other neural network figures (i.e., \cref{Rplot1,rplot2,rplot3,rplot4,rplot5,rplot2-1,rplot2-2,rplot2-3,rplot2-4,rplot2-5}) correspond to the values found by training the respective networks over all known data (i.e., in these no data were removed as in cross-validation tests). 

\begin{table*}[t]
\begin{center}
\caption{The training and cross-validation accuracy data for the neural net model of figure \ref{Table-1}. Herein, we analyzed two-feature (tolerance and octahedral factor) 
cubic perovskite \cite{formability} and double perovskite data, see Section \ref{NND2}.  \label{tab:3-nod-algorithm-result} }
\begin{tabular}{ |c|c|c| }
\hline
 & Cross-validated Training Accuracy & Cross-validated Testing Accuracy \\

\hline
Number of repetitions  &  &  \\
\hline
1 & 0.928& 0.899 \\
\hline
2 & 0.926& 0.909 \\
\hline
3 & 0.925& 0.906 \\
\hline
4 & 0.927&0.889 \\
\hline
5 & 0.927&0.899 \\
\hline
Average  &0.926 & 0.900 \\

\hline
\end{tabular}
\end{center}
\end{table*}

\begin{table*}[t]
\begin{center}
\caption{The training accuracies of different neural net models (\cref{Table-1,Rplot1,rplot2,rplot3,rplot4,rplot5,rplot2-1,rplot2-2,rplot2-3,rplot2-4,rplot2-5}) used for the two-feature (tolerance and octahedral factor) 
cubic perovskite \cite{formability} and double perovskite data, see Section \ref{NND2}.  \label{tab:neural net voting} }
\begin{tabular}{ |c|c|c|c|c|c|c| }
\hline
&&\multicolumn{5}{p{2cm}|}{\centering  Number of nodes in the second layer }\\

\hline
No. of layers & No. of nodes in the first layer &1&2&3&4&5  \\
\hline
1 &  &0.862&0.929&0.929& 0.929&0.919 \\
\hline
2 &2  & 0.925&0.936&0.929&0.929&0.929 \\
\hline
2 &3 &  0.932&0.929&0.946&0.932&0.936 \\
\hline
2&4  &0.939&0.936&0.939&0.936& \\
\hline
2& 5 & 0.936& 0.936& 0.936& 0.939& \\
\hline

\hline
\end{tabular}
\end{center}
\end{table*}

\begin{table*}[t]

\caption{The individual 5-fold cross validation testing accuracies of different neural net models (\cref{Table-1,Rplot1,rplot2,rplot3,rplot4,rplot5,rplot2-1,rplot2-2,rplot2-3,rplot2-4,rplot2-5}) for the combined formability \cite{formability} and double perovskite data (Appendix \ref{Sec:D}) sets, see Section \ref{NND2}. The first row is for a single hidden layer network of, respectively, $1 \le n \le 5$ nodes.   \label{tab:neural net voting examination} }

\newcommand*{\TitleParbox}[1]{\parbox[c]{1.75cm}{\raggedright #1}}%
\begin{tabular}{ |c|c|c|c|c|c|c|c|c|}
\hline
&&\multicolumn{6}{p{1cm}|}{  \centering Number of nodes in the second layer }&  \TitleParbox{Averaged voting cross-validation accuracy}\\

\hline
\parbox[c]{1.5cm}{No. of layers} &\parbox[c]{2cm}{\raggedright No. of nodes in the first layer}  &1&2&3&4&5&6&  \\
\hline
1 &  &0.86&0.895&0.9& 0.91&0.88&& \\
\hline
2 &2  & 0.858&0.898&0.9&0.9&0.9&0.87& \\
\hline
2 &3 &  0.9&0.9&0.9&0.89&0.89& 0.87&\\
\hline
2&5  &0.89&0.9&0.88&0.88& 0.88&&\\
\hline
\multicolumn{2}{p{5cm}|}{  \centering {Averaged voting testing accuracy} }&  &&&&&&0.88\\

\hline

\hline
\end{tabular}

\end{table*}

We next comment on the results obtained with different neural net architectures. Figure \ref{Rplot1} displays the smallest neural net that we employed. We applied (as outlined in Section \ref{NN:sec}) SRVM to neural net type functions. Specifically, we constructed 21 different neural net models- ``replicas''. A further subset of these models is depicted in  \cref{Rplot1,rplot2,rplot3,rplot4,rplot5} for single hidden layer neural networks and in \cref{rplot2-1,rplot2-2,rplot2-3,rplot2-4,rplot2-5} for two-hidden layer neural nets. The predictions and accuracies of each individual neural net model have been recorded for comparison purposes. As in the earlier Sections, the data examined had two features (the tolerance and octahedral factors) for the cubic perovskite formability \cite{formability} and the double perovskite (Appendix \ref{Sec:D})  datasets.  In Table \ref{tab:neural net voting}, we report on the neural net models with their corresponding training accuracies. More important than training alone, the predictions for the test data amongst the different classifiers were seen to be stable. Cross-validation accuracy results for the 21 different neural nets for the testing data sets are provided in table \ref{tab:neural net voting examination}. Without applying overlap measures (as in Section \ref{Gaussian}), the average accuracy for the testing data set after voting amongst the replicas is 0.88.


\clearpage

\section{Probability for candidate double perovskites based on the tolerance and octahedral factors}
\label{Sec:C}

Below we expand on the summary of our reported findings in Section \ref{srvm-predict-2} and list the probabilities that different double perovskites will constitute new stable compounds. These probabilities are computed as an average over the predictions of individual neural network solvers. The two features employed in these neural networks are the tolerance and octahedral factors of Eqs. (\ref{x1eq},\ref{x2eq},\ref{rabab}).  There are, as in highlighted in Section \ref{srvm-predict-2}, eight candidate compounds that we predict to be stable perovskite with high confidence (unity when averaged over the different neural networks). 

\begin{longtable*}[htbp]{|c|c|c|c|c|c|}
 \caption{Predictions for candidate double perovskite based on the combined cubic perovskite \cite{formability} and double perovskite (Appendix \ref{Sec:D}) data with only  two features: the tolerance and octahedral factors, see Section \ref{srvm-predict-2}. The probability that each individual composition A'A''B'B''O$_{6}$ will realize a stable perovskite structure is given as the fraction of the neural net models  (\cref{Table-1,Rplot1,rplot2,rplot3,rplot4,rplot5,rplot2-1,rplot2-2,rplot2-3,rplot2-4,rplot2-5}) that predict a stable double perovskite.  \label{Stable double perovskite }}\\

\hline

 \multicolumn{5}{| c |}{\centering Double perovskite with BI at site B }    & Probability to make a stable perovskite\\ 
 \hline
 \endfirsthead
 
 \hline

 \multicolumn{5}{| c |}{\centering Double perovskite with BI at site B }    & Probability to make a stable perovskite \\ 
 \hline
 \endhead
 
 \hline
 \endfoot
 
 \hline

 \hline
\hline
 \endlastfoot

    Ag:   & Cs:   & Mn:   & Bi:   & O:    & 0 \\
    Ag:   & Cs:   & Tc:   & Bi:   & O:    & 0.875 \\
    Ba:   & Ba:   & Nb:   & Bi:   & O:    & 1 \\
    Ba:   & Tl:   & Pb:   & Bi:   & O:    & 0 \\
    Ba:   & Ba:   & Sb:   & Bi:   & O:    & 1 \\
    Ba:   & Ba:   & Ta:   & Bi:   & O:    & 1 \\
    Ba:   & Ba:   & V:    & Bi:   & O:    & 0.125 \\
    Ca:   & Cs:   & Cr:   & Bi:   & O:    & 0 \\
    Ca:   & Rb:   & Cr:   & Bi:   & O:    & 0 \\
    Ca:   & Cs:   & Fe:   & Bi:   & O:    & 0 \\
    Ca:   & Rb:   & Fe:   & Bi:   & O:    & 0 \\
    Ca:   & Cs:   & Mo:   & Bi:   & O:    & 1 \\
    Ca:   & Rb:   & Mo:   & Bi:   & O:    & 0.125 \\
    Ca:   & Cs:   & Te:   & Bi:   & O:    & 0.625 \\
    Ca:   & Rb:   & Te:   & Bi:   & O:    & 0 \\
    Ca:   & Rb:   & W:    & Bi:   & O:    & 0.125 \\
    Cd:   & Cs:   & Cr:   & Bi:   & O:    & 0 \\
    Cd:   & Cs:   & Fe:   & Bi:   & O:    & 0 \\
    Cd:   & Cs:   & Mo:   & Bi:   & O:    & 1 \\
    Cd:   & Cs:   & Te:   & Bi:   & O:    & 0.75 \\
    Cd:   & Cs:   & W:    & Bi:   & O:    & 1 \\
    Cs:   & Pb:   & Cr:   & Bi:   & O:    & 0.125 \\
    Cs:   & Pb:   & Fe:   & Bi:   & O:    & 0 \\
    Cs:   & Hf:   & Ge:   & Bi:   & O:    & 0 \\
    Cs:   & Hf:   & Hf:   & Bi:   & O:    & 0 \\
    Cs:   & Pb:   & Hf:   & Bi:   & O:    & 0 \\
    Cs:   & Hf:   & Ir:   & Bi:   & O:    & 0 \\
    Cs:   & Hf:   & Mn:   & Bi:   & O:    & 0 \\
    Cs:   & Pb:   & Mn:   & Bi:   & O:    & 0 \\
    Cs:   & Hf:   & Mo:   & Bi:   & O:    & 0 \\
    Cs:   & Hg:   & Mo:   & Bi:   & O:    & 0 \\
    Cs:   & Pb:   & Mo:   & Bi:   & O:    & 0.125 \\
    Cs:   & Tl:   & Nb:   & Bi:   & O:    & 0 \\
    Cs:   & Hf:   & Os:   & Bi:   & O:    & 0 \\
    Cs:   & Hf:   & Pb:   & Bi:   & O:    & 0 \\
    Cs:   & Pb:   & Pb:   & Bi:   & O:    & 0 \\
    Cs:   & Hf:   & Pd:   & Bi:   & O:    & 0 \\
    Cs:   & Pb:   & Pd:   & Bi:   & O:    & 0 \\
    Cs:   & Hf:   & Pt:   & Bi:   & O:    & 0 \\
    Cs:   & Hf:   & Re:   & Bi:   & O:    & 0 \\
    Cs:   & Hf:   & Ru:   & Bi:   & O:    & 0 \\
    Cs:   & Tl:   & Sb:   & Bi:   & O:    & 0 \\
    Cs:   & Hf:   & Sn:   & Bi:   & O:    & 0 \\
    Cs:   & Pb:   & Sn:   & Bi:   & O:    & 0 \\
    Cs:   & Tl:   & Ta:   & Bi:   & O:    & 0 \\
    Cs:   & Hf:   & Tc:   & Bi:   & O:    & 0 \\
    Cs:   & Hf:   & Te:   & Bi:   & O:    & 0.125 \\
    Cs:   & Pb:   & Te:   & Bi:   & O:    & 0.125 \\
    Cs:   & Hf:   & Ti:   & Bi:   & O:    & 0 \\
    Cs:   & Pb:   & Ti:   & Bi:   & O:    & 0 \\
    Cs:   & Hf:   & W:    & Bi:   & O:    & 0 \\
    Cs:   & Ta:   & Y:    & Bi:   & O:    & 0 \\
    Cs:   & Hf:   & Zr:   & Bi:   & O:    & 0 \\
    Cs:   & Pb:   & Zr:   & Bi:   & O:    & 0 \\
    In:   & Cs:   & Nb:   & Bi:   & O:    & 0 \\
    In:   & Cs:   & Sb:   & Bi:   & O:    & 0 \\
    In:   & Cs:   & Ta:   & Bi:   & O:    & 0 \\
    In:   & Cs:   & V:    & Bi:   & O:    & 0 \\
    K:    & Ba:   & Cr:   & Bi:   & O:    & 0 \\
    K:    & Sr:   & Cr:   & Bi:   & O:    & 0 \\
    K:    & Ba:   & Fe:   & Bi:   & O:    & 0 \\
    K:    & Sr:   & Fe:   & Bi:   & O:    & 0 \\
    K:    & K:    & Mn:   & Bi:   & O:    & 0.125 \\
    K:    & Ba:   & Mo:   & Bi:   & O:    & 1 \\
    K:    & Sr:   & Mo:   & Bi:   & O:    & 0.875 \\
    K:    & Tl:   & Nb:   & Bi:   & O:    & 0 \\
    K:    & Tl:   & Sb:   & Bi:   & O:    & 0 \\
    K:    & Tl:   & Ta:   & Bi:   & O:    & 0 \\
    K:    & Ba:   & Te:   & Bi:   & O:    & 0.125 \\
    K:    & Sr:   & Te:   & Bi:   & O:    & 0 \\
    K:    & Ba:   & W:    & Bi:   & O:    & 1 \\
    K:    & Sr:   & W:    & Bi:   & O:    & 0.875 \\
    Mg:   & Cs:   & Mo:   & Bi:   & O:    & 0 \\
    Mo:   & Cs:   & Ca:   & Bi:   & O:    & 0.125 \\
    Mo:   & Cs:   & Pb:   & Bi:   & O:    & 0 \\
    Na:   & Cs:   & Mn:   & Bi:   & O:    & 0.125 \\
    Na:   & Rb:   & Mn:   & Bi:   & O:    & 0 \\
    Na:   & Rb:   & Tc:   & Bi:   & O:    & 0.625 \\
    Nb:   & Cs:   & Y:    & Bi:   & O:    & 0 \\
    Rb:   & Pb:   & Cr:   & Bi:   & O:    & 0 \\
    Rb:   & Sr:   & Cr:   & Bi:   & O:    & 0 \\
    Rb:   & Pb:   & Fe:   & Bi:   & O:    & 0 \\
    Rb:   & Sr:   & Fe:   & Bi:   & O:    & 0 \\
    Rb:   & Pb:   & Ge:   & Bi:   & O:    & 0 \\
    Rb:   & Pb:   & Hf:   & Bi:   & O:    & 0 \\
    Rb:   & Pb:   & Ir:   & Bi:   & O:    & 0 \\
    Rb:   & Pb:   & Mn:   & Bi:   & O:    & 0 \\
    Rb:   & Tl:   & Mn:   & Bi:   & O:    & 0.125 \\
    Rb:   & Ba:   & Mo:   & Bi:   & O:    & 0.25 \\
    Rb:   & Pb:   & Mo:   & Bi:   & O:    & 1 \\
    Rb:   & Sr:   & Mo:   & Bi:   & O:    & 1 \\
    Rb:   & Tl:   & Nb:   & Bi:   & O:    & 0 \\
    Rb:   & Y:    & Nb:   & Bi:   & O:    & 0 \\
    Rb:   & Pb:   & Os:   & Bi:   & O:    & 0 \\
    Rb:   & Pb:   & Pb:   & Bi:   & O:    & 0 \\
    Rb:   & Pb:   & Pd:   & Bi:   & O:    & 0 \\
    Rb:   & Pb:   & Pt:   & Bi:   & O:    & 0 \\
    Rb:   & Pb:   & Re:   & Bi:   & O:    & 0 \\
    Rb:   & Pb:   & Ru:   & Bi:   & O:    & 0 \\
    Rb:   & Tl:   & Sb:   & Bi:   & O:    & 0 \\
    Rb:   & Y:    & Sb:   & Bi:   & O:    & 0 \\
    Rb:   & Pb:   & Sn:   & Bi:   & O:    & 0 \\
    Rb:   & Tl:   & Ta:   & Bi:   & O:    & 0 \\
    Rb:   & Y:    & Ta:   & Bi:   & O:    & 0 \\
    Rb:   & Pb:   & Tc:   & Bi:   & O:    & 0 \\
    Rb:   & Tl:   & Tc:   & Bi:   & O:    & 0.125 \\
    Rb:   & Pb:   & Te:   & Bi:   & O:    & 0.625 \\
    Rb:   & Sr:   & Te:   & Bi:   & O:    & 0.875 \\
    Rb:   & Pb:   & Ti:   & Bi:   & O:    & 0 \\
    Rb:   & Tl:   & V:    & Bi:   & O:    & 0 \\
    Rb:   & Y:    & V:    & Bi:   & O:    & 0 \\
    Rb:   & Pb:   & W:    & Bi:   & O:    & 1 \\
    Rb:   & Sr:   & W:    & Bi:   & O:    & 1 \\
    Rb:   & Pb:   & Zr:   & Bi:   & O:    & 0 \\
    Sb:   & Cs:   & Y:    & Bi:   & O:    & 0 \\
    Sc:   & Cs:   & Nb:   & Bi:   & O:    & 0 \\
    Sc:   & Cs:   & Sb:   & Bi:   & O:    & 0 \\
    Sc:   & Cs:   & Ta:   & Bi:   & O:    & 0 \\
    Sc:   & Cs:   & V:    & Bi:   & O:    & 0 \\
    Sn:   & Cs:   & Pb:   & Bi:   & O:    & 0 \\
    Sr:   & Ba:   & Nb:   & Bi:   & O:    & 0.625 \\
    Sr:   & Ba:   & Sb:   & Bi:   & O:    & 0.125 \\
    Sr:   & Ba:   & Ta:   & Bi:   & O:    & 0.625 \\
    Sr:   & Ba:   & V:    & Bi:   & O:    & 0 \\
    Y:    & Cs:   & Nb:   & Bi:   & O:    & 0 \\
    Y:    & Cs:   & Sb:   & Bi:   & O:    & 0 \\
    Y:    & Cs:   & Ta:   & Bi:   & O:    & 0 \\
    Zr:   & Cs:   & Ge:   & Bi:   & O:    & 0 \\
    Zr:   & Cs:   & Hf:   & Bi:   & O:    & 0 \\
    Zr:   & Cs:   & Ir:   & Bi:   & O:    & 0 \\
    Zr:   & Cs:   & Mn:   & Bi:   & O:    & 0 \\
    Zr:   & Cs:   & Mo:   & Bi:   & O:    & 0 \\
    Zr:   & Cs:   & Os:   & Bi:   & O:    & 0 \\
    Zr:   & Cs:   & Pb:   & Bi:   & O:    & 0 \\
    Zr:   & Cs:   & Pd:   & Bi:   & O:    & 0 \\
    Zr:   & Cs:   & Pt:   & Bi:   & O:    & 0 \\
    Zr:   & Cs:   & Re:   & Bi:   & O:    & 0 \\
    Zr:   & Cs:   & Ru:   & Bi:   & O:    & 0 \\
    Zr:   & Cs:   & Sn:   & Bi:   & O:    & 0 \\
    Zr:   & Cs:   & Tc:   & Bi:   & O:    & 0 \\
    Zr:   & Cs:   & Te:   & Bi:   & O:    & 0.125 \\
    Zr:   & Cs:   & Ti:   & Bi:   & O:    & 0 \\
    Zr:   & Cs:   & W:    & Bi:   & O:    & 0 \\
    Zr:   & Cs:   & Zr:   & Bi:   & O:    & 0 \\

\hline
   
 \end{longtable*}%

\section{Five Feature Empirical data and our predictions for stable double perovskite materials}
\label{sec:e}

Below we give predictions for stability (See Section \ref{Section:5feature}) based on all five features data appearing in provide Appendix \ref{Sec:D}. Listed below are predictions made with different algorithms- 
SVM, Gaussian kernel SRVM, and neural networks of different topologies (having a total of three- and five- hidden neurons). In Appendix \ref{Sec:E} we provide further details on these neural networks. 
Appendix \ref{Sec:F} provide the explicit forms of some of the SRVM functions employed. 

\begin{longtable*}[htbp]{| c | c | c | c | c | c | c | c | c |}
 \caption{Predictions for stable double perovskite A'A''B'B''O$_{6}$ compounds based on the data of Appendix \ref{Sec:D}. These predictions were performed with four different algorithms: SVM, SRVM and two neural networks with three and five hidden neurons (N-N-3 and N-N-5). The predictions are based on the full 5 feature double perovskite training data of Appendix \ref{Sec:D}. The detailed neural network and Gaussian kernel SRVM functions are, respectively, given in the \cref{SRVM1,SRVM2,SRVM3,SRVM4,SRVM5,SRVM6,SRVM7,SRVM8,SRVM9} and \cref{5-nod-neural-function,3-nod-neural-function}. A value of ``1'' in any of the columns corresponds to a stable perovskite structure while ``0'' indicates an unstable structure. \label{Screened Predictions}}\\
 
 \hline

 \hline

 \hline
    \multicolumn{5}{| c |}{\centering Double Perovskite with BI at site B }  &{N-N-3} & {N-N-5} & {SRVM} & {SVM} \\
 \hline
 \endhead
 
 \hline
 \endfoot
  
 \hline

 \hline\hline
 \endlastfoot

         \multicolumn{5}{| c |}{\centering Double Perovskite with BI at site B }  & {N-N-3} & {N-N-5} & {SRVM} &{SVM} \\
      Ag: & Bi: & Co: & Mo: & O: & 1  & 1  & 1  & 0 \\
      Ag: & Bi: & Co: & Nb: & O: & 1  & 1  & 1  & 1 \\
      Ag: & Bi: & Co: & Sb: & O: & 1  & 1  & 1  & 0 \\
      Ag: & Bi: & Co: & Ta: & O: & 1  & 1  & 1  & 1 \\
      Ag: & Bi: & Cr: & Fe: & O: & 1  & 1  & 1  & 1 \\
      Ag: & Bi: & Cr: & Mn: & O: & 1  & 1  & 1  & 1 \\
      Ag: & Bi: & Cr: & Pd: & O: & 1  & 1  & 1  & 0 \\
      Ag: & Bi: & Cr: & Te: & O: & 0  & 0  & 1  & 0 \\
      Ag: & Bi: & Cu: & Mo: & O: & 1  & 1  & 1  & 0 \\
      Ag: & Bi: & Fe: & Fe: & O: & 1  & 1  & 1  & 1 \\
      Ag: & Bi: & Fe: & Pd: & O: & 1  & 1  & 1  & 0 \\
      Ag: & Bi: & Fe: & Te: & O: & 1  & 1  & 1  & 0 \\
      Ag: & Bi: & Ge: & Hf: & O: & 1  & 1  & 1  & 1 \\
      Ag: & Bi: & Ge: & Zr: & O: & 1  & 1  & 1  & 1 \\
      Ag: & Bi: & Li: & Tc: & O: & 1  & 1  & 1  & 1 \\
      Ag: & Bi: & Mn: & Fe: & O: & 1  & 1  & 1  & 1 \\
      Ag: & Bi: & Mn: & Mn: & O: & 1  & 1  & 1  & 1 \\
      Ag: & Bi: & Mn: & Pd: & O: & 1  & 1  & 1  & 0 \\
      Ag: & Bi: & Mn: & Te: & O: & 0  & 0  & 1  & 0 \\
      Ag: & Bi: & Mo: & Ir: & O: & 1  & 1  & 0  & 1 \\
      Ag: & Bi: & Mo: & Pt: & O: & 1  & 0  & 0  & 1 \\
      Ag: & Bi: & Mo: & Re: & O: & 1  & 1  & 1  & 0 \\
      Ag: & Bi: & Mo: & Tc: & O: & 1  & 1  & 1  & 0 \\
      Ag: & Bi: & Mo: & W: & O: & 1  & 1  & 0  & 1 \\
      Ag: & Bi: & Nb: & Au: & O: & 0  & 0  & 0  & 0 \\
      Ag: & Bi: & Nb: & Ir: & O: & 1  & 1  & 1  & 0 \\
      Ag: & Bi: & Nb: & Rh: & O: & 1  & 1  & 1  & 0 \\
      Ag: & Bi: & Ni: & Mo: & O: & 1  & 1  & 1  & 0 \\
      Ag: & Bi: & Pd: & Pd: & O: & 1  & 1  & 0  & 1 \\
      Ag: & Bi: & Pd: & Te: & O: & 1  & 0  & 0  & 1 \\
      Ag: & Bi: & Rh: & Sb: & O: & 1  & 1  & 0  & 1 \\
      Ag: & Bi: & Rh: & Ta: & O: & 1  & 1  & 1  & 0 \\
      Ag: & Bi: & Sb: & Au: & O: & 1  & 0  & 0  & 1 \\
      Ag: & Bi: & Sb: & Ir: & O: & 1  & 1  & 1  & 1 \\
      Ag: & Bi: & Sn: & Ir: & O: & 1  & 1  & 1  & 1 \\
      Ag: & Bi: & Sn: & Pt: & O: & 1  & 1  & 1  & 1 \\
      Ag: & Bi: & Sn: & Re: & O: & 1  & 1  & 1  & 0 \\
      Ag: & Bi: & Sn: & W: & O: & 1  & 1  & 0  & 1 \\
      Ag: & Bi: & Ta: & Au: & O: & 0  & 0  & 1  & 0 \\
      Ag: & Bi: & Ta: & Ir: & O: & 1  & 1  & 1  & 0 \\
      Ag: & Bi: & Tc: & Sn: & O: & 1  & 1  & 1  & 0 \\
      Ag: & Bi: & Te: & Te: & O: & 0  & 0  & 0  & 0 \\
      Ag: & Bi: & Ti: & Mn: & O: & 1  & 1  & 1  & 1 \\
      Ag: & Bi: & Ti: & Pd: & O: & 1  & 1  & 1  & 0 \\
      Ag: & Bi: & Ti: & Te: & O: & 0  & 0  & 1  & 0 \\
      Ag: & Bi: & Ti: & Ti: & O: & 1  & 1  & 1  & 1 \\
      Ag: & Bi: & V: & Sb: & O: & 1  & 1  & 1  & 0 \\
      Ag: & Bi: & Zn: & Mo: & O: & 1  & 1  & 1  & 0 \\
      Bi: & Bi: & Al: & In: & O: & 1  & 1  & 1  & 1 \\
      Bi: & Bi: & Co: & Mo: & O: & 0  & 0  & 1  & 0 \\
      Bi: & Bi: & Co: & Sb: & O: & 0  & 1  & 1  & 0 \\
      Bi: & Bi: & Co: & Sn: & O: & 0  & 0  & 1  & 0 \\
      Bi: & Bi: & Cr: & Cr: & O: & 1  & 1  & 1  & 1 \\
      Bi: & Bi: & Cr: & Fe: & O: & 1  & 1  & 1  & 1 \\
      Bi: & Bi: & Cu: & Mo: & O: & 1  & 0  & 1  & 0 \\
      Bi: & Bi: & Cu: & Sn: & O: & 0  & 0  & 1  & 0 \\
      Bi: & Bi: & Fe: & Fe: & O: & 1  & 1  & 1  & 0 \\
      Bi: & Bi: & Fe: & Pd: & O: & 0  & 0  & 1  & 0 \\
      Bi: & Bi: & Fe: & Te: & O: & 0  & 0  & 0  & 0 \\
      Bi: & Bi: & Ge: & Cd: & O: & 0  & 0  & 1  & 0 \\
      Bi: & Bi: & Ge: & Hf: & O: & 1  & 0  & 1  & 0 \\
      Bi: & Bi: & Ge: & Zr: & O: & 0  & 0  & 1  & 0 \\
      Bi: & Bi: & Li: & V: & O: & 1  & 1  & 1  & 1 \\
      Bi: & Bi: & Mg: & Os: & O: & 0  & 1  & 1  & 0 \\
      Bi: & Bi: & Mg: & Ru: & O: & 0  & 1  & 1  & 0 \\
      Bi: & Bi: & Mn: & Fe: & O: & 1  & 1  & 1  & 1 \\
      Bi: & Bi: & Mn: & Mn: & O: & 1  & 1  & 1  & 1 \\
      Bi: & Bi: & Mn: & Pd: & O: & 0  & 0  & 1  & 0 \\
      Bi: & Bi: & Mn: & Te: & O: & 0  & 0  & 0  & 0 \\
      Bi: & Bi: & Mo: & Pt: & O: & 1  & 0  & 0  & 1 \\
      Bi: & Bi: & Ni: & Mo: & O: & 1  & 1  & 1  & 0 \\
      Bi: & Bi: & Ni: & Sn: & O: & 0  & 1  & 1  & 0 \\
      Bi: & Bi: & Os: & Hg: & O: & 0  & 0  & 0  & 0 \\
      Bi: & Bi: & Pd: & Pd: & O: & 1  & 0  & 0  & 1 \\
      Bi: & Bi: & Pd: & Te: & O: & 0  & 0  & 0  & 0 \\
      Bi: & Bi: & Rh: & Sb: & O: & 1  & 0  & 0  & 1 \\
      Bi: & Bi: & Ru: & Hg: & O: & 0  & 0  & 0  & 0 \\
      Bi: & Bi: & Sb: & Au: & O: & 0  & 0  & 0  & 1 \\
      Bi: & Bi: & Sb: & Ir: & O: & 1  & 0  & 1  & 1 \\
      Bi: & Bi: & Sn: & Ir: & O: & 0  & 0  & 0  & 0 \\
      Bi: & Bi: & Sn: & Pt: & O: & 0  & 0  & 0  & 1 \\
      Bi: & Bi: & Sn: & Re: & O: & 0  & 0  & 0  & 0 \\
      Bi: & Bi: & Sn: & W: & O: & 0  & 0  & 0  & 0 \\
      Bi: & Bi: & Tc: & Sn: & O: & 0  & 0  & 0  & 0 \\
      Bi: & Bi: & Ti: & Fe: & O: & 1  & 1  & 1  & 0 \\
      Bi: & Bi: & Ti: & Mn: & O: & 1  & 0  & 1  & 1 \\
      Bi: & Bi: & Ti: & Pd: & O: & 0  & 0  & 1  & 0 \\
      Bi: & Bi: & Ti: & Te: & O: & 0  & 0  & 0  & 0 \\
      Bi: & Bi: & Zn: & Mo: & O: & 0  & 0  & 1  & 0 \\
      Bi: & Bi: & Zn: & Sn: & O: & 0  & 0  & 1  & 0 \\
      Ca: & Bi: & Al: & Hf: & O: & 1  & 1  & 1  & 1 \\
      Ca: & Bi: & Al: & Zr: & O: & 1  & 1  & 1  & 1 \\
      Ca: & Bi: & Co: & Mo: & O: & 0  & 0  & 0  & 0 \\
      Ca: & Bi: & Co: & Nb: & O: & 0  & 1  & 1  & 0 \\
      Ca: & Bi: & Co: & Sb: & O: & 0  & 0  & 0  & 0 \\
      Ca: & Bi: & Co: & Sn: & O: & 0  & 0  & 0  & 0 \\
      Ca: & Bi: & Co: & Ta: & O: & 1  & 1  & 1  & 1 \\
      Ca: & Bi: & Cr: & Mn: & O: & 1  & 1  & 1  & 1 \\
      Ca: & Bi: & Cr: & Pd: & O: & 0  & 0  & 0  & 0 \\
      Ca: & Bi: & Cr: & Te: & O: & 0  & 0  & 1  & 0 \\
      Ca: & Bi: & Cu: & Nb: & O: & 0  & 1  & 1  & 0 \\
      Ca: & Bi: & Cu: & Sb: & O: & 0  & 0  & 0  & 0 \\
      Ca: & Bi: & Cu: & Ta: & O: & 1  & 1  & 1  & 1 \\
      Ca: & Bi: & Fe: & Pd: & O: & 0  & 0  & 0  & 0 \\
      Ca: & Bi: & Fe: & Te: & O: & 0  & 0  & 0  & 0 \\
      Ca: & Bi: & Ge: & In: & O: & 0  & 1  & 1  & 0 \\
      Ca: & Bi: & Li: & W: & O: & 1  & 1  & 1  & 1 \\
      Ca: & Bi: & Mn: & Fe: & O: & 1  & 1  & 1  & 1 \\
      Ca: & Bi: & Mo: & Au: & O: & 0  & 0  & 0  & 1 \\
      Ca: & Bi: & Mo: & Ir: & O: & 0  & 0  & 0  & 1 \\
      Ca: & Bi: & Mo: & Rh: & O: & 0  & 0  & 0  & 1 \\
      Ca: & Bi: & Nb: & Pt: & O: & 0  & 0  & 1  & 0 \\
      Ca: & Bi: & Ni: & Nb: & O: & 0  & 1  & 1  & 1 \\
      Ca: & Bi: & Ni: & Sb: & O: & 0  & 0  & 0  & 0 \\
      Ca: & Bi: & Ni: & Ta: & O: & 1  & 1  & 1  & 1 \\
      Ca: & Bi: & Rh: & Sn: & O: & 0  & 0  & 0  & 1 \\
      Ca: & Bi: & Sb: & Ir: & O: & 0  & 0  & 0  & 1 \\
      Ca: & Bi: & Sb: & Pt: & O: & 0  & 0  & 0  & 1 \\
      Ca: & Bi: & Sb: & Re: & O: & 0  & 0  & 0  & 0 \\
      Ca: & Bi: & Sb: & W: & O: & 0  & 0  & 0  & 1 \\
      Ca: & Bi: & Sn: & Au: & O: & 0  & 0  & 0  & 1 \\
      Ca: & Bi: & Sn: & Ir: & O: & 0  & 0  & 0  & 0 \\
      Ca: & Bi: & Ta: & Pt: & O: & 0  & 1  & 1  & 0 \\
      Ca: & Bi: & Tc: & Sb: & O: & 0  & 0  & 0  & 0 \\
      Ca: & Bi: & Ti: & Cr: & O: & 1  & 1  & 1  & 1 \\
      Ca: & Bi: & Ti: & Fe: & O: & 1  & 1  & 1  & 1 \\
      Ca: & Bi: & V: & Sn: & O: & 0  & 1  & 1  & 0 \\
      Ca: & Bi: & Zn: & Nb: & O: & 1  & 1  & 1  & 1 \\
      Ca: & Bi: & Zn: & Sb: & O: & 0  & 1  & 1  & 0 \\
      Ca: & Bi: & Zn: & Ta: & O: & 1  & 1  & 1  & 1 \\
      Cd: & Bi: & Al: & Hf: & O: & 1  & 1  & 1  & 1 \\
      Cd: & Bi: & Al: & Zr: & O: & 1  & 1  & 1  & 1 \\
      Cd: & Bi: & Co: & Mo: & O: & 1  & 1  & 1  & 0 \\
      Cd: & Bi: & Co: & Nb: & O: & 1  & 1  & 1  & 0 \\
      Cd: & Bi: & Co: & Sb: & O: & 1  & 0  & 1  & 0 \\
      Cd: & Bi: & Co: & Sn: & O: & 1  & 1  & 1  & 0 \\
      Cd: & Bi: & Co: & Ta: & O: & 1  & 1  & 1  & 1 \\
      Cd: & Bi: & Cr: & Mn: & O: & 1  & 1  & 1  & 1 \\
      Cd: & Bi: & Cr: & Pd: & O: & 1  & 1  & 1  & 0 \\
      Cd: & Bi: & Cr: & Te: & O: & 0  & 0  & 1  & 0 \\
      Cd: & Bi: & Cu: & Nb: & O: & 1  & 1  & 1  & 0 \\
      Cd: & Bi: & Cu: & Sb: & O: & 1  & 0  & 1  & 0 \\
      Cd: & Bi: & Cu: & Ta: & O: & 1  & 1  & 1  & 1 \\
      Cd: & Bi: & Fe: & Pd: & O: & 1  & 1  & 1  & 0 \\
      Cd: & Bi: & Fe: & Te: & O: & 0  & 0  & 0  & 0 \\
      Cd: & Bi: & Ge: & In: & O: & 1  & 0  & 1  & 0 \\
      Cd: & Bi: & Li: & W: & O: & 1  & 1  & 1  & 1 \\
      Cd: & Bi: & Mn: & Fe: & O: & 1  & 1  & 1  & 1 \\
      Cd: & Bi: & Mo: & Au: & O: & 1  & 0  & 0  & 1 \\
      Cd: & Bi: & Mo: & Ir: & O: & 1  & 0  & 0  & 1 \\
      Cd: & Bi: & Mo: & Rh: & O: & 1  & 0  & 0  & 1 \\
      Cd: & Bi: & Nb: & Pt: & O: & 1  & 0  & 1  & 0 \\
      Cd: & Bi: & Ni: & Nb: & O: & 1  & 1  & 1  & 1 \\
      Cd: & Bi: & Ni: & Sb: & O: & 1  & 1  & 1  & 0 \\
      Cd: & Bi: & Ni: & Ta: & O: & 1  & 1  & 1  & 1 \\
      Cd: & Bi: & Rh: & Sn: & O: & 1  & 0  & 0  & 1 \\
      Cd: & Bi: & Sb: & Ir: & O: & 1  & 0  & 0  & 1 \\
      Cd: & Bi: & Sb: & Pt: & O: & 1  & 0  & 0  & 1 \\
      Cd: & Bi: & Sb: & Re: & O: & 1  & 0  & 1  & 0 \\
      Cd: & Bi: & Sb: & W: & O: & 1  & 0  & 0  & 1 \\
      Cd: & Bi: & Sn: & Au: & O: & 1  & 0  & 0  & 1 \\
      Cd: & Bi: & Sn: & Ir: & O: & 1  & 0  & 1  & 0 \\
      Cd: & Bi: & Ta: & Pt: & O: & 1  & 0  & 1  & 0 \\
      Cd: & Bi: & Tc: & Sb: & O: & 1  & 0  & 1  & 0 \\
      Cd: & Bi: & Ti: & Cr: & O: & 1  & 1  & 1  & 1 \\
      Cd: & Bi: & Ti: & Fe: & O: & 1  & 1  & 1  & 1 \\
      Cd: & Bi: & V: & Sn: & O: & 1  & 1  & 1  & 0 \\
      Cd: & Bi: & Zn: & Nb: & O: & 1  & 1  & 1  & 1 \\
      Cd: & Bi: & Zn: & Sb: & O: & 1  & 1  & 1  & 0 \\
      Cd: & Bi: & Zn: & Ta: & O: & 1  & 1  & 1  & 1 \\
      Hf: & Bi: & Al: & Cd: & O: & 0  & 0  & 1  & 0 \\
      Hf: & Bi: & Co: & Sb: & O: & 0  & 0  & 1  & 0 \\
      Hf: & Bi: & Co: & Sn: & O: & 0  & 0  & 0  & 0 \\
      Hf: & Bi: & Cr: & Fe: & O: & 0  & 0  & 1  & 0 \\
      Hf: & Bi: & Cr: & Mn: & O: & 0  & 0  & 1  & 0 \\
      Hf: & Bi: & Cr: & Pd: & O: & 0  & 0  & 1  & 0 \\
      Hf: & Bi: & Cu: & Sb: & O: & 0  & 0  & 1  & 0 \\
      Hf: & Bi: & Fe: & Fe: & O: & 0  & 0  & 1  & 0 \\
      Hf: & Bi: & Fe: & Pd: & O: & 0  & 0  & 1  & 0 \\
      Hf: & Bi: & Ga: & Hg: & O: & 0  & 0  & 1  & 0 \\
      Hf: & Bi: & Ge: & In: & O: & 0  & 0  & 1  & 0 \\
      Hf: & Bi: & Li: & Ir: & O: & 0  & 0  & 1  & 0 \\
      Hf: & Bi: & Li: & Pt: & O: & 0  & 0  & 1  & 0 \\
      Hf: & Bi: & Li: & Re: & O: & 1  & 0  & 1  & 1 \\
      Hf: & Bi: & Li: & Tc: & O: & 1  & 0  & 1  & 1 \\
      Hf: & Bi: & Li: & W: & O: & 0  & 0  & 1  & 0 \\
      Hf: & Bi: & Mg: & Ga: & O: & 0  & 1  & 1  & 1 \\
      Hf: & Bi: & Mg: & Ru: & O: & 0  & 0  & 1  & 0 \\
      Hf: & Bi: & Mn: & Fe: & O: & 0  & 0  & 1  & 0 \\
      Hf: & Bi: & Ni: & Sb: & O: & 0  & 0  & 1  & 0 \\
      Hf: & Bi: & Os: & Hg: & O: & 0  & 0  & 0  & 0 \\
      Hf: & Bi: & Rh: & Sn: & O: & 0  & 0  & 0  & 0 \\
      Hf: & Bi: & Ru: & Hg: & O: & 0  & 0  & 0  & 0 \\
      Hf: & Bi: & Sb: & Pt: & O: & 0  & 0  & 0  & 0 \\
      Hf: & Bi: & Sn: & Au: & O: & 0  & 0  & 0  & 0 \\
      Hf: & Bi: & Sn: & Ir: & O: & 0  & 0  & 0  & 0 \\
      Hf: & Bi: & Zn: & Sb: & O: & 0  & 0  & 1  & 0 \\
      Hg: & Bi: & Al: & Hf: & O: & 1  & 1  & 1  & 1 \\
      Hg: & Bi: & Al: & Zr: & O: & 1  & 1  & 1  & 1 \\
      Hg: & Bi: & Co: & Mo: & O: & 1  & 1  & 1  & 0 \\
      Hg: & Bi: & Co: & Nb: & O: & 0  & 1  & 1  & 0 \\
      Hg: & Bi: & Co: & Sb: & O: & 1  & 1  & 1  & 0 \\
      Hg: & Bi: & Co: & Sn: & O: & 1  & 1  & 1  & 0 \\
      Hg: & Bi: & Co: & Ta: & O: & 0  & 1  & 1  & 0 \\
      Hg: & Bi: & Cr: & Fe: & O: & 1  & 1  & 1  & 1 \\
      Hg: & Bi: & Cr: & Mn: & O: & 1  & 1  & 1  & 1 \\
      Hg: & Bi: & Cr: & Pd: & O: & 1  & 1  & 1  & 0 \\
      Hg: & Bi: & Cr: & Te: & O: & 0  & 0  & 1  & 0 \\
      Hg: & Bi: & Cu: & Mo: & O: & 1  & 1  & 1  & 0 \\
      Hg: & Bi: & Cu: & Nb: & O: & 0  & 1  & 1  & 0 \\
      Hg: & Bi: & Cu: & Sb: & O: & 1  & 1  & 1  & 0 \\
      Hg: & Bi: & Cu: & Ta: & O: & 0  & 1  & 1  & 0 \\
      Hg: & Bi: & Fe: & Fe: & O: & 1  & 1  & 1  & 1 \\
      Hg: & Bi: & Fe: & Pd: & O: & 1  & 1  & 1  & 0 \\
      Hg: & Bi: & Fe: & Te: & O: & 0  & 0  & 0  & 0 \\
      Hg: & Bi: & Ge: & Hf: & O: & 1  & 1  & 1  & 1 \\
      Hg: & Bi: & Ge: & In: & O: & 0  & 1  & 1  & 0 \\
      Hg: & Bi: & Ge: & Zr: & O: & 1  & 1  & 1  & 1 \\
      Hg: & Bi: & Li: & Tc: & O: & 1  & 1  & 1  & 1 \\
      Hg: & Bi: & Li: & W: & O: & 1  & 1  & 1  & 1 \\
      Hg: & Bi: & Mn: & Fe: & O: & 1  & 1  & 1  & 1 \\
      Hg: & Bi: & Mn: & Mn: & O: & 1  & 1  & 1  & 1 \\
      Hg: & Bi: & Mn: & Pd: & O: & 1  & 1  & 1  & 0 \\
      Hg: & Bi: & Mn: & Te: & O: & 0  & 0  & 1  & 0 \\
      Hg: & Bi: & Mo: & Au: & O: & 1  & 0  & 0  & 1 \\
      Hg: & Bi: & Mo: & Ir: & O: & 1  & 1  & 1  & 1 \\
      Hg: & Bi: & Mo: & Pt: & O: & 1  & 0  & 0  & 1 \\
      Hg: & Bi: & Mo: & Re: & O: & 1  & 1  & 1  & 0 \\
      Hg: & Bi: & Mo: & Rh: & O: & 1  & 1  & 1  & 1 \\
      Hg: & Bi: & Mo: & Tc: & O: & 1  & 1  & 1  & 0 \\
      Hg: & Bi: & Mo: & W: & O: & 1  & 1  & 0  & 1 \\
      Hg: & Bi: & Nb: & Au: & O: & 0  & 0  & 0  & 0 \\
      Hg: & Bi: & Nb: & Ir: & O: & 1  & 1  & 1  & 0 \\
      Hg: & Bi: & Nb: & Pt: & O: & 0  & 0  & 1  & 0 \\
      Hg: & Bi: & Nb: & Rh: & O: & 1  & 1  & 1  & 0 \\
      Hg: & Bi: & Ni: & Mo: & O: & 1  & 1  & 1  & 0 \\
      Hg: & Bi: & Ni: & Nb: & O: & 0  & 1  & 1  & 0 \\
      Hg: & Bi: & Ni: & Sb: & O: & 1  & 1  & 1  & 0 \\
      Hg: & Bi: & Ni: & Ta: & O: & 1  & 1  & 1  & 1 \\
      Hg: & Bi: & Pd: & Pd: & O: & 1  & 1  & 1  & 1 \\
      Hg: & Bi: & Pd: & Te: & O: & 1  & 0  & 0  & 1 \\
      Hg: & Bi: & Rh: & Sb: & O: & 1  & 1  & 1  & 1 \\
      Hg: & Bi: & Rh: & Sn: & O: & 1  & 1  & 1  & 1 \\
      Hg: & Bi: & Rh: & Ta: & O: & 1  & 1  & 1  & 0 \\
      Hg: & Bi: & Sb: & Au: & O: & 1  & 0  & 0  & 1 \\
      Hg: & Bi: & Sb: & Ir: & O: & 1  & 1  & 1  & 1 \\
      Hg: & Bi: & Sb: & Pt: & O: & 1  & 0  & 1  & 1 \\
      Hg: & Bi: & Sb: & Re: & O: & 0  & 0  & 1  & 0 \\
      Hg: & Bi: & Sb: & W: & O: & 1  & 0  & 0  & 1 \\
      Hg: & Bi: & Sn: & Au: & O: & 0  & 0  & 0  & 1 \\
      Hg: & Bi: & Sn: & Ir: & O: & 1  & 0  & 1  & 0 \\
      Hg: & Bi: & Sn: & Pt: & O: & 1  & 1  & 1  & 1 \\
      Hg: & Bi: & Sn: & Re: & O: & 1  & 1  & 1  & 0 \\
      Hg: & Bi: & Sn: & W: & O: & 1  & 1  & 1  & 1 \\
      Hg: & Bi: & Ta: & Au: & O: & 0  & 0  & 1  & 0 \\
      Hg: & Bi: & Ta: & Ir: & O: & 1  & 1  & 1  & 0 \\
      Hg: & Bi: & Ta: & Pt: & O: & 0  & 0  & 1  & 0 \\
      Hg: & Bi: & Tc: & Sb: & O: & 0  & 0  & 1  & 0 \\
      Hg: & Bi: & Tc: & Sn: & O: & 1  & 1  & 1  & 0 \\
      Hg: & Bi: & Te: & Te: & O: & 0  & 0  & 0  & 0 \\
      Hg: & Bi: & Ti: & Cr: & O: & 1  & 1  & 1  & 1 \\
      Hg: & Bi: & Ti: & Fe: & O: & 1  & 1  & 1  & 1 \\
      Hg: & Bi: & Ti: & Mn: & O: & 1  & 1  & 1  & 1 \\
      Hg: & Bi: & Ti: & Pd: & O: & 1  & 1  & 1  & 0 \\
      Hg: & Bi: & Ti: & Te: & O: & 0  & 0  & 1  & 0 \\
      Hg: & Bi: & Ti: & Ti: & O: & 1  & 1  & 1  & 1 \\
      Hg: & Bi: & V: & Sb: & O: & 1  & 1  & 1  & 0 \\
      Hg: & Bi: & V: & Sn: & O: & 0  & 1  & 1  & 0 \\
      Hg: & Bi: & Zn: & Mo: & O: & 1  & 1  & 1  & 0 \\
      Hg: & Bi: & Zn: & Nb: & O: & 1  & 1  & 1  & 1 \\
      Hg: & Bi: & Zn: & Sb: & O: & 0  & 1  & 1  & 0 \\
      Hg: & Bi: & Zn: & Ta: & O: & 1  & 1  & 1  & 1 \\
      In: & Bi: & Al: & In: & O: & 0  & 0  & 1  & 0 \\
      In: & Bi: & Co: & Mo: & O: & 0  & 0  & 1  & 0 \\
      In: & Bi: & Co: & Sb: & O: & 0  & 0  & 1  & 0 \\
      In: & Bi: & Co: & Sn: & O: & 0  & 0  & 1  & 0 \\
      In: & Bi: & Cr: & Cr: & O: & 0  & 1  & 1  & 1 \\
      In: & Bi: & Cr: & Fe: & O: & 0  & 1  & 1  & 0 \\
      In: & Bi: & Cu: & Mo: & O: & 0  & 0  & 1  & 0 \\
      In: & Bi: & Cu: & Sn: & O: & 0  & 0  & 1  & 0 \\
      In: & Bi: & Fe: & Fe: & O: & 0  & 1  & 1  & 0 \\
      In: & Bi: & Fe: & Pd: & O: & 0  & 0  & 1  & 0 \\
      In: & Bi: & Fe: & Te: & O: & 0  & 0  & 0  & 0 \\
      In: & Bi: & Ge: & Cd: & O: & 0  & 0  & 1  & 0 \\
      In: & Bi: & Ge: & Hf: & O: & 0  & 0  & 1  & 0 \\
      In: & Bi: & Ge: & Zr: & O: & 0  & 0  & 1  & 0 \\
      In: & Bi: & Li: & V: & O: & 1  & 1  & 1  & 1 \\
      In: & Bi: & Mg: & Os: & O: & 0  & 0  & 1  & 0 \\
      In: & Bi: & Mg: & Ru: & O: & 0  & 0  & 1  & 0 \\
      In: & Bi: & Mn: & Fe: & O: & 0  & 1  & 1  & 0 \\
      In: & Bi: & Mn: & Mn: & O: & 1  & 0  & 1  & 1 \\
      In: & Bi: & Mn: & Pd: & O: & 0  & 0  & 1  & 0 \\
      In: & Bi: & Mn: & Te: & O: & 0  & 0  & 0  & 0 \\
      In: & Bi: & Mo: & Pt: & O: & 0  & 0  & 0  & 1 \\
      In: & Bi: & Ni: & Mo: & O: & 0  & 0  & 1  & 0 \\
      In: & Bi: & Ni: & Sn: & O: & 0  & 0  & 1  & 0 \\
      In: & Bi: & Os: & Hg: & O: & 0  & 0  & 0  & 0 \\
      In: & Bi: & Pd: & Pd: & O: & 0  & 0  & 0  & 1 \\
      In: & Bi: & Pd: & Te: & O: & 0  & 0  & 0  & 0 \\
      In: & Bi: & Rh: & Sb: & O: & 0  & 0  & 1  & 1 \\
      In: & Bi: & Ru: & Hg: & O: & 0  & 0  & 0  & 0 \\
      In: & Bi: & Sb: & Au: & O: & 0  & 0  & 0  & 1 \\
      In: & Bi: & Sb: & Ir: & O: & 0  & 0  & 1  & 0 \\
      In: & Bi: & Sn: & Ir: & O: & 0  & 0  & 0  & 0 \\
      In: & Bi: & Sn: & Pt: & O: & 0  & 0  & 0  & 0 \\
      In: & Bi: & Sn: & Re: & O: & 0  & 0  & 0  & 0 \\
      In: & Bi: & Sn: & W: & O: & 0  & 0  & 0  & 0 \\
      In: & Bi: & Tc: & Sn: & O: & 0  & 0  & 0  & 0 \\
      In: & Bi: & Ti: & Fe: & O: & 0  & 0  & 1  & 0 \\
      In: & Bi: & Ti: & Mn: & O: & 1  & 0  & 1  & 0 \\
      In: & Bi: & Ti: & Pd: & O: & 0  & 0  & 1  & 0 \\
      In: & Bi: & Ti: & Te: & O: & 0  & 0  & 0  & 0 \\
      In: & Bi: & Zn: & Mo: & O: & 0  & 0  & 1  & 0 \\
      In: & Bi: & Zn: & Sn: & O: & 0  & 0  & 1  & 0 \\
      Mg: & Bi: & Al: & Hf: & O: & 1  & 1  & 1  & 1 \\
      Mg: & Bi: & Al: & Zr: & O: & 1  & 1  & 1  & 1 \\
      Mg: & Bi: & Co: & Mo: & O: & 0  & 0  & 1  & 0 \\
      Mg: & Bi: & Co: & Nb: & O: & 0  & 0  & 1  & 0 \\
      Mg: & Bi: & Co: & Sb: & O: & 0  & 0  & 1  & 0 \\
      Mg: & Bi: & Co: & Sn: & O: & 0  & 0  & 1  & 0 \\
      Mg: & Bi: & Co: & Ta: & O: & 0  & 0  & 1  & 0 \\
      Mg: & Bi: & Cr: & Mn: & O: & 1  & 1  & 1  & 1 \\
      Mg: & Bi: & Cr: & Pd: & O: & 0  & 0  & 1  & 0 \\
      Mg: & Bi: & Cr: & Te: & O: & 0  & 0  & 1  & 0 \\
      Mg: & Bi: & Cu: & Nb: & O: & 0  & 0  & 1  & 0 \\
      Mg: & Bi: & Cu: & Sb: & O: & 0  & 0  & 1  & 0 \\
      Mg: & Bi: & Cu: & Ta: & O: & 0  & 0  & 1  & 0 \\
      Mg: & Bi: & Fe: & Pd: & O: & 0  & 0  & 1  & 0 \\
      Mg: & Bi: & Fe: & Te: & O: & 0  & 0  & 1  & 0 \\
      Mg: & Bi: & Ge: & In: & O: & 0  & 0  & 1  & 0 \\
      Mg: & Bi: & Li: & W: & O: & 0  & 1  & 1  & 0 \\
      Mg: & Bi: & Mn: & Fe: & O: & 0  & 1  & 1  & 1 \\
      Mg: & Bi: & Mo: & Au: & O: & 0  & 0  & 0  & 1 \\
      Mg: & Bi: & Mo: & Ir: & O: & 0  & 0  & 1  & 1 \\
      Mg: & Bi: & Mo: & Rh: & O: & 0  & 0  & 1  & 1 \\
      Mg: & Bi: & Nb: & Pt: & O: & 0  & 0  & 1  & 0 \\
      Mg: & Bi: & Ni: & Nb: & O: & 0  & 0  & 1  & 0 \\
      Mg: & Bi: & Ni: & Sb: & O: & 0  & 0  & 1  & 0 \\
      Mg: & Bi: & Ni: & Ta: & O: & 0  & 0  & 1  & 0 \\
      Mg: & Bi: & Rh: & Sn: & O: & 0  & 0  & 1  & 0 \\
      Mg: & Bi: & Sb: & Ir: & O: & 0  & 0  & 1  & 0 \\
      Mg: & Bi: & Sb: & Pt: & O: & 0  & 0  & 1  & 1 \\
      Mg: & Bi: & Sb: & Re: & O: & 0  & 0  & 1  & 0 \\
      Mg: & Bi: & Sb: & W: & O: & 0  & 0  & 1  & 1 \\
      Mg: & Bi: & Sn: & Au: & O: & 0  & 0  & 0  & 1 \\
      Mg: & Bi: & Sn: & Ir: & O: & 0  & 0  & 1  & 0 \\
      Mg: & Bi: & Ta: & Pt: & O: & 0  & 0  & 1  & 0 \\
      Mg: & Bi: & Tc: & Sb: & O: & 0  & 0  & 1  & 0 \\
      Mg: & Bi: & Ti: & Cr: & O: & 1  & 1  & 1  & 1 \\
      Mg: & Bi: & Ti: & Fe: & O: & 0  & 0  & 1  & 1 \\
      Mg: & Bi: & V: & Sn: & O: & 0  & 0  & 1  & 0 \\
      Mg: & Bi: & Zn: & Nb: & O: & 0  & 1  & 1  & 0 \\
      Mg: & Bi: & Zn: & Sb: & O: & 0  & 0  & 1  & 0 \\
      Mg: & Bi: & Zn: & Ta: & O: & 0  & 1  & 1  & 0 \\
      Na: & Bi: & Co: & Mo: & O: & 0  & 0  & 0  & 0 \\
      Na: & Bi: & Co: & Nb: & O: & 1  & 1  & 1  & 1 \\
      Na: & Bi: & Co: & Sb: & O: & 0  & 0  & 0  & 0 \\
      Na: & Bi: & Co: & Ta: & O: & 1  & 1  & 1  & 1 \\
      Na: & Bi: & Cr: & Fe: & O: & 1  & 1  & 0  & 1 \\
      Na: & Bi: & Cr: & Mn: & O: & 1  & 1  & 1  & 1 \\
      Na: & Bi: & Cr: & Pd: & O: & 0  & 0  & 0  & 0 \\
      Na: & Bi: & Cr: & Te: & O: & 0  & 0  & 1  & 0 \\
      Na: & Bi: & Cu: & Mo: & O: & 0  & 0  & 0  & 0 \\
      Na: & Bi: & Fe: & Fe: & O: & 1  & 1  & 0  & 1 \\
      Na: & Bi: & Fe: & Pd: & O: & 0  & 1  & 0  & 0 \\
      Na: & Bi: & Fe: & Te: & O: & 0  & 0  & 0  & 0 \\
      Na: & Bi: & Ge: & Hf: & O: & 1  & 1  & 1  & 1 \\
      Na: & Bi: & Ge: & Zr: & O: & 1  & 1  & 1  & 1 \\
      Na: & Bi: & Li: & Tc: & O: & 1  & 1  & 1  & 1 \\
      Na: & Bi: & Mn: & Fe: & O: & 1  & 1  & 0  & 1 \\
      Na: & Bi: & Mn: & Mn: & O: & 1  & 1  & 0  & 1 \\
      Na: & Bi: & Mn: & Pd: & O: & 0  & 1  & 0  & 0 \\
      Na: & Bi: & Mn: & Te: & O: & 0  & 1  & 1  & 0 \\
      Na: & Bi: & Mo: & Ir: & O: & 0  & 0  & 0  & 1 \\
      Na: & Bi: & Mo: & Pt: & O: & 0  & 0  & 0  & 1 \\
      Na: & Bi: & Mo: & Re: & O: & 0  & 0  & 0  & 0 \\
      Na: & Bi: & Mo: & Tc: & O: & 0  & 0  & 0  & 0 \\
      Na: & Bi: & Mo: & W: & O: & 0  & 0  & 0  & 1 \\
      Na: & Bi: & Nb: & Au: & O: & 0  & 0  & 0  & 0 \\
      Na: & Bi: & Nb: & Ir: & O: & 0  & 1  & 0  & 0 \\
      Na: & Bi: & Nb: & Rh: & O: & 0  & 0  & 0  & 0 \\
      Na: & Bi: & Ni: & Mo: & O: & 0  & 0  & 0  & 0 \\
      Na: & Bi: & Pd: & Pd: & O: & 0  & 0  & 0  & 1 \\
      Na: & Bi: & Pd: & Te: & O: & 0  & 0  & 0  & 1 \\
      Na: & Bi: & Rh: & Sb: & O: & 0  & 0  & 0  & 1 \\
      Na: & Bi: & Rh: & Ta: & O: & 0  & 1  & 0  & 0 \\
      Na: & Bi: & Sb: & Au: & O: & 0  & 0  & 0  & 1 \\
      Na: & Bi: & Sb: & Ir: & O: & 0  & 0  & 0  & 1 \\
      Na: & Bi: & Sn: & Ir: & O: & 0  & 0  & 0  & 0 \\
      Na: & Bi: & Sn: & Pt: & O: & 0  & 0  & 0  & 1 \\
      Na: & Bi: & Sn: & Re: & O: & 0  & 0  & 0  & 0 \\
      Na: & Bi: & Sn: & W: & O: & 0  & 0  & 0  & 1 \\
      Na: & Bi: & Ta: & Au: & O: & 0  & 0  & 0  & 0 \\
      Na: & Bi: & Ta: & Ir: & O: & 0  & 1  & 0  & 0 \\
      Na: & Bi: & Tc: & Sn: & O: & 0  & 0  & 0  & 0 \\
      Na: & Bi: & Te: & Te: & O: & 0  & 0  & 0  & 0 \\
      Na: & Bi: & Ti: & Mn: & O: & 1  & 1  & 1  & 1 \\
      Na: & Bi: & Ti: & Pd: & O: & 0  & 1  & 0  & 0 \\
      Na: & Bi: & Ti: & Te: & O: & 0  & 0  & 1  & 0 \\
      Na: & Bi: & Ti: & Ti: & O: & 1  & 1  & 1  & 1 \\
      Na: & Bi: & V: & Sb: & O: & 0  & 1  & 0  & 0 \\
      Na: & Bi: & Zn: & Mo: & O: & 0  & 1  & 0  & 0 \\
      Pb: & Bi: & Al: & Cd: & O: & 0  & 0  & 1  & 0 \\
      Pb: & Bi: & Al: & Hf: & O: & 1  & 1  & 1  & 1 \\
      Pb: & Bi: & Al: & Zr: & O: & 1  & 1  & 1  & 1 \\
      Pb: & Bi: & Co: & Mo: & O: & 1  & 1  & 1  & 0 \\
      Pb: & Bi: & Co: & Nb: & O: & 1  & 1  & 1  & 0 \\
      Pb: & Bi: & Co: & Sb: & O: & 1  & 1  & 1  & 0 \\
      Pb: & Bi: & Co: & Sn: & O: & 1  & 1  & 1  & 0 \\
      Pb: & Bi: & Co: & Ta: & O: & 1  & 1  & 1  & 0 \\
      Pb: & Bi: & Cr: & Fe: & O: & 0  & 1  & 1  & 0 \\
      Pb: & Bi: & Cr: & Mn: & O: & 1  & 1  & 1  & 1 \\
      Pb: & Bi: & Cr: & Pd: & O: & 1  & 1  & 1  & 0 \\
      Pb: & Bi: & Cr: & Te: & O: & 0  & 0  & 1  & 0 \\
      Pb: & Bi: & Cu: & Nb: & O: & 1  & 1  & 1  & 0 \\
      Pb: & Bi: & Cu: & Sb: & O: & 1  & 1  & 1  & 0 \\
      Pb: & Bi: & Cu: & Ta: & O: & 1  & 1  & 1  & 0 \\
      Pb: & Bi: & Fe: & Fe: & O: & 0  & 1  & 1  & 0 \\
      Pb: & Bi: & Fe: & Pd: & O: & 1  & 1  & 1  & 0 \\
      Pb: & Bi: & Fe: & Te: & O: & 0  & 0  & 0  & 0 \\
      Pb: & Bi: & Ga: & Hg: & O: & 0  & 0  & 0  & 0 \\
      Pb: & Bi: & Ge: & In: & O: & 1  & 1  & 1  & 0 \\
      Pb: & Bi: & Li: & Ir: & O: & 1  & 1  & 1  & 0 \\
      Pb: & Bi: & Li: & Pt: & O: & 1  & 1  & 1  & 0 \\
      Pb: & Bi: & Li: & Re: & O: & 1  & 1  & 1  & 1 \\
      Pb: & Bi: & Li: & Tc: & O: & 1  & 1  & 1  & 1 \\
      Pb: & Bi: & Li: & W: & O: & 1  & 1  & 1  & 1 \\
      Pb: & Bi: & Mg: & Ga: & O: & 1  & 1  & 1  & 1 \\
      Pb: & Bi: & Mg: & Ru: & O: & 0  & 1  & 1  & 0 \\
      Pb: & Bi: & Mn: & Fe: & O: & 1  & 1  & 1  & 1 \\
      Pb: & Bi: & Mo: & Au: & O: & 1  & 0  & 0  & 1 \\
      Pb: & Bi: & Mo: & Ir: & O: & 1  & 1  & 0  & 1 \\
      Pb: & Bi: & Mo: & Rh: & O: & 1  & 1  & 0  & 1 \\
      Pb: & Bi: & Nb: & Pt: & O: & 0  & 1  & 1  & 0 \\
      Pb: & Bi: & Ni: & Nb: & O: & 1  & 1  & 1  & 0 \\
      Pb: & Bi: & Ni: & Sb: & O: & 1  & 1  & 1  & 0 \\
      Pb: & Bi: & Ni: & Ta: & O: & 1  & 1  & 1  & 1 \\
      Pb: & Bi: & Os: & Hg: & O: & 0  & 0  & 0  & 0 \\
      Pb: & Bi: & Rh: & Sn: & O: & 1  & 1  & 1  & 1 \\
      Pb: & Bi: & Ru: & Hg: & O: & 0  & 0  & 0  & 0 \\
      Pb: & Bi: & Sb: & Ir: & O: & 1  & 1  & 0  & 1 \\
      Pb: & Bi: & Sb: & Pt: & O: & 1  & 1  & 0  & 1 \\
      Pb: & Bi: & Sb: & Re: & O: & 1  & 1  & 1  & 0 \\
      Pb: & Bi: & Sb: & W: & O: & 1  & 1  & 0  & 1 \\
      Pb: & Bi: & Sn: & Au: & O: & 1  & 0  & 0  & 1 \\
      Pb: & Bi: & Sn: & Ir: & O: & 1  & 1  & 1  & 0 \\
      Pb: & Bi: & Ta: & Pt: & O: & 0  & 1  & 1  & 0 \\
      Pb: & Bi: & Tc: & Sb: & O: & 1  & 1  & 1  & 0 \\
      Pb: & Bi: & Ti: & Cr: & O: & 1  & 1  & 1  & 1 \\
      Pb: & Bi: & Ti: & Fe: & O: & 1  & 1  & 1  & 1 \\
      Pb: & Bi: & V: & Sn: & O: & 1  & 1  & 1  & 0 \\
      Pb: & Bi: & Zn: & Nb: & O: & 1  & 1  & 1  & 1 \\
      Pb: & Bi: & Zn: & Sb: & O: & 1  & 1  & 1  & 0 \\
      Pb: & Bi: & Zn: & Ta: & O: & 1  & 1  & 1  & 1 \\
      Sc: & Bi: & Al: & In: & O: & 0  & 0  & 1  & 0 \\
      Sc: & Bi: & Co: & Mo: & O: & 0  & 0  & 1  & 0 \\
      Sc: & Bi: & Co: & Sb: & O: & 0  & 0  & 1  & 0 \\
      Sc: & Bi: & Co: & Sn: & O: & 0  & 0  & 1  & 0 \\
      Sc: & Bi: & Cr: & Cr: & O: & 0  & 1  & 1  & 1 \\
      Sc: & Bi: & Cr: & Fe: & O: & 0  & 0  & 1  & 0 \\
      Sc: & Bi: & Cu: & Mo: & O: & 0  & 0  & 1  & 0 \\
      Sc: & Bi: & Cu: & Sn: & O: & 0  & 0  & 1  & 0 \\
      Sc: & Bi: & Fe: & Fe: & O: & 0  & 0  & 1  & 0 \\
      Sc: & Bi: & Fe: & Pd: & O: & 0  & 0  & 1  & 0 \\
      Sc: & Bi: & Fe: & Te: & O: & 0  & 0  & 0  & 0 \\
      Sc: & Bi: & Ge: & Cd: & O: & 0  & 0  & 1  & 0 \\
      Sc: & Bi: & Ge: & Hf: & O: & 0  & 0  & 1  & 0 \\
      Sc: & Bi: & Ge: & Zr: & O: & 0  & 0  & 1  & 0 \\
      Sc: & Bi: & Li: & V: & O: & 1  & 1  & 1  & 1 \\
      Sc: & Bi: & Mg: & Os: & O: & 0  & 0  & 1  & 0 \\
      Sc: & Bi: & Mg: & Ru: & O: & 0  & 0  & 1  & 0 \\
      Sc: & Bi: & Mn: & Fe: & O: & 0  & 0  & 1  & 0 \\
      Sc: & Bi: & Mn: & Mn: & O: & 0  & 1  & 1  & 1 \\
      Sc: & Bi: & Mn: & Pd: & O: & 0  & 0  & 1  & 0 \\
      Sc: & Bi: & Mn: & Te: & O: & 0  & 0  & 0  & 0 \\
      Sc: & Bi: & Mo: & Pt: & O: & 0  & 0  & 1  & 1 \\
      Sc: & Bi: & Ni: & Mo: & O: & 0  & 0  & 1  & 0 \\
      Sc: & Bi: & Ni: & Sn: & O: & 0  & 0  & 1  & 0 \\
      Sc: & Bi: & Os: & Hg: & O: & 0  & 0  & 0  & 0 \\
      Sc: & Bi: & Pd: & Pd: & O: & 0  & 0  & 1  & 1 \\
      Sc: & Bi: & Pd: & Te: & O: & 0  & 0  & 0  & 0 \\
      Sc: & Bi: & Rh: & Sb: & O: & 0  & 0  & 1  & 1 \\
      Sc: & Bi: & Ru: & Hg: & O: & 0  & 0  & 0  & 0 \\
      Sc: & Bi: & Sb: & Au: & O: & 0  & 0  & 0  & 1 \\
      Sc: & Bi: & Sb: & Ir: & O: & 0  & 0  & 1  & 0 \\
      Sc: & Bi: & Sn: & Pt: & O: & 0  & 0  & 1  & 0 \\
      Sc: & Bi: & Sn: & Re: & O: & 0  & 0  & 0  & 0 \\
      Sc: & Bi: & Sn: & W: & O: & 0  & 0  & 0  & 0 \\
      Sc: & Bi: & Tc: & Sn: & O: & 0  & 0  & 0  & 0 \\
      Sc: & Bi: & Ti: & Fe: & O: & 0  & 0  & 1  & 0 \\
      Sc: & Bi: & Ti: & Mn: & O: & 0  & 0  & 1  & 0 \\
      Sc: & Bi: & Ti: & Pd: & O: & 0  & 0  & 1  & 0 \\
      Sc: & Bi: & Ti: & Te: & O: & 0  & 0  & 0  & 0 \\
      Sc: & Bi: & Zn: & Mo: & O: & 0  & 0  & 1  & 0 \\
      Sc: & Bi: & Zn: & Sn: & O: & 0  & 0  & 1  & 0 \\
      Tl: & Bi: & Al: & In: & O: & 0  & 0  & 1  & 0 \\
      Tl: & Bi: & Co: & Mo: & O: & 0  & 0  & 1  & 0 \\
      Tl: & Bi: & Co: & Nb: & O: & 1  & 1  & 1  & 1 \\
      Tl: & Bi: & Co: & Sb: & O: & 1  & 1  & 0  & 0 \\
      Tl: & Bi: & Co: & Sn: & O: & 0  & 0  & 1  & 0 \\
      Tl: & Bi: & Co: & Ta: & O: & 1  & 1  & 1  & 1 \\
      Tl: & Bi: & Cr: & Cr: & O: & 1  & 1  & 1  & 1 \\
      Tl: & Bi: & Cr: & Fe: & O: & 1  & 1  & 1  & 1 \\
      Tl: & Bi: & Cr: & Mn: & O: & 1  & 1  & 1  & 1 \\
      Tl: & Bi: & Cr: & Pd: & O: & 1  & 1  & 0  & 0 \\
      Tl: & Bi: & Cr: & Te: & O: & 1  & 1  & 1  & 0 \\
      Tl: & Bi: & Cu: & Mo: & O: & 1  & 1  & 0  & 0 \\
      Tl: & Bi: & Cu: & Sn: & O: & 0  & 0  & 1  & 0 \\
      Tl: & Bi: & Fe: & Fe: & O: & 0  & 0  & 1  & 0 \\
      Tl: & Bi: & Fe: & Fe: & O: & 1  & 1  & 0  & 0 \\
      Tl: & Bi: & Fe: & Pd: & O: & 0  & 0  & 1  & 0 \\
      Tl: & Bi: & Fe: & Te: & O: & 1  & 1  & 0  & 0 \\
      Tl: & Bi: & Ge: & Cd: & O: & 0  & 0  & 1  & 0 \\
      Tl: & Bi: & Ge: & Hf: & O: & 1  & 1  & 1  & 1 \\
      Tl: & Bi: & Ge: & Zr: & O: & 1  & 1  & 1  & 1 \\
      Tl: & Bi: & Li: & Tc: & O: & 1  & 1  & 1  & 1 \\
      Tl: & Bi: & Li: & V: & O: & 1  & 1  & 1  & 1 \\
      Tl: & Bi: & Mg: & Os: & O: & 0  & 0  & 1  & 0 \\
      Tl: & Bi: & Mg: & Ru: & O: & 0  & 0  & 1  & 0 \\
      Tl: & Bi: & Mn: & Fe: & O: & 0  & 0  & 1  & 1 \\
      Tl: & Bi: & Mn: & Mn: & O: & 1  & 1  & 0  & 1 \\
      Tl: & Bi: & Mn: & Mn: & O: & 1  & 1  & 1  & 1 \\
      Tl: & Bi: & Mn: & Pd: & O: & 1  & 1  & 0  & 0 \\
      Tl: & Bi: & Mn: & Te: & O: & 1  & 1  & 1  & 0 \\
      Tl: & Bi: & Mo: & Ir: & O: & 1  & 1  & 0  & 1 \\
      Tl: & Bi: & Mo: & Pt: & O: & 1  & 1  & 0  & 1 \\
      Tl: & Bi: & Mo: & Re: & O: & 1  & 1  & 0  & 0 \\
      Tl: & Bi: & Mo: & Tc: & O: & 1  & 1  & 0  & 0 \\
      Tl: & Bi: & Mo: & W: & O: & 1  & 1  & 0  & 1 \\
      Tl: & Bi: & Nb: & Au: & O: & 1  & 1  & 0  & 0 \\
      Tl: & Bi: & Nb: & Ir: & O: & 1  & 1  & 0  & 0 \\
      Tl: & Bi: & Nb: & Rh: & O: & 1  & 1  & 0  & 0 \\
      Tl: & Bi: & Ni: & Mo: & O: & 1  & 1  & 0  & 0 \\
      Tl: & Bi: & Ni: & Sn: & O: & 0  & 0  & 1  & 0 \\
      Tl: & Bi: & Os: & Hg: & O: & 0  & 0  & 0  & 0 \\
      Tl: & Bi: & Pd: & Pd: & O: & 1  & 1  & 0  & 1 \\
      Tl: & Bi: & Pd: & Pd: & O: & 0  & 0  & 0  & 1 \\
      Tl: & Bi: & Pd: & Te: & O: & 1  & 1  & 0  & 1 \\
      Tl: & Bi: & Rh: & Sb: & O: & 1  & 1  & 0  & 1 \\
      Tl: & Bi: & Rh: & Ta: & O: & 1  & 1  & 0  & 0 \\
      Tl: & Bi: & Ru: & Hg: & O: & 0  & 0  & 0  & 0 \\
      Tl: & Bi: & Sb: & Au: & O: & 1  & 1  & 0  & 1 \\
      Tl: & Bi: & Sb: & Ir: & O: & 1  & 1  & 0  & 1 \\
      Tl: & Bi: & Sn: & Ir: & O: & 1  & 1  & 0  & 0 \\
      Tl: & Bi: & Sn: & Pt: & O: & 1  & 1  & 0  & 1 \\
      Tl: & Bi: & Sn: & Re: & O: & 1  & 1  & 0  & 0 \\
      Tl: & Bi: & Sn: & W: & O: & 1  & 0  & 0  & 1 \\
      Tl: & Bi: & Ta: & Au: & O: & 1  & 1  & 0  & 0 \\
      Tl: & Bi: & Ta: & Ir: & O: & 1  & 1  & 1  & 0 \\
      Tl: & Bi: & Tc: & Sn: & O: & 1  & 1  & 0  & 0 \\
      Tl: & Bi: & Te: & Te: & O: & 0  & 0  & 0  & 0 \\
      Tl: & Bi: & Ti: & Fe: & O: & 0  & 0  & 1  & 0 \\
      Tl: & Bi: & Ti: & Mn: & O: & 1  & 1  & 1  & 1 \\
      Tl: & Bi: & Ti: & Pd: & O: & 1  & 1  & 0  & 0 \\
      Tl: & Bi: & Ti: & Te: & O: & 1  & 1  & 1  & 0 \\
      Tl: & Bi: & Ti: & Ti: & O: & 1  & 1  & 1  & 1 \\
      Tl: & Bi: & V: & Sb: & O: & 1  & 1  & 1  & 0 \\
      Tl: & Bi: & Zn: & Mo: & O: & 1  & 1  & 0  & 0 \\
      Tl: & Bi: & Zn: & Sn: & O: & 0  & 0  & 1  & 0 \\
      Y: & Bi: & Al: & In: & O: & 0  & 1  & 1  & 1 \\
      Y: & Bi: & Co: & Mo: & O: & 0  & 0  & 1  & 0 \\
      Y: & Bi: & Co: & Sb: & O: & 0  & 0  & 1  & 0 \\
      Y: & Bi: & Co: & Sn: & O: & 0  & 0  & 1  & 0 \\
      Y: & Bi: & Cr: & Cr: & O: & 1  & 1  & 1  & 1 \\
      Y: & Bi: & Cr: & Fe: & O: & 0  & 1  & 1  & 1 \\
      Y: & Bi: & Cu: & Mo: & O: & 0  & 0  & 1  & 0 \\
      Y: & Bi: & Cu: & Sn: & O: & 0  & 0  & 1  & 0 \\
      Y: & Bi: & Fe: & Fe: & O: & 0  & 0  & 1  & 0 \\
      Y: & Bi: & Fe: & Pd: & O: & 0  & 0  & 1  & 0 \\
      Y: & Bi: & Fe: & Te: & O: & 0  & 0  & 0  & 0 \\
      Y: & Bi: & Ge: & Cd: & O: & 0  & 0  & 1  & 0 \\
      Y: & Bi: & Ge: & Hf: & O: & 0  & 1  & 1  & 0 \\
      Y: & Bi: & Ge: & Zr: & O: & 0  & 0  & 1  & 0 \\
      Y: & Bi: & Li: & V: & O: & 1  & 1  & 1  & 1 \\
      Y: & Bi: & Mg: & Os: & O: & 0  & 1  & 1  & 0 \\
      Y: & Bi: & Mg: & Ru: & O: & 0  & 1  & 1  & 0 \\
      Y: & Bi: & Mn: & Fe: & O: & 0  & 1  & 1  & 1 \\
      Y: & Bi: & Mn: & Mn: & O: & 0  & 1  & 1  & 1 \\
      Y: & Bi: & Mn: & Pd: & O: & 0  & 0  & 1  & 0 \\
      Y: & Bi: & Mn: & Te: & O: & 0  & 0  & 0  & 0 \\
      Y: & Bi: & Mo: & Pt: & O: & 0  & 0  & 0  & 1 \\
      Y: & Bi: & Ni: & Mo: & O: & 0  & 0  & 1  & 0 \\
      Y: & Bi: & Ni: & Sn: & O: & 0  & 0  & 1  & 0 \\
      Y: & Bi: & Os: & Hg: & O: & 0  & 0  & 0  & 0 \\
      Y: & Bi: & Pd: & Pd: & O: & 0  & 0  & 0  & 1 \\
      Y: & Bi: & Pd: & Te: & O: & 0  & 0  & 0  & 0 \\
      Y: & Bi: & Rh: & Sb: & O: & 0  & 0  & 0  & 1 \\
      Y: & Bi: & Ru: & Hg: & O: & 0  & 0  & 0  & 0 \\
      Y: & Bi: & Sb: & Au: & O: & 0  & 0  & 0  & 1 \\
      Y: & Bi: & Sb: & Ir: & O: & 0  & 0  & 1  & 0 \\
      Y: & Bi: & Sn: & Ir: & O: & 0  & 0  & 0  & 0 \\
      Y: & Bi: & Sn: & Pt: & O: & 0  & 0  & 1  & 0 \\
      Y: & Bi: & Sn: & Re: & O: & 0  & 0  & 0  & 0 \\
      Y: & Bi: & Sn: & W: & O: & 0  & 0  & 0  & 0 \\
      Y: & Bi: & Tc: & Sn: & O: & 0  & 0  & 0  & 0 \\
      Y: & Bi: & Ti: & Fe: & O: & 0  & 1  & 1  & 0 \\
      Y: & Bi: & Ti: & Mn: & O: & 0  & 1  & 1  & 1 \\
      Y: & Bi: & Ti: & Pd: & O: & 0  & 0  & 1  & 0 \\
      Y: & Bi: & Ti: & Te: & O: & 0  & 0  & 0  & 0 \\
      Y: & Bi: & Zn: & Mo: & O: & 0  & 0  & 1  & 0 \\
      Y: & Bi: & Zn: & Sn: & O: & 0  & 0  & 1  & 0 \\
      Zr: & Bi: & Al: & Cd: & O: & 0  & 0  & 1  & 0 \\
      Zr: & Bi: & Co: & Sb: & O: & 0  & 0  & 1  & 0 \\
      Zr: & Bi: & Co: & Sn: & O: & 0  & 0  & 0  & 0 \\
      Zr: & Bi: & Cr: & Fe: & O: & 0  & 0  & 1  & 0 \\
      Zr: & Bi: & Cr: & Mn: & O: & 0  & 0  & 1  & 0 \\
      Zr: & Bi: & Cr: & Pd: & O: & 0  & 0  & 1  & 0 \\
      Zr: & Bi: & Cu: & Sb: & O: & 0  & 0  & 1  & 0 \\
      Zr: & Bi: & Fe: & Fe: & O: & 0  & 0  & 1  & 0 \\
      Zr: & Bi: & Fe: & Pd: & O: & 0  & 0  & 1  & 0 \\
      Zr: & Bi: & Ga: & Hg: & O: & 0  & 0  & 1  & 0 \\
      Zr: & Bi: & Ge: & In: & O: & 0  & 0  & 1  & 0 \\
      Zr: & Bi: & Li: & Ir: & O: & 0  & 1  & 1  & 0 \\
      Zr: & Bi: & Li: & Pt: & O: & 0  & 0  & 1  & 0 \\
      Zr: & Bi: & Li: & Re: & O: & 1  & 1  & 1  & 1 \\
      Zr: & Bi: & Li: & Tc: & O: & 1  & 0  & 1  & 1 \\
      Zr: & Bi: & Li: & W: & O: & 0  & 0  & 1  & 0 \\
      Zr: & Bi: & Mg: & Ga: & O: & 0  & 1  & 1  & 1 \\
      Zr: & Bi: & Mg: & Ru: & O: & 0  & 0  & 1  & 0 \\
      Zr: & Bi: & Mn: & Fe: & O: & 0  & 0  & 1  & 0 \\
      Zr: & Bi: & Ni: & Sb: & O: & 0  & 0  & 1  & 0 \\
      Zr: & Bi: & Os: & Hg: & O: & 0  & 0  & 0  & 0 \\
      Zr: & Bi: & Rh: & Sn: & O: & 0  & 0  & 0  & 0 \\
      Zr: & Bi: & Ru: & Hg: & O: & 0  & 0  & 0  & 0 \\
      Zr: & Bi: & Sb: & Pt: & O: & 0  & 0  & 0  & 0 \\
      Zr: & Bi: & Sn: & Au: & O: & 0  & 0  & 0  & 0 \\
      Zr: & Bi: & Sn: & Ir: & O: & 0  & 0  & 0  & 0 \\
      Zr: & Bi: & Zn: & Sb: & O: & 0  & 0  & 1  & 0 \\
      Ag: & Cs: & Mn: & Bi: & O: & 1  & 1  & 0  & 1 \\
      Ag: & Cs: & Tc: & Bi: & O: & 0  & 1  & 0  & 0 \\
      Ba: & Ba: & Nb: & Bi: & O: & 0  & 1  & 0  & 0 \\
      Ba: & Tl: & Pb: & Bi: & O: & 0  & 0  & 0  & 0 \\
      Ba: & Ba: & Sb: & Bi: & O: & 0  & 1  & 0  & 0 \\
      Ba: & Ba: & Ta: & Bi: & O: & 0  & 1  & 0  & 0 \\
      Ba: & Ba: & V: & Bi: & O: & 0  & 1  & 0  & 0 \\
      Ca: & Cs: & Cr: & Bi: & O: & 0  & 1  & 0  & 0 \\
      Ca: & Rb: & Cr: & Bi: & O: & 0  & 1  & 0  & 0 \\
      Ca: & Cs: & Fe: & Bi: & O: & 1  & 1  & 0  & 0 \\
      Ca: & Rb: & Fe: & Bi: & O: & 0  & 0  & 0  & 0 \\
      Ca: & Cs: & Mo: & Bi: & O: & 0  & 0  & 0  & 0 \\
      Ca: & Rb: & Mo: & Bi: & O: & 0  & 0  & 0  & 0 \\
      Ca: & Cs: & Te: & Bi: & O: & 0  & 0  & 0  & 0 \\
      Ca: & Rb: & Te: & Bi: & O: & 0  & 0  & 0  & 0 \\
      Ca: & Rb: & W: & Bi: & O: & 0  & 0  & 0  & 0 \\
      Cd: & Cs: & Cr: & Bi: & O: & 0  & 1  & 0  & 0 \\
      Cd: & Cs: & Fe: & Bi: & O: & 0  & 1  & 0  & 0 \\
      Cd: & Cs: & Mo: & Bi: & O: & 0  & 1  & 0  & 0 \\
      Cd: & Cs: & Te: & Bi: & O: & 0  & 1  & 0  & 0 \\
      Cd: & Cs: & W: & Bi: & O: & 0  & 1  & 0  & 0 \\
      Cs: & Pb: & Cr: & Bi: & O: & 1  & 1  & 0  & 0 \\
      Cs: & Pb: & Fe: & Bi: & O: & 1  & 1  & 0  & 0 \\
      Cs: & Hf: & Ge: & Bi: & O: & 0  & 0  & 0  & 0 \\
      Cs: & Hf: & Hf: & Bi: & O: & 0  & 0  & 0  & 0 \\
      Cs: & Pb: & Hf: & Bi: & O: & 0  & 1  & 1  & 0 \\
      Cs: & Hf: & Ir: & Bi: & O: & 0  & 0  & 0  & 0 \\
      Cs: & Hf: & Mn: & Bi: & O: & 0  & 1  & 0  & 0 \\
      Cs: & Pb: & Mn: & Bi: & O: & 1  & 1  & 1  & 0 \\
      Cs: & Hf: & Mo: & Bi: & O: & 0  & 0  & 0  & 0 \\
      Cs: & Hg: & Mo: & Bi: & O: & 0  & 1  & 0  & 0 \\
      Cs: & Pb: & Mo: & Bi: & O: & 1  & 1  & 0  & 0 \\
      Cs: & Tl: & Nb: & Bi: & O: & 0  & 0  & 0  & 0 \\
      Cs: & Hf: & Os: & Bi: & O: & 0  & 0  & 0  & 0 \\
      Cs: & Hf: & Pb: & Bi: & O: & 0  & 0  & 0  & 0 \\
      Cs: & Pb: & Pb: & Bi: & O: & 1  & 0  & 0  & 0 \\
      Cs: & Hf: & Pd: & Bi: & O: & 0  & 0  & 0  & 0 \\
      Cs: & Pb: & Pd: & Bi: & O: & 1  & 0  & 0  & 0 \\
      Cs: & Hf: & Pt: & Bi: & O: & 0  & 0  & 0  & 0 \\
      Cs: & Hf: & Re: & Bi: & O: & 0  & 0  & 0  & 0 \\
      Cs: & Hf: & Ru: & Bi: & O: & 0  & 0  & 0  & 0 \\
      Cs: & Tl: & Sb: & Bi: & O: & 0  & 0  & 0  & 0 \\
      Cs: & Hf: & Sn: & Bi: & O: & 0  & 0  & 0  & 0 \\
      Cs: & Pb: & Sn: & Bi: & O: & 1  & 0  & 0  & 0 \\
      Cs: & Tl: & Ta: & Bi: & O: & 0  & 0  & 0  & 0 \\
      Cs: & Hf: & Tc: & Bi: & O: & 0  & 0  & 0  & 0 \\
      Cs: & Hf: & Te: & Bi: & O: & 0  & 0  & 0  & 0 \\
      Cs: & Pb: & Te: & Bi: & O: & 1  & 1  & 0  & 0 \\
      Cs: & Hf: & Ti: & Bi: & O: & 0  & 0  & 0  & 0 \\
      Cs: & Pb: & Ti: & Bi: & O: & 1  & 0  & 1  & 0 \\
      Cs: & Hf: & W: & Bi: & O: & 0  & 0  & 0  & 0 \\
      Cs: & Ta: & Y: & Bi: & O: & 0  & 1  & 0  & 0 \\
      Cs: & Hf: & Zr: & Bi: & O: & 0  & 0  & 0  & 0 \\
      Cs: & Pb: & Zr: & Bi: & O: & 0  & 1  & 1  & 0 \\
      In: & Cs: & Nb: & Bi: & O: & 0  & 0  & 0  & 0 \\
      In: & Cs: & Sb: & Bi: & O: & 0  & 0  & 0  & 0 \\
      In: & Cs: & Ta: & Bi: & O: & 0  & 0  & 0  & 0 \\
      In: & Cs: & V: & Bi: & O: & 0  & 1  & 0  & 0 \\
      K: & Ba: & Cr: & Bi: & O: & 1  & 1  & 0  & 0 \\
      K: & Sr: & Cr: & Bi: & O: & 0  & 1  & 0  & 0 \\
      K: & Ba: & Fe: & Bi: & O: & 1  & 1  & 0  & 1 \\
      K: & Sr: & Fe: & Bi: & O: & 0  & 0  & 0  & 0 \\
      K: & K: & Mn: & Bi: & O: & 1  & 1  & 0  & 1 \\
      K: & Ba: & Mo: & Bi: & O: & 0  & 0  & 0  & 0 \\
      K: & Sr: & Mo: & Bi: & O: & 0  & 0  & 0  & 0 \\
      K: & Tl: & Nb: & Bi: & O: & 0  & 0  & 0  & 0 \\
      K: & Tl: & Sb: & Bi: & O: & 0  & 0  & 0  & 0 \\
      K: & Tl: & Ta: & Bi: & O: & 0  & 0  & 1  & 0 \\
      K: & Ba: & Te: & Bi: & O: & 0  & 0  & 0  & 0 \\
      K: & Sr: & Te: & Bi: & O: & 0  & 0  & 0  & 0 \\
      K: & Ba: & W: & Bi: & O: & 0  & 0  & 0  & 0 \\
      K: & Sr: & W: & Bi: & O: & 0  & 0  & 0  & 0 \\
      Mg: & Cs: & Mo: & Bi: & O: & 0  & 0  & 0  & 0 \\
      Mo: & Cs: & Ca: & Bi: & O: & 0  & 1  & 0  & 0 \\
      Mo: & Cs: & Pb: & Bi: & O: & 0  & 0  & 0  & 0 \\
      Na: & Cs: & Mn: & Bi: & O: & 1  & 1  & 0  & 1 \\
      Na: & Rb: & Mn: & Bi: & O: & 0  & 1  & 0  & 0 \\
      Na: & Rb: & Tc: & Bi: & O: & 0  & 1  & 0  & 0 \\
      Nb: & Cs: & Y: & Bi: & O: & 0  & 1  & 0  & 0 \\
      Rb: & Pb: & Cr: & Bi: & O: & 1  & 1  & 0  & 0 \\
      Rb: & Sr: & Cr: & Bi: & O: & 0  & 1  & 0  & 0 \\
      Rb: & Pb: & Fe: & Bi: & O: & 1  & 1  & 0  & 0 \\
      Rb: & Sr: & Fe: & Bi: & O: & 1  & 1  & 0  & 0 \\
      Rb: & Pb: & Ge: & Bi: & O: & 1  & 0  & 0  & 0 \\
      Rb: & Pb: & Hf: & Bi: & O: & 0  & 1  & 1  & 0 \\
      Rb: & Pb: & Ir: & Bi: & O: & 1  & 0  & 0  & 0 \\
      Rb: & Pb: & Mn: & Bi: & O: & 1  & 0  & 1  & 0 \\
      Rb: & Tl: & Mn: & Bi: & O: & 1  & 1  & 0  & 1 \\
      Rb: & Ba: & Mo: & Bi: & O: & 0  & 1  & 0  & 0 \\
      Rb: & Pb: & Mo: & Bi: & O: & 1  & 1  & 0  & 0 \\
      Rb: & Sr: & Mo: & Bi: & O: & 0  & 0  & 0  & 0 \\
      Rb: & Tl: & Nb: & Bi: & O: & 0  & 0  & 0  & 0 \\
      Rb: & Y: & Nb: & Bi: & O: & 0  & 0  & 0  & 0 \\
      Rb: & Pb: & Os: & Bi: & O: & 1  & 0  & 0  & 0 \\
      Rb: & Pb: & Pb: & Bi: & O: & 0  & 0  & 0  & 0 \\
      Rb: & Pb: & Pd: & Bi: & O: & 1  & 0  & 0  & 0 \\
      Rb: & Pb: & Pt: & Bi: & O: & 1  & 0  & 0  & 0 \\
      Rb: & Pb: & Re: & Bi: & O: & 1  & 0  & 0  & 0 \\
      Rb: & Pb: & Ru: & Bi: & O: & 1  & 0  & 0  & 0 \\
      Rb: & Tl: & Sb: & Bi: & O: & 0  & 0  & 0  & 0 \\
      Rb: & Y: & Sb: & Bi: & O: & 0  & 0  & 0  & 0 \\
      Rb: & Pb: & Sn: & Bi: & O: & 0  & 0  & 0  & 0 \\
      Rb: & Tl: & Ta: & Bi: & O: & 0  & 0  & 0  & 0 \\
      Rb: & Y: & Ta: & Bi: & O: & 0  & 0  & 0  & 0 \\
      Rb: & Pb: & Tc: & Bi: & O: & 1  & 0  & 0  & 0 \\
      Rb: & Tl: & Tc: & Bi: & O: & 0  & 1  & 0  & 0 \\
      Rb: & Pb: & Te: & Bi: & O: & 1  & 1  & 0  & 0 \\
      Rb: & Sr: & Te: & Bi: & O: & 0  & 0  & 0  & 0 \\
      Rb: & Pb: & Ti: & Bi: & O: & 0  & 0  & 1  & 0 \\
      Rb: & Tl: & V: & Bi: & O: & 0  & 0  & 0  & 0 \\
      Rb: & Y: & V: & Bi: & O: & 0  & 1  & 0  & 0 \\
      Rb: & Pb: & W: & Bi: & O: & 1  & 1  & 0  & 1 \\
      Rb: & Sr: & W: & Bi: & O: & 0  & 0  & 0  & 0 \\
      Rb: & Pb: & Zr: & Bi: & O: & 0  & 1  & 1  & 0 \\
      Sb: & Cs: & Y: & Bi: & O: & 0  & 1  & 0  & 0 \\
      Sc: & Cs: & Nb: & Bi: & O: & 0  & 0  & 0  & 0 \\
      Sc: & Cs: & Sb: & Bi: & O: & 0  & 0  & 0  & 0 \\
      Sc: & Cs: & Ta: & Bi: & O: & 0  & 0  & 0  & 0 \\
      Sc: & Cs: & V: & Bi: & O: & 0  & 1  & 0  & 0 \\
      Sn: & Cs: & Pb: & Bi: & O: & 0  & 0  & 0  & 0 \\
      Sr: & Ba: & Nb: & Bi: & O: & 0  & 0  & 0  & 0 \\
      Sr: & Ba: & Sb: & Bi: & O: & 0  & 0  & 0  & 0 \\
      Sr: & Ba: & Ta: & Bi: & O: & 0  & 0  & 0  & 0 \\
      Sr: & Ba: & V: & Bi: & O: & 0  & 1  & 0  & 0 \\
      Y: & Cs: & Nb: & Bi: & O: & 0  & 0  & 0  & 0 \\
      Y: & Cs: & Sb: & Bi: & O: & 0  & 0  & 0  & 0 \\
      Y: & Cs: & Ta: & Bi: & O: & 0  & 0  & 0  & 0 \\
      Zr: & Cs: & Ge: & Bi: & O: & 0  & 0  & 0  & 0 \\
      Zr: & Cs: & Hf: & Bi: & O: & 0  & 0  & 0  & 0 \\
      Zr: & Cs: & Ir: & Bi: & O: & 0  & 0  & 0  & 0 \\
      Zr: & Cs: & Mn: & Bi: & O: & 0  & 1  & 0  & 0 \\
      Zr: & Cs: & Mo: & Bi: & O: & 0  & 0  & 0  & 0 \\
      Zr: & Cs: & Os: & Bi: & O: & 0  & 0  & 0  & 0 \\
      Zr: & Cs: & Pb: & Bi: & O: & 0  & 0  & 0  & 0 \\
      Zr: & Cs: & Pd: & Bi: & O: & 0  & 0  & 0  & 0 \\
      Zr: & Cs: & Pt: & Bi: & O: & 0  & 0  & 0  & 0 \\
      Zr: & Cs: & Re: & Bi: & O: & 0  & 0  & 0  & 0 \\
      Zr: & Cs: & Ru: & Bi: & O: & 0  & 0  & 0  & 0 \\
      Zr: & Cs: & Sn: & Bi: & O: & 0  & 0  & 0  & 0 \\
      Zr: & Cs: & Tc: & Bi: & O: & 0  & 0  & 0  & 0 \\
      Zr: & Cs: & Te: & Bi: & O: & 0  & 0  & 0  & 0 \\
      Zr: & Cs: & Ti: & Bi: & O: & 0  & 0  & 0  & 0 \\
      Zr: & Cs: & W: & Bi: & O: & 0  & 0  & 0  & 0 \\
      Zr: & Cs: & Zr: & Bi: & O: & 0  & 0  & 0  & 0 \\

\label{Screened Predictions}
   
 \end{longtable*}%

\section{Neural network for five feature analysis of double perovskites}
\label{Sec:E}

\begin{figure}[htbp]
 \includegraphics[width=\linewidth]{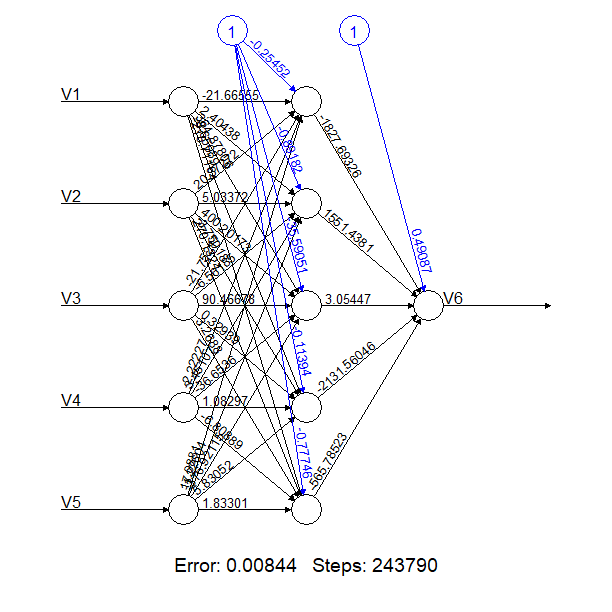}
 \caption{A neural network for the analysis of the full five feature double perovskite data of Appendix \ref{Sec:D}. The 3-fold cross validated accuracy for the double perovskite data of Appendix \ref{Sec:D} is 0.89.  The parameters underlying this network are listed in table \ref{5-nod-neural-function}.}
  \label{fig 5-node-function}
 \end{figure}

\begin{figure}[htbp]
 \includegraphics[width=\linewidth]{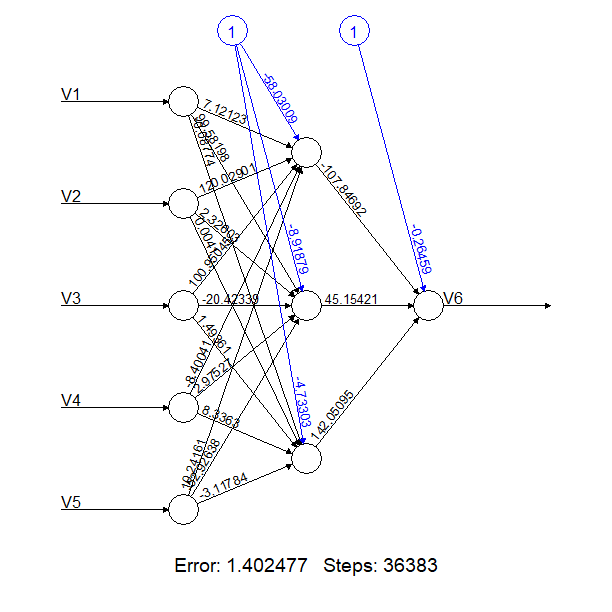}
 \caption{A neural network with 3 hidden nodes that we constructed for the analysis of the full five feature double perovskite data of Appendix \ref{Sec:D}. The 3-fold cross validated accuracy for this data set is 0.88. The optimized parameters are provided in table \ref{3-nod-neural-function}.}
  \label{fig 3-node-function}
 \end{figure}

In this Appendix, we provide technical details underlying the neural networks employed in Section \ref{Section:5feature}. In Figures (\ref{fig 5-node-function}, \ref{fig 3-node-function}), we constructed two single hidden layer neural nets with, respectively, five and three neurons in the hidden layer. The five hidden neuron net of Figure \ref{fig 5-node-function} yielded a 3-fold cross validation accuracy for the double perovskite data  (Appendix \ref{Sec:D}) equal to 0.89; the corresponding weights are shown in table \ref{5-nod-neural-function}. The simpler three hidden neuron model of 
Figure \ref{fig 3-node-function} (with the parameters given in table \ref{3-nod-neural-function}) exhibited a nearly identical average 3-fold cross validation accuracy (0.88).  In Table \ref{Screened Predictions} we provide predictions for the stability of the screened candidate compounds (Appendix \ref{Sec:D}) when combining these two neural net predictive functions. The two columns labelled ``N-N-3'' and ``N-N-5'' list the stability (``1'') or instability
(``0'') predictions of the three- and five- hidden neuron networks for the individual double perovskites. These neural networks were training on the full data set of Appendix \ref{Sec:D} (no data were removed for cross-validation purposes).

\begin{table*}[htbp]
  \centering
  \caption{The below are the optimized parameters  (the weights $w^{(\alpha-1)}_{kk'}$ and ``intercepts'' $c^{(\alpha -1)}_{k}$ (see Eq. (\ref{ek}))) defining the neural network of Figure \ref{fig 5-node-function}. The 3-fold cross validation accuracy for the double perovskite data of Appendix \ref{Sec:D} is 0.89.}
    \begin{tabular}{| c | c |}
\hline
Nodes& Weights  \\
\hline
    Intercept.to.1layhid1   &  0.322486349496      \\
    V2.to.1layhid1        &   -0.689027834161  \\
    V3.to.1layhid1  &         -1.881094208765  \\
    V4.to.1layhid1    &       -1.283319229033  \\
    V5.to.1layhid1      &     -1.760572408905  \\
    V6.to.1layhid1        &    4.256473574810  \\
\hline
    Intercept.to.1layhid2 &    1.527796108323     \\
    V2.to.1layhid2            &3.713727732068  \\
    V3.to.1layhid2 &         -15.955311688727  \\
    V4.to.1layhid2   &         0.158374107691 \\
    V5.to.1layhid2     &       2.688213052680 \\
    V6.to.1layhid2       &     1.672333791950  \\
\hline
    Intercept.to.1layhid3&    -0.611631744186   \\
    V2.to.1layhid3           & 0.332633208229 \\
    V3.to.1layhid3 &           3.427019036381 \\
    V4.to.1layhid3   &        -0.959850896562 \\
    V5.to.1layhid3     &       0.283990792640 \\
    V6.to.1layhid3       &    -0.108144207645 \\
\hline
    Intercept.to.1layhid4&   -15.881926385322   \\
    V2.to.1layhid4&          -29.081443388777  \\
    V3.to.1layhid4  &       -222.081995317924   \\
    V4.to.1layhid4    &      246.801133531051   \\
    V5.to.1layhid4      &    -13.904339124611  \\
    V6.to.1layhid4        &   99.991540326203  \\
\hline
    Intercept.to.1layhid5 &    0.782569281777   \\
    V2.to.1layhid5&           -4.313479926015 \\
    V3.to.1layhid5  &         -2.743165886209 \\
    V4.to.1layhid5    &       -1.302274139636 \\
    V5.to.1layhid5      &     -0.045243735295 \\
    V6.to.1layhid5        &    0.999617101153 \\
\hline
    Intercept.to.V1          &-0.51987838479\\
    1layhid.1.to.V1 &       -184.781490031687   \\
    1layhid.2.to.V1   &      172.889349939525   \\
    1layhid.3.to.V1     &     28.260610237903  \\
    1layhid.4.to.V1       &   22.203951059008  \\
    1layhid.5.to.V1         &-42.817893264326  \\
\hline
\hline
    \end{tabular}%
  \label{5-nod-neural-function}%
\end{table*}%

\begin{table*}[htbp]
  \centering
  \caption{The optimized parameters- the weights $w^{(\alpha-1)}_{kk'}$ and ``intercepts'' $c^{(\alpha -1)}_{k}$ (see Eq. (\ref{ek})) in the neural network of Figure \ref{fig 3-node-function}. The 3-fold cross validation accuracy for the double perovskite data  of Appendix \ref{Sec:D}  is 0.88.}
    \begin{tabular}{|c|c|}
\hline
Nodes& Weights  \\
\hline
    Intercept.to.1layhid1&    -4.0816259810747 \\
    V2.to.1layhid1&           -1.7554149769600 \\
    V3.to.1layhid1  &         -0.5309951002429 \\
    V4.to.1layhid1    &        2.0301683064776 \\
    V5.to.1layhid1      &      9.8417366048861 \\
    V6.to.1layhid1        &   -4.3782850399626 \\
\hline
    Intercept.to.1layhid2 &   -0.0008847242465 \\
    V2.to.1layhid2&            2.7668989847385 \\
    V3.to.1layhid2  &         -5.4616040061074 \\
    V4.to.1layhid2    &       -0.1471232362196 \\
    V5.to.1layhid2      &     -2.9824586267727 \\
    V6.to.1layhid2        &    1.4124389665471 \\
\hline
    Intercept.to.1layhid3 &    2.0014164526234 \\
    V2.to.1layhid3&           -8.1217363487714 \\
    V3.to.1layhid3  &         -0.7281802203924 \\
    V4.to.1layhid3    &        1.5240379652984 \\
    V5.to.1layhid3      &      0.1158046947227 \\
    V6.to.1layhid3        &    0.1549360633580 \\
\hline
    Intercept.to.V1         &-21.0272192266685 \\
    1layhid.1.to.V1&          73.4894503885432 \\
    1layhid.2.to.V1  &        51.6976458939264 \\
    1layhid.3.to.V1    &    -178.5299735691168 \\
\hline
\hline

    \end{tabular}%

  \label{3-nod-neural-function}%
\end{table*}%

\newpage

\section{Specific SRVM-Gaussian functions used for the five predictions of stable double perovskites}
\label{Sec:F}

In \cref{SRVM1,SRVM2,SRVM3,SRVM4,SRVM5,SRVM6,SRVM7,SRVM8,SRVM9}, we provide the parameters underlying nine of the SRVM Gaussian functions that we employed for the predictions of Section \ref{Section:5feature}. In general, given any new compound with the five known features described in Section \ref{Section:5feature}, we may substitute these features into the functions of Eq. (\ref{GaussG}). These functions were found using all of the double perovskite data of Appendix  \ref{Sec:D} (no data were removed for cross-validation purposes). 

\begin{table*}[htbp]
  \centering
  \caption{The below are the specific parameters defining the first replica function $G_{a=1}$ (see Eq. (\ref{kernelG})) used for the analysis of the five-feature double perovskite data of Appendix \ref{Sec:D} 
  (see Section \ref{Section:5feature}). The five parameters define a 5-dimensional Euclidean space $\vec{v}$. The values $v^{i}_{j1}$ (with $i=1,2, 3,4, 5$) are the Euclidean components of the $R=17$ randomly chosen anchor points $\vec{v}_{j}$ in this five dimensional space that appear in Eq. (\ref{kernelG}). The constant $c_{j1}$ is the relative weight of the individual Gaussians in the sum of Eq. (\ref{kernelG}).} 
    \begin{tabular}{|r|r|r|r|r|r|r|}
\hline

{j}&{$c_{j1}$} & {$v^1_{j1}$} & {$v^2_{j1}$} & {$v^3_{j1}$} &{$v^4_{j1}$} & {$v^5_{j1}$} \\
\hline
\hline
    1     & 42.66265 & 0.301224 & 0.187041 & 0.303544 & -0.17677 & -0.09425 \\
\hline
    2     & -52.9432 & 0.511416 & -0.23729 & -0.0682 & 0.750532 & 0.04155 \\
\hline
    3     & -208.178 & -0.83097 & -0.67721 & -0.05884 & -0.96602 & 0.744377 \\
\hline
    4     & -800.605 & -0.3761 & 0.739197 & -0.49887 & -0.35582 & -0.57134 \\
\hline
    5     & 216.1146 & -0.17156 & -0.40313 & -0.76467 & -0.37163 & 0.262623 \\
\hline
    6     & 5.018424 & -0.31824 & 0.699813 & 0.265285 & -0.48924 & 0.495064 \\
\hline
    7     & 998.4435 & -0.75582 & 0.942943 & -0.6613 & -0.13563 & -0.76752 \\
\hline
    8     & 96.36753 & 0.7201 & 0.15729 & -0.37134 & 0.67279 & -0.35576 \\
\hline
    9     & -3.84684 & 0.133245 & -0.97737 & 0.835982 & 0.896302 & 0.223061 \\
\hline
    10    & 152.6827 & 0.948785 & 0.544076 & -0.01372 & -0.70999 & 0.025669 \\
\hline
    11    & 137.2292 & 0.398505 & -0.95262 & 0.401984 & 0.820616 & -0.6941 \\
\hline
    12    & 89.95792 & -0.41978 & 0.510591 & 0.214192 & 0.983804 & -0.94872 \\
\hline
    13    & -63.9454 & -0.63401 & 0.405116 & -0.65393 & 0.509293 & -0.59201 \\
\hline
    14    & -75.0786 & 0.404624 & 0.361196 & 0.565233 & -0.5864 & -0.02424 \\
\hline
    15    & 925.4074 & -0.72941 & -0.26476 & -0.35278 & -0.81482 & -0.85427 \\
\hline
    16    & -2.86607 & -0.52848 & -0.23627 & -0.80696 & 0.057165 & 0.42997 \\
\hline
    17    & -406.901 & 0.886022 & 0.342068 & -0.5412 & -0.32859 & -0.55296 \\

   \hline
\hline
 \end{tabular}%

  \label{SRVM1}%

\end{table*}%

  \begin{table*}[htbp]
    \centering
      \caption{Parameters defining the second replica function $G_{a=2}$ (Eq. (\ref{kernelG})) used for the analysis of the five-feature double perovskite data of Appendix \ref{Sec:D} (see Section \ref{Section:5feature}).}
    \begin{tabular}{|r|r|r|r|r|r|r|}
\hline

{j}&{$c_{j2}$} & {$v^1_{j2}$} & {$v^2_{j2}$} & {$v^3_{j2}$} &{$v^4_{j2}$} & {$v^5_{j2}$} \\
\hline

      \hline
      1  & 85.30229 & -0.88458 & 0.992656 & 0.887611 & -0.62213 & 0.232414 \\
      \hline
      2  & 8.977308 & 0.637475 & 0.275766 & -0.75056 & 0.734639 & 0.687448 \\
      \hline
      3  & -328.5 & -0.78661 & -0.41947 & -0.9392 & -0.3905 & 0.534033 \\
      \hline
      4  & 9.86238 & -0.83599 & 0.945115 & -0.29954 & 0.946104 & 0.01971 \\
      \hline
      5  & 294.6979 & -0.84643 & -0.29492 & -0.19101 & -0.37982 & 0.463973 \\
      \hline
      6  & 183.9229 & 0.943353 & -0.97148 & 0.052025 & 0.273835 & -0.91011 \\
      \hline
      7  & 270.3541 & -0.01669 & -0.25925 & -0.636 & -0.34293 & -0.42158 \\
      \hline
      8  & -83.7778 & 0.20874 & -0.92793 & -0.16796 & -0.1622 & -0.32029 \\
      \hline
      9  & -4.74883 & 0.511514 & -0.27868 & 0.929049 & 0.738096 & 0.085939 \\
      \hline
      10 & -86.2916 & -0.26816 & 0.889562 & -0.68677 & -0.34934 & 0.517183 \\
      \hline
      11 & -25.5756 & 0.978946 & -0.12299 & -0.10717 & 0.127257 & 0.07633 \\
      \hline
      12 & 90.3408 & 0.158801 & 0.034894 & 0.691243 & 0.228302 & -0.40564 \\
      \hline
      13 & 26.40855 & 0.958969 & 0.415524 & 0.523732 & 0.16636 & -0.42367 \\
      \hline
      14 & -21.1956 & -0.42286 & 0.099672 & 0.268904 & 0.226927 & -0.16622 \\
      \hline
      15 & -768.363 & -0.79922 & -0.49003 & -0.26705 & 0.434031 & -0.91316 \\
      \hline
      16 & -42.5849 & -0.02577 & 0.032557 & 0.917936 & 0.004445 & 0.280619 \\
      \hline
      17 & -28.9923 & 0.382228 & 0.518098 & 0.325924 & -0.35554 & -0.11598 \\
      \hline
\hline
      \end{tabular}%
    \label{SRVM2}%
  \end{table*}%

  \begin{table*}[htbp]
    \centering
      \caption{Parameters defining the third replica function $G_{a=3}$ (Eq. (\ref{kernelG})) used for the analysis of the five-feature double perovskite data of Appendix \ref{Sec:D} (see Section \ref{Section:5feature}).}
    \begin{tabular}{|r|r|r|r|r|r|r|}
\hline

{j}&{$c_{j3}$} & {$v^1_{j3}$} & {$v^2_{j3}$} & {$v^3_{j3}$} &{$v^4_{j3}$} & {$v^5_{j3}$} \\
\hline

      \hline
   1  & 1.927016 & -0.11434 & -0.10393 & 0.172173 & 0.992977 & -0.97304 \\
      \hline
      2  & -444.309 & -0.23341 & 0.877449 & 0.64954 & -0.82771 & -0.7032 \\
      \hline
      3  & -27.1444 & 0.609797 & 0.133753 & 0.775568 & -0.58632 & -0.35086 \\
      \hline
      4  & 0.938015 & 0.262505 & 0.402545 & -0.19557 & 0.919419 & 0.126204 \\
      \hline
      5  & 2.233768 & 0.91498 & -0.22567 & 0.272144 & 0.507126 & 0.603167 \\
      \hline
      6  & -9.07432 & 0.262141 & -0.48964 & 0.736083 & 0.462494 & -0.61093 \\
      \hline
      7  & -70.846 & -0.8542 & -0.23374 & 0.910938 & -0.75779 & 0.162057 \\
      \hline
      8  & -31.2954 & -0.5752 & 0.050162 & -0.80534 & -0.94189 & 0.302472 \\
      \hline
      9  & -5.5822 & -0.44971 & -0.15624 & 0.26295 & 0.823308 & 0.936288 \\
      \hline
      10 & 2.569315 & 0.88697 & 0.779808 & 0.388021 & 0.960101 & -0.13499 \\
      \hline
      11 & 24.99833 & 0.903649 & 0.963347 & 0.760198 & -0.66875 & 0.715841 \\
      \hline
      12 & -3.41891 & -0.31179 & 0.86282 & -0.11072 & 0.359071 & -0.19603 \\
      \hline
      13 & 58.32798 & -0.23517 & 0.266026 & 0.505038 & 0.19355 & -0.70428 \\
      \hline
      14 & 9.199793 & -0.95239 & -0.48032 & -0.8379 & -0.4091 & 0.543475 \\
      \hline
      15 & -299.164 & 0.503709 & 0.675397 & -0.07433 & -0.48666 & -0.89583 \\
      \hline
      16 & 933.926 & -0.33648 & 0.052186 & -0.06967 & -0.90551 & -0.7389 \\
      \hline
      17 & 10.13302 & 0.562424 & 0.613796 & -0.18703 & -0.48355 & -0.65861 \\
\hline
      \end{tabular}%
    \label{SRVM3}%
  \end{table*}%

  \begin{table*}[htbp]
    \centering
      \caption{Parameters defining the second replica function $G_{a=4}$ (Eq. (\ref{kernelG})) used for the analysis of the five-feature double perovskite data of Appendix \ref{Sec:D} (see Section \ref{Section:5feature}).}
    \begin{tabular}{|r|r|r|r|r|r|r|}
\hline

{j}&{$c_{j4}$} & {$v^1_{j4}$} & {$v^2_{j4}$} & {$v^3_{j4}$} &{$v^4_{j4}$} & {$v^5_{j4}$} \\
\hline

      \hline

      1  & -168.171 & 0.171587 & -0.20569 & -0.71739 & -0.75953 & 0.630486 \\
      \hline
      2  & 78.45192 & 0.084468 & 0.112939 & 0.109303 & -0.0706 & -0.46403 \\
      \hline
      3  & 51.25864 & 0.51755 & -0.49657 & 0.920592 & -0.62878 & 0.311399 \\
      \hline
      4  & -32.5909 & -0.99869 & -0.12392 & 0.86037 & -0.02546 & 0.858242 \\
      \hline
      5  & -16.5471 & -0.17723 & -0.96499 & 0.277164 & 0.854763 & 0.892638 \\
      \hline
      6  & 43.83384 & -0.26173 & 0.076033 & -0.31144 & 0.324417 & -0.31024 \\
      \hline
      7  & 163.0846 & 0.104459 & -0.34328 & 0.673199 & -0.86735 & 0.236553 \\
      \hline
      8  & 30.02841 & -0.95296 & -0.81587 & -0.59089 & 0.734958 & 0.425885 \\
      \hline
      9  & 1003.332 & -0.02897 & -0.56445 & -0.73684 & -0.90759 & 0.750804 \\
      \hline
      10 & 1.019157 & -0.18472 & 0.977644 & 0.922615 & 0.315497 & 0.763191 \\
      \hline
      11 & -168.057 & -0.32625 & 0.145492 & -0.26187 & 0.124391 & -0.47039 \\
      \hline
      12 & 48.87624 & 0.312238 & 0.323741 & -0.11828 & 0.413613 & -0.77225 \\
      \hline
      13 & -213.597 & 0.23861 & -0.24927 & 0.83616 & -0.95324 & 0.354061 \\
      \hline
      14 & 104.6992 & -0.85023 & 0.556416 & 0.017257 & -0.59031 & -0.0739 \\
      \hline
      15 & 0.240399 & 0.975459 & 0.066515 & -0.12161 & -0.5274 & 0.61754 \\
      \hline
      16 & -207.508 & 0.086153 & 0.22469 & -0.45578 & -0.80889 & 0.059676 \\
      \hline
      17 & -73.7415 & 0.35169 & -0.22372 & 0.205103 & -0.36355 & -0.39363 \\

\hline
      \end{tabular}%
    \label{SRVM4}%
  \end{table*}%

  \begin{table*}[htbp]
    \centering
      \caption{Parameters defining the fifth replica function $G_{a=5}$ (Eq. (\ref{kernelG})) used for the analysis of the five-feature double perovskite data of Appendix \ref{Sec:D} (see Section \ref{Section:5feature}).}
    \begin{tabular}{|r|r|r|r|r|r|r|}
\hline

{j}&{$c_{j5}$} & {$v^1_{j5}$} & {$v^2_{j5}$} & {$v^3_{j5}$} &{$v^4_{j5}$} & {$v^5_{j5}$} \\
\hline

      \hline

    1  & 3.16044 & 0.325102 & 0.921106 & -0.66222 & 0.725417 & 0.180718 \\
      \hline
      2  & 39.66355 & 0.469994 & 0.888736 & 0.811437 & -0.80195 & 0.221599 \\
      \hline
      3  & -378.895 & -0.41361 & 0.398454 & -0.20969 & -0.75904 & 0.014306 \\
      \hline
      4  & -502.312 & 0.666659 & -0.02381 & 0.779051 & -0.29487 & -0.81904 \\
      \hline
      5  & -68.583 & 0.604385 & 0.152124 & -0.81912 & -0.3502 & 0.361085 \\
      \hline
      6  & 268.5355 & -0.32703 & 0.338031 & -0.06398 & -0.78351 & -0.94651 \\
      \hline
      7  & 827.0132 & 0.898839 & -0.60506 & -0.48349 & -0.61651 & -0.1769 \\
      \hline
      8  & 32.14631 & 0.686537 & 0.241474 & 0.071333 & -0.31396 & 0.683823 \\
      \hline
      9  & 1.762983 & -0.37029 & 0.792009 & 0.934086 & 0.171015 & 0.486793 \\
      \hline
      10 & 28.76404 & 0.886283 & -0.19376 & 0.957817 & -0.09277 & 0.554183 \\
      \hline
      11 & -1270.95 & -0.63349 & -0.6386 & -0.04373 & 0.177619 & -0.62642 \\
      \hline
      12 & -452.472 & 0.811499 & -0.54878 & -0.25592 & -0.38987 & -0.20785 \\
      \hline
      13 & 221.533 & 0.523579 & -0.89885 & -0.61481 & -0.32429 & -0.32046 \\
      \hline
      14 & 1194.729 & -0.22541 & -0.45507 & 0.310312 & -0.22194 & -0.70526 \\
      \hline
      15 & 24.56107 & -0.60659 & -0.06317 & -0.68843 & 0.794131 & 0.05439 \\
      \hline
      16 & -78.818 & -0.4356 & -0.08225 & 0.789817 & -0.4306 & 0.339743 \\
      \hline
      17 & 359.2178 & -0.91212 & -0.25995 & -0.29756 & 0.217808 & -0.20629 \\
\hline
      \end{tabular}%
    \label{SRVM5}%
  \end{table*}%

  \begin{table*}[htbp]
    \centering
      \caption{Parameters defining the sixth replica function $G_{a=6}$ (Eq. (\ref{kernelG})) used for the analysis of the five-feature double perovskite data of Appendix \ref{Sec:D} (see Section \ref{Section:5feature}).}
    \begin{tabular}{|r|r|r|r|r|r|r|}
\hline

{j}&{$c_{j6}$} & {$v^1_{j6}$} & {$v^2_{j6}$} & {$v^3_{j6}$} &{$v^4_{j6}$} & {$v^5_{j6}$} \\
\hline

      \hline

    1  & 571.6779 & 0.692162 & -0.60735 & -0.99309 & -0.37182 & -0.31621 \\
      \hline
      2  & -30.5383 & -0.5543 & -0.27285 & 0.185263 & 0.505977 & 0.23979 \\
      \hline
      3  & -518.747 & -0.83196 & -0.50696 & 0.748583 & -0.95675 & 0.513509 \\
      \hline
      4  & 165.2649 & -0.94751 & 0.085417 & 0.729866 & 0.122304 & -0.41937 \\
      \hline
      5  & 56.32682 & 0.743902 & 0.237139 & 0.365586 & -0.63276 & 0.476396 \\
      \hline
      6  & 16.29653 & 0.582831 & -0.07971 & -0.1417 & 0.446608 & 0.369132 \\
      \hline
      7  & 173.4209 & -0.60379 & 0.248175 & -0.46902 & 0.555994 & 0.940274 \\
      \hline
      8  & -43.4043 & 0.036049 & 0.394897 & -0.69574 & 0.767074 & 0.935861 \\
      \hline
      9  & -69.5107 & 0.301979 & 0.735006 & -0.17429 & 0.361781 & 0.88549 \\
      \hline
      10 & 19.85028 & 0.49414 & -0.36807 & 0.468725 & -0.78611 & 0.351469 \\
      \hline
      11 & -146.916 & -0.9605 & -0.10731 & -0.24537 & -0.35603 & 0.622912 \\
      \hline
      12 & -1.36709 & 0.160567 & -0.5578 & 0.905541 & 0.972754 & 0.723689 \\
      \hline
      13 & -332.229 & 0.607216 & 0.036596 & 0.28232 & -0.67553 & -0.56247 \\
      \hline
      14 & 115.0979 & 0.525087 & 0.924594 & -0.39539 & 0.621958 & -0.00386 \\
      \hline
      15 & -73.8918 & 0.640419 & 0.223398 & -0.89091 & 0.395518 & 0.411668 \\
      \hline
      16 & -341.457 & -0.69946 & 0.911862 & -0.6587 & 0.320826 & -0.06014 \\
      \hline
      17 & 114.7313 & -0.26991 & 0.628984 & -0.42824 & -0.21141 & 0.302937 \\
\hline
\hline
      \end{tabular}%
    \label{SRVM6}%
  \end{table*}%

  \begin{table*}[htbp]
    \centering
      \caption{Parameters defining the seventh replica function $G_{a=7}$ (Eq. (\ref{kernelG})) used for the analysis of the five-feature double perovskite data of Appendix \ref{Sec:D} (see Section \ref{Section:5feature}).}
    \begin{tabular}{|r|r|r|r|r|r|r|}
\hline

{j}&{$c_{j7}$} & {$v^1_{j7}$} & {$v^2_{j7}$} & {$v^3_{j7}$} &{$v^4_{j7}$} & {$v^5_{j7}$} \\
\hline

      \hline

   1  & -39.4596 & -0.9051 & -0.6677 & 0.936483 & -0.55319 & 0.028574 \\
      \hline
      2  & -12.4545 & -0.94464 & 0.695545 & 0.465852 & 0.722544 & 0.682123 \\
      \hline
      3  & -75.4599 & -0.65781 & -0.85426 & -0.28401 & 0.814389 & 0.104258 \\
      \hline
      4  & 32.40792 & -0.81338 & 0.233216 & -0.32108 & 0.754617 & 0.874158 \\
      \hline
      5  & -9.08658 & 0.427231 & 0.774879 & -0.88542 & 0.15405 & 0.093796 \\
      \hline
      6  & -12.3737 & 0.395852 & -0.15932 & 0.145294 & 0.221031 & -0.79238 \\
      \hline
      7  & 1.603347 & 0.670894 & 0.422267 & 0.039185 & -0.16679 & 0.413948 \\
      \hline
      8  & -18.9682 & -0.02131 & 0.570193 & 0.247668 & -0.30061 & 0.867162 \\
      \hline
      9  & -15.4713 & 0.888067 & 0.423962 & -0.79956 & 0.814243 & -0.64256 \\
      \hline
      10 & 6.118748 & 0.987868 & 0.532833 & -0.1118 & 0.725714 & -0.01262 \\
      \hline
      11 & -59.5019 & -0.04564 & -0.95266 & 0.284183 & 0.352254 & -0.43871 \\
      \hline
      12 & 42.48926 & -0.30867 & 0.888436 & 0.314393 & -0.50268 & -0.08076 \\
      \hline
      13 & 201.3999 & -0.32932 & -0.6991 & -0.33236 & -0.48805 & 0.189914 \\
      \hline
      14 & 2.929333 & -0.15359 & 0.908867 & 0.907969 & 0.808864 & 0.29167 \\
      \hline
      15 & -238.777 & -0.30315 & 0.972761 & -0.34049 & -0.95493 & 0.285743 \\
      \hline
      16 & 17.01863 & 0.345794 & -0.61359 & 0.473074 & 0.974844 & -0.52963 \\
      \hline
      17 & 29.40171 & -0.05927 & 0.097739 & 0.302651 & -0.84021 & -0.04148 \\
\hline
\hline
      \end{tabular}%
    \label{SRVM7}%
  \end{table*}%

  \begin{table*}[htbp]
    \centering
      \caption{Parameters defining the eighth replica function $G_{a=8}$ (Eq. (\ref{kernelG})) used for the analysis of the five-feature double perovskite data of Appendix \ref{Sec:D} (see Section \ref{Section:5feature}).}
    \begin{tabular}{|r|r|r|r|r|r|r|}
\hline

{j}&{$c_{j8}$} & {$v^1_{j8}$} & {$v^2_{j8}$} & {$v^3_{j8}$} &{$v^4_{j8}$} & {$v^5_{j8}$} \\
\hline

      \hline

      1  & 88.84422 & 0.571364 & 0.119427 & -0.24771 & 0.691267 & -0.2988 \\
      \hline
      2  & 174.2853 & -0.06807 & -0.6586 & 0.498736 & -0.84785 & 0.232166 \\
      \hline
      3  & 10.13218 & 0.908712 & 0.607267 & 0.607713 & -0.20869 & 0.934951 \\
      \hline
      4  & -22.0976 & 0.240986 & -0.09054 & 0.256491 & 0.850468 & -0.53986 \\
      \hline
      5  & 25.84403 & -0.39764 & 0.271124 & -0.12159 & -0.0905 & 0.771413 \\
      \hline
      6  & -51.0625 & 0.967531 & 0.141153 & 0.725322 & -0.92992 & 0.814647 \\
      \hline
      7  & 7.698623 & 0.666298 & -0.07534 & 0.42188 & 0.575636 & -0.31152 \\
      \hline
      8  & 126.7194 & -0.33566 & -0.46319 & -0.85048 & 0.853413 & 0.025779 \\
      \hline
      9  & 93.31989 & -0.9104 & 0.313965 & -0.19069 & -0.45344 & -0.72681 \\
      \hline
      10 & -19.3418 & 0.382058 & 0.067816 & 0.126465 & 0.684678 & 0.230108 \\
      \hline
      11 & -110.187 & -0.73836 & 0.962948 & 0.336621 & 0.140849 & 0.306034 \\
      \hline
      12 & -159.445 & -0.05326 & 0.152936 & -0.69868 & 0.151915 & -0.63191 \\
      \hline
      13 & -389.25 & 0.394415 & -0.30683 & -0.77989 & 0.671314 & -0.4532 \\
      \hline
      14 & -121.822 & -0.38425 & -0.11579 & -0.02001 & -0.51612 & -0.51162 \\
      \hline
      15 & 172.2679 & -0.91631 & 0.982199 & 0.591577 & 0.326072 & -0.15971 \\
      \hline
      16 & 315.2304 & 0.211636 & -0.44996 & -0.79067 & 0.561918 & -0.47945 \\
      \hline
      17 & -5.18003 & 0.601396 & 0.116586 & -0.16331 & -0.20782 & 0.798875 \\
\hline
\hline
      \end{tabular}%
    \label{SRVM8}%
  \end{table*}%

  \begin{table*}[htbp]
    \centering
      \caption{Parameters defining the ninth replica function $G_{a=9}$ (Eq. (\ref{kernelG})) used for the analysis of the five-feature double perovskite data of Appendix \ref{Sec:D} (see Section \ref{Section:5feature}).}
    \begin{tabular}{|r|r|r|r|r|r|r|}
\hline

{j}&{$c_{j9}$} & {$v^1_{j9}$} & {$v^2_{j9}$} & {$v^3_{j9}$} &{$v^4_{j9}$} & {$v^5_{j9}$} \\
\hline

      \hline

      1  & 88.5165 & 0.904113 & -0.88306 & -0.01856 & -0.79141 & 0.780351 \\
      \hline
      2  & 7.032538 & -0.83204 & 0.412497 & 0.626756 & 0.823564 & -0.43631 \\
      \hline
      3  & -11.1869 & -0.55311 & 0.983024 & -0.64543 & 0.81566 & -0.07414 \\
      \hline
      4  & -67.643 & 0.9374 & -0.72864 & -0.68673 & 0.27049 & -0.29005 \\
      \hline
      5  & 1.599603 & -0.7853 & -0.2384 & 0.308389 & 0.160927 & 0.685789 \\
      \hline
      6  & -40.4704 & -0.24336 & 0.112004 & 0.84368 & -0.70199 & 0.449621 \\
      \hline
      7  & 2.803726 & 0.13383 & -0.40025 & 0.582284 & 0.428634 & 0.506175 \\
      \hline
      8  & 18.07024 & 0.845427 & 0.303671 & 0.604487 & -0.49294 & -0.10087 \\
      \hline
      9  & 3.189425 & 0.486972 & 0.552616 & 0.80801 & 0.752221 & 0.181653 \\
      \hline
      10 & -55.7322 & 0.607613 & -0.32863 & 0.539215 & -0.09131 & 0.449838 \\
      \hline
      11 & -19.9 & -0.60879 & -0.42733 & -0.52628 & 0.164851 & 0.973381 \\
      \hline
      12 & -206.011 & 0.347209 & -0.99434 & 0.109674 & -0.50631 & -0.53011 \\
      \hline
      13 & -38.7461 & 0.15525 & -0.59238 & -0.5731 & -0.82764 & 0.863834 \\
      \hline
      14 & -7.04589 & 0.0938 & 0.25432 & 0.484615 & 0.23171 & -0.36139 \\
      \hline
      15 & 65.80984 & -0.1512 & -0.17237 & 0.905818 & -0.51815 & -0.37394 \\
      \hline
      16 & 53.77325 & 0.768871 & -0.44064 & 0.394008 & -0.06863 & 0.351904 \\
      \hline
      17 & 106.224 & -0.36221 & -0.8989 & -0.30546 & 0.069453 & 0.544122 \\
\hline
\hline
      \end{tabular}%
    \label{SRVM9}%
  \end{table*}%


\begin{thebibliography}{9}
\bibitem{explain-Gauss} A trivial extension of these functions is one in which the Cartesian coordinate axis are 
rotated and dilated (allowing for a sum of general multivariate Gaussian distributions with a non-diagonal kernel). For simplicity, in the current
work, we will contend ourselves with the simple diagonal covariant form of Eq. (\ref{GaussG}).
\end{thebibliography}
\end{document}